%% file: mainfile.tex
\documentclass[conference,compsoc]{IEEEtran}

\newcommand{\myparatight}[1]{\smallskip\noindent{\bf {#1}:}~}

\usepackage[skip=0pt]{caption}

\usepackage{amsthm}
\usepackage{amsmath}
\usepackage{amssymb}
\usepackage{shortcut}
\usepackage{adjustbox}
\usepackage{booktabs}
\usepackage{array} 
\usepackage{multirow}
\usepackage{graphics}
\usepackage{graphicx}
\usepackage{algorithm}
\usepackage{algorithmic}
\usepackage{amsfonts}
\usepackage{color, url}

\usepackage{balance}

\usepackage[hidelinks]{hyperref}
\hypersetup{
    colorlinks,
    linkcolor={red!50!black},
    citecolor={green!50!black},
    urlcolor={black!80!black}
}

\usepackage{caption}
\usepackage{subcaption}

\usepackage{tabto}
\usepackage{xcolor, colortbl}
\usepackage{enumitem}
\usepackage{bbm}
\usepackage{dsfont}
\usepackage{tcolorbox}
\usepackage{hhline}

\usepackage{bm}
\usepackage[mathscr]{eucal}
\usepackage[normalem]{ulem}
\useunder{\uline}{\ul}{}
\usepackage{bigstrut,multirow,rotating}
\usepackage{placeins}
\usepackage{pgfplots}
\pgfplotsset{compat=1.17}
\makeatletter
\@namedef{r@tocindent4}{0pt}
\makeatother

\usepackage{color, colortbl}
\definecolor{greyL}{RGB}{230,248,255}

\definecolor{customblue}{HTML}{243885}

\usepackage{xspace}
\newcommand{\alg}{RAGOrigin\xspace}

\usepackage{cite}

\newcommand{\argmax}{\operatornamewithlimits{argmax}}

\usepackage[T1]{fontenc}
\usepackage[hidelinks]{hyperref}

\usepackage[misc,geometry]{ifsym}

\usepackage{eso-pic}
\usepackage[absolute,overlay]{textpos}

\begin{document}

\AddToShipoutPictureBG*{%
	\AtPageUpperLeft{%
		\setlength\unitlength{1in}%
		\hspace*{\dimexpr0.5\paperwidth\relax}
		\makebox(0,-0.75)[c]{To appear in the IEEE Symposium on Security and Privacy, 2026}%
}}

\title{Who Taught the Lie? Responsibility Attribution for Poisoned Knowledge in Retrieval-Augmented Generation}

\author{ 
\IEEEauthorblockN{
Baolei Zhang\IEEEauthorrefmark{2}\textsuperscript{*},
Haoran Xin\IEEEauthorrefmark{2}\textsuperscript{*},
Yuxi Chen\IEEEauthorrefmark{3},
Zhuqing Liu\IEEEauthorrefmark{4}, 
Biao Yi\IEEEauthorrefmark{2},\\
Tong Li\IEEEauthorrefmark{2},
Lihai Nie\IEEEauthorrefmark{2}\textsuperscript{\Letter},
Zheli Liu\IEEEauthorrefmark{2},
Minghong Fang\IEEEauthorrefmark{5}\textsuperscript{\Letter}
}
\IEEEauthorblockA{
\IEEEauthorrefmark{2}CS\&CCS, DISSec, AAIS, Nankai University,\\
\{zhangbaolei, haoranxin, yibiao\}@mail.nankai.edu.cn, \{tongli, NLH, liuzheli\}@nankai.edu.cn;\\
\IEEEauthorrefmark{3}Independent Researcher, chenyuxi030810@gmail.com;\\
\IEEEauthorrefmark{4}University of North Texas, zhuqing.liu@unt.edu;\\
\IEEEauthorrefmark{5}University of Louisville, minghong.fang@louisville.edu}
}
\maketitle

\makeatletter
\renewcommand{\@makefntext}[1]{\noindent#1}
\makeatother
\footnotetext{\textsuperscript{*} Equal Contribution.}
\footnotetext{\textsuperscript{\Letter} Corresponding Authors.}

\begin{abstract}

Retrieval-Augmented Generation (RAG) integrates external knowledge into large language models to improve response quality. However, recent work has shown that RAG systems are highly vulnerable to poisoning attacks, where malicious texts are inserted into the knowledge database to influence model outputs. While several defenses have been proposed, they are often circumvented by more adaptive or sophisticated attacks.

This paper presents RAGOrigin, a black-box responsibility attribution framework designed to identify which texts in the knowledge database are responsible for misleading or incorrect generations. Our method constructs a focused attribution scope tailored to each misgeneration event and assigns a responsibility score to each candidate text by evaluating its retrieval ranking, semantic relevance, and influence on the generated response. The system then isolates poisoned texts using an unsupervised clustering method. We evaluate RAGOrigin across seven datasets and fifteen poisoning attacks, including newly developed adaptive poisoning strategies and multi-attacker scenarios. Our approach outperforms existing baselines in identifying poisoned content and remains robust under dynamic and noisy conditions. These results suggest that RAGOrigin provides a practical and effective solution for tracing the origins of corrupted knowledge in RAG systems.
Our code is available at: \url{https://github.com/zhangbl6618/RAG-Responsibility-Attribution}

\end{abstract}

\input{introduction}
\input{related}

\input{threatModel}

\input{method}

\input{experiments}

\input{discussion}

\input{conclusion}

\bibliographystyle{IEEEtran}
\bibliography{IEEEabrv,refs}

\input{appendix}

\input{meta_review}

\end{document}

%% file: introduction.tex

\section{Introduction} 
\label{sec:intro}

Large language models (LLMs)~\cite{brown2020language, achiam2023gpt, anil2023palm, dubey2024llama, bai2023qwen, team2024gemma} have demonstrated exceptional capabilities in text generation, but they often lack access to up-to-date or task-specific knowledge. To overcome this limitation, Retrieval-Augmented Generation (RAG)~\cite{karpukhin2020dense, lewis2020retrieval, borgeaud2022improving, chen2024benchmarking, gao2023retrieval} has emerged as a dominant paradigm, allowing LLMs to retrieve relevant texts from external knowledge sources and use them as grounding for generation. In a typical RAG pipeline, a retriever selects the top-$K$ documents based on their similarity to the user question, and the retrieved texts are fed into the LLM alongside the question to produce a response.

However, recent studies~\cite{zou2024poisonedrag,zhang2024hijackrag,shafran2024machine,tan2024glue,xue2024badrag,chaudhari2024phantom} have demonstrated that RAG systems are highly vulnerable to poisoning attacks. Since the knowledge database typically incorporates texts from various sources (such as Wikipedia~\cite{thakur2021beir} and FactoidWiki~\cite{chen2023dense}), the attacker can exploit this by injecting carefully crafted poisoned texts into these sources, thereby contaminating the knowledge database, causing the RAG system to generate attacker-desired target answers for specific target questions. While several defenses have been proposed~\cite{zhong2023poisoning,zou2024poisonedrag,shafran2024machine,xue2024badrag,xiang2024certifiably,asai2023self}, they often prove inadequate against sophisticated attacks. For instance, RobustRAG~\cite{xiang2024certifiably} fails when an attacker poisons more than half of the retrieved texts for a target question.

This paper addresses a fundamental research problem: responsibility attribution in RAG systems. Rather than asking how to prevent poisoning attacks, we ask: {\em who taught the lie?} That is, given a misgeneration event caused by a poisoning attack, can we identify which texts in the knowledge database are most likely responsible for the misgeneration? Motivated by the limitations of existing defenses and the inevitability of sophisticated attacks, we focus on post-attack responsibility attribution. This is a critical first step toward understanding the root causes of model misbehavior, enabling system designers to localize poisoned knowledge, audit data pipelines, and trace attacks back to their sources.

Unfortunately, existing forensics techniques~\cite{cheng2023beagle, hammoudeh2022identifying, jia2024tracing, rose2024utrace} fall short in this setting. These methods typically assume access to model parameters or gradients, which are not available in commercial or black-box deployments. Moreover, responsibility attribution in RAG systems presents unique challenges due to three factors. First, poisoned texts are extremely sparse; only a few entries may be sufficient to trigger malicious behavior, yet they must be identified from among millions of benign documents. Second, the interaction between the retriever and the LLM is highly nonlinear, making it difficult to isolate causal influence from the observed outputs. Third, the knowledge database is often constructed from heterogeneous sources with diverse content distributions, rendering many statistical assumptions brittle and vulnerable to evasion by the attacker.
RAGForensics~\cite{zhang2025traceback}, the most related prior work, identifies poisoned content by detecting direct semantic overlap with generated responses. Extensive experiments show that this approach fails when the connection is intentionally disguised, reducing its reliability in adversarial settings.

To address these challenges, we present \alg, a black-box responsibility attribution framework for RAG systems.
Our approach is based on a key observation: for a poisoning attack to succeed, a poisoned text must satisfy two essential conditions. First, it must rank highly in the retriever’s results for the target question. Second, it must influence the language model to produce the attacker’s desired response. \alg leverages these properties to isolate and identify the texts most likely responsible for misgeneration.

The core of \alg consists of two main components: adaptive scope construction and responsibility attribution. In the first stage, we construct an adaptive attribution scope, which is a focused subset of the knowledge database that is most likely to contain the poisoned texts. To achieve this, we partition the knowledge database into segments of size $K$, ordered by retrieval similarity to the target question. For each segment, we simulate the RAG process by using the segment’s texts as input and observe whether the generated output matches the misgenerated response. This procedure continues until a sufficient number of segments yield divergent outputs. This ensures that the selected scope captures the most relevant and potentially malicious texts.

In the second stage, we attribute responsibility by computing a responsibility score for each text within the attribution scope. This score reflects how likely each text is to have caused the misgeneration. The scoring framework incorporates three complementary signals. The first is embedding similarity, which quantifies how well the text aligns with the user question in the retriever's embedding space.
The second is semantic correlation, which measures how closely the text content matches the user question in terms of semantics.
The third is generation influence, which evaluates the extent to which a text contributes to the LLM producing the incorrect response when included in the input context. These signals are normalized and averaged to form a final responsibility score. A clustering method is then applied to separate poisoned texts from benign ones without requiring labeled data or manually defined thresholds.
Together, these components make \alg an effective and scalable tool for responsibility attribution in RAG systems.

We conduct extensive evaluations of our proposed method across seven datasets, including five widely used question-answering benchmarks, one large-scale dataset containing 16.7 million texts, and one open-ended dataset. Our experiments cover 15 poisoning attacks, including those designed for RAG systems, attacks tailored for agents, and three newly developed adaptive attacks. We also compare against eight state-of-the-art baselines. Furthermore, we explore a range of practical RAG scenarios, such as cases where a single target question is poisoned by multiple attack methods, or where multiple independent attackers launch separate poisoning attempts. We also evaluate conditions involving noisy user feedback and dynamic changes to the knowledge database. Across all settings, \alg consistently identifies the poisoned texts responsible for misgeneration events with high accuracy. Finally, we show that \alg introduces only slight computational overhead and incurs negligible monetary cost, making it suitable for real-world applications.
We believe this work represents a significant step toward making RAG systems more transparent and trustworthy. As RAG deployment expands in high-stakes domains, tools like \alg are essential for understanding and mitigating the impact of poisoned knowledge.

Our contributions are summarized as follows:

\begin{list}{\labelitemi}{\leftmargin=1em \itemindent=-0.0em \itemsep=.2em}

    \item We propose \alg, a novel responsibility attribution system for RAG, which accurately traces the poisoned texts responsible for misgeneration events.
    
    \item Through comprehensive evaluation against 15 poisoning attacks across 7 datasets, we demonstrate \alg's effectiveness with high accuracy and low false positive/negative rates. 
    
    \item 
    We evaluate our \alg and demonstrate its robustness against three adaptive attacks specifically crafted to bypass our proposed responsibility attribution system.

 \end{list}

%% file: related.tex

\section{Preliminaries and Related Work} \label{sec:related}

\subsection{Retrieval-Augmented Generation (RAG)} 
\label{RAG_intro}

Retrieval-Augmented Generation (RAG)~\cite{zou2024poisonedrag, shafran2024machine, zhong2023poisoning, chaudhari2024phantom, deng2024pandora, tan2024glue} enhances the response quality of LLMs by incorporating relevant texts retrieved from an external knowledge database. 
A typical RAG system consists of a knowledge database $\mathcal{D}$, a retriever, and an LLM. Given a user question $q$, the retriever selects relevant texts from $\mathcal{D}$, which are then used by the LLM to generate a final answer. The inference process proceeds in two steps:

\begin{list}{\labelitemi}{\leftmargin=1em \itemindent=-0.0em \itemsep=.2em}

\item \textbf{Step I (Retrieving relevant knowledge):} 
Initially, the retriever encodes the user question $q$ through the question encoder, producing the question embedding $E(q)$. Subsequently, it employs a similarity metric to identify and retrieve the top-$K$ texts most closely related to the question embedding from the knowledge database. The embeddings for these texts are pre-calculated and maintained in a vector database using the question encoder.

\item \textbf{Step II (Knowledge-augmented generation):} 
The top-$K$ texts retrieved as relevant are provided as context to the LLM together with the user question $q$. The LLM integrates this contextual data with the user question to produce the final response.

\end{list}

\subsection{Related Work}

\myparatight{Poisoning attacks to RAG}%
Recent studies have highlighted the growing threat of poisoning attacks on RAG systems~\cite{zou2024poisonedrag,zhang2024hijackrag,shafran2024machine,tan2024glue,xue2024badrag,chaudhari2024phantom}, where the attacker injects crafted texts into the knowledge database to manipulate the system's output for target questions. For instance, PoisonedRAG~\cite{zou2024poisonedrag} splits poisoned content to influence both the retriever and the LLM, using different optimization strategies for black-box and white-box settings. LIAR~\cite{tan2024glue} jointly optimizes for both components in white-box environments. Other attacks like BadRAG~\cite{xue2024badrag} and Phantom~\cite{chaudhari2024phantom} further refine trigger-based poisoning to produce attacker-specified outputs.

\myparatight{Defenses against poisoning attacks to RAG}%
Several defenses~\cite{zhong2023poisoning,zou2024poisonedrag,shafran2024machine,xue2024badrag,xiang2024certifiably,asai2023self} have been proposed to detect and mitigate poisoned texts in RAG systems. Norm-based methods~\cite{zhong2023poisoning} flag texts with unusually large embedding norms, while perplexity-based approaches~\cite{zou2024poisonedrag,shafran2024machine,xue2024badrag} identify outliers using a proxy model. RobustRAG~\cite{xiang2024certifiably} reduces poisoning impact by aggregating outputs generated from individual retrieved texts.

\myparatight{Responsibility attribution}%
Responsibility attribution, also referred to as forensics in some studies~\cite{shan2022poison,cheng2023beagle,hammoudeh2022identifying,jia2024tracing,rose2024utrace}, aims to trace the origin of poisoning attacks after they occur. In centralized learning, the focus is on identifying poisoned training samples that lead to misclassification~\cite{shan2022poison,cheng2023beagle}, often by clustering and filtering out benign examples. In federated learning, the goal shifts to identifying malicious clients responsible for the attack~\cite{jia2024tracing,rose2024utrace}.

\myparatight{Responsibility attribution vs. Defenses}%
Responsibility attribution complements defense by focusing on identifying the root causes of attacks rather than preventing them. Although advanced defenses exist, past work~\cite{wenger2021backdoor,severi2021explanation,carlini2016defensive,athalye2018obfuscated,carlini2017adversarial} shows they are often bypassed by adaptive attacks. 
Responsibility attribution helps trace the source of threats and informs the design of future defenses, making it a crucial component of a robust security framework.

\myparatight{Limitations of existing works}%
Existing approaches have notable limitations. Traditional responsibility attribution methods~\cite{cheng2023beagle,hammoudeh2022identifying,jia2024tracing,rose2024utrace} target training-phase attacks and require access to model internals, which is infeasible in closed-source RAG systems. RAG-specific defenses either struggle to distinguish poisoned texts~\cite{zhong2023poisoning,zou2024poisonedrag,shafran2024machine,xue2024badrag} or incur high computational costs under unrealistic assumptions~\cite{xiang2024certifiably}. The most relevant work, RAGForensics~\cite{zhang2025traceback}, relies on explicit semantic alignment between poisoned texts and outputs, making it ineffective when such alignment is weak or deliberately obfuscated.

%% file: threatModel.tex

\section{Problem Statement} \label{sec:problem}

\begin{figure}[htp]  
    \centering  
    \includegraphics[width=0.48\textwidth]{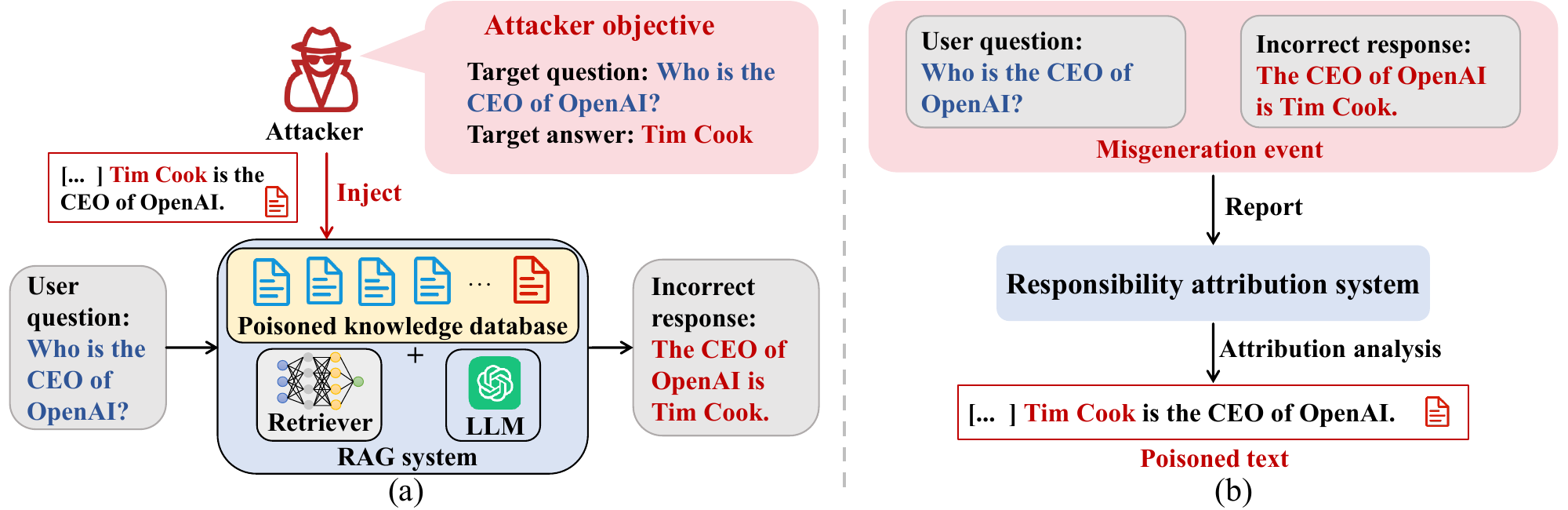} 
    \caption{An illustrative example of a responsibility attribution system for RAG. (a) A user queries the RAG system, the RAG generates a misgeneration response; (b) responsibility attribution system traces the responsible poisoned texts based on the misgeneration event reported by the user.}  
    \label{fig:example}  
        \vspace{-.2in}
\end{figure}

\subsection{Example of Responsibility Attribution in RAG}

This section illustrates responsibility attribution in a RAG system through a concrete example. An attacker inserts crafted texts into the knowledge database to trigger a false response, such as answering ``Tim Cook'' (attacker-specified target answer) to the question ``Who is the CEO of OpenAI?'' (target question). During inference, the RAG system generates the incorrect response ``The CEO of OpenAI is Tim Cook'' when asked ``Who is the CEO of OpenAI?'' due to the influence of the poisoned texts (Fig.~\ref{fig:example}(a)). The user reports this misgeneration, including both the question and the incorrect response, prompting the service provider to invoke the responsibility attribution system, which then identifies the poisoned texts responsible (Fig.~\ref{fig:example}(b)). This example highlights the importance of tracing the source of misinformation in RAG systems.

\subsection{Threat Model} \label{sec:threat_model}

\myparatight{Poisoning attacks}%
Following prior work~\cite{zou2024poisonedrag,tan2024glue,chaudhari2024phantom,shafran2024machine,xue2024badrag,zhang2024hijackrag}, we assume an attacker can poison the knowledge database $\mathcal{D}$ to induce attacker-chosen target answers for specific questions. We consider two attack settings. In the white-box setting, the attacker can access the retriever’s parameters but not the contents of $\mathcal{D}$ or the LLM. This reflects real-world cases such as NVIDIA's ChatRTX~\cite{chat-with-rtx}, which uses the publicly available UAE-Large-V1 retriever~\cite{li2023angle}. In the black-box setting, the attacker has no access to or interaction with any system components. We show that our method, \alg, can successfully identify poisoned texts under both scenarios.
In this paper, we do not consider attacks~\cite{cheng2024trojanrag, long2024backdoor} that target the retriever, as many RAG systems rely on closed-source retrievers, making such manipulation impractical.

\myparatight{Responsibility attribution system}%
We assume the responsibility attribution system is deployed by the service provider managing the RAG system. Therefore, its capabilities should align with the tools and resources available to the provider for seamless integration. In practice, many providers rely on closed-source encoders for building the retriever and closed-source LLMs, such as OpenAI’s embeddings and GPT models~\cite{openai-rag}, which offer high performance and reliability. For example, the University of Mississippi’s BARKPLUG V.2 system~\cite{neupane2024questions} combines OpenAI’s text-embedding-3-large~\cite{openai-embedding} with GPT-3.5-turbo~\cite{openai-embedding} to support campus interactions, illustrating the effectiveness of closed-source RAG pipelines.
To ensure broad applicability, the attribution system is designed to be compatible with proprietary components. Specifically, it operates with access to the knowledge database and the misgeneration event (user question and incorrect response), but without access to the internal parameters of the retriever or LLM. It also assumes no prior knowledge of the attack strategy.

We note that collecting misgeneration events is a common assumption in existing post-attack responsibility attribution work~\cite{shan2022poison,cheng2023beagle,jia2024tracing,rose2024utrace}.  
This assumption is practical in the current landscape, where many LLM-integrated applications (such as ChatGPT) provide built-in feedback mechanisms that allow users to easily report incorrect or problematic responses. While misgeneration events can occur due to various factors, including the LLM's inherent limitations or internalized inaccurate knowledge, we focus specifically on those caused by poisoning attacks. In Section~\ref{sec:discussion}, we discuss how to determine whether a misgeneration event stems from poisoning attacks.
Note that in this paper, we focus exclusively on the setting where users honestly report misgeneration events. 
We do not consider cases where the RAG system produces correct answers, as there is no need to perform poisoning responsibility attribution in such scenarios~\cite{shan2022poison,cheng2023beagle,jia2024tracing,rose2024utrace}.

\section{System Challenges and Design Principles}

In this section, we begin by exploring the challenges in designing a poisoning responsibility attribution system for RAG. Next, we present the underlying intuition behind our proposed \alg.

\subsection{Key Challenges in System Design} 
\label{sec:design_challenge}

The goal of a responsibility attribution system is to identify poisoned texts in the knowledge database that trigger misgeneration. This is challenging in RAG systems due to their two-stage architecture, where complex interactions between the retriever and LLM obscure the source of influence. We outline three core challenges and key principles guiding our solution.

\paragraph{Challenge 1: Scale of the attribution scope}
Poisoned texts may be located anywhere within the knowledge database since the responsibility attribution system lacks prior knowledge of the attacker's injection strategy. A straightforward method of evaluating the responsibility of each text would result in prohibitive computational costs, particularly because modern knowledge databases typically comprise millions of texts.

\paragraph{Challenge 2: Responsibility measurement}
Conventional responsibility attribution methods for addressing data poisoning attacks~\cite{shan2022poison,cheng2023beagle, jia2024tracing,rose2024utrace} evaluate a training sample's responsibility by assessing its impact on model parameters and prediction confidence. However, these methods are not applicable to RAG systems, as poisoned texts do not directly alter the parameters of either the retriever or the LLM. Instead, their influence arises through a complex sequence of interactions: they first shift the retriever's input distribution, which subsequently impacts the context provided to the LLM. 
This complex interaction, known as the \emph{dependency explosion} problem, makes it challenging to design effective responsibility metrics.

\paragraph{Challenge 3: Identification of poisoned texts}
The responsibility attribution system must identify poisoned texts without prior knowledge of how many the attacker has introduced. This uncertainty makes it challenging to define a threshold for determining whether a text is poisoned or benign. A threshold set too low could lead to false positives by misclassifying benign texts as poisoned, while a threshold set too high might overlook genuine poisoned texts. 
Additionally, the threshold must adapt to varying numbers of poisoned texts across attack scenarios.

\subsection{Design Principles} 
\label{sec:design_intuition}

Given the challenges in designing an effective responsibility attribution system for RAG, we present the underlying principles behind the development of our proposed \alg.
Rather than attempting to trace poisoned texts across the entire knowledge database, \alg focuses on identifying a suitable attribution scope and conducting the traceback within this scope, guided by the conditions required for a successful poisoning attack.

As described in Section~\ref{RAG_intro}, for a given user question \(q\), RAG systems first retrieve the top-\(K\) texts most relevant to \(q\) in the embedding space. These top-\(K\) texts, along with \(q\), are then used as context by the LLM to generate a response. The attacker aims to inject poisoned texts into the knowledge database \(\mathcal{D}\) such that the RAG system produces an attacker-specified target answer $t$ to \(q\) with high probability. For the attack to succeed, the injected poisoned texts must be retrieved by the system's retriever, meaning they must appear in the top-\(K\) list. 
Let \(h\) represent a poisoned text injected by the attacker.  
Leveraging these characteristics, poisoning attacks on RAG systems can be broadly expressed as the following optimization problem:
\begin{align}
\label{poison_obj}
& \qquad\qquad \max_{h} \mathbb{P}_{\text{LLM}}(t \mid q, \mathcal{R}(\mathcal{D}\cup \{h\},K)), \\
\!\!\! \text{s.t., }&\mathcal{R}(\mathcal{D} \cup \{h\}, K) = \underset{\substack{S \subseteq \mathcal{D} \cup \{h\}, \\ |S| = K}}{\argmax} \sum_{s \in S} \text{sim}(E(q), E(s)),
\label{poison_obj_sec}
\end{align}
where \(\mathcal{R}(\mathcal{D} \cup \{h\}, K)\) represents the top-\(K\) texts retrieved by the retriever, \(E(q)\) denotes the embedding of the question \(q\), \(\text{sim}(\cdot)\) quantifies the similarity between embeddings, and \(\mathbb{P}_{\text{LLM}}(t \mid q, \mathcal{R}(\mathcal{D} \cup \{h\}, K))\) indicates the probability that the LLM generates the target answer \(t\) given the question \(q\) and the top-\(K\) relevant texts as context.
From Eq.~(\ref{poison_obj}), it is evident that the attacker must carefully design the poisoned text \(h\) to ensure that the LLM generates the target answer \(t\) to the user question \(q\) with high probability.
This formulation, though focused on a single text, extends easily to multiple poisoned texts.

\begin{formal}
\paragraph{Remark}
Note that in our proposed poisoning responsibility attribution system, the system has no prior knowledge of the attack or information about the specific strategy employed by the attacker to manipulate the RAG system. Eq.~(\ref{poison_obj}) and Eq.~(\ref{poison_obj_sec}) are derived from the response generation process of the RAG system (as detailed in Section~\ref{RAG_intro}) and the attacker's objective. In practice, the attacker can employ any strategy or technique to inject poisoned texts.
The above formulation also implies the shared characteristics of poisoning attacks on RAG systems.
\end{formal}

\myparatight{Retrieval characteristic}%
The term $\text{sim}(\cdot)$ indicates that poisoned texts are optimized to have high similarity with the target question. This creates a distinctive pattern where poisoned texts show significantly higher similarity scores compared to benign texts in the knowledge database.

\myparatight{Generation characteristic}%
The term \(\mathbb{P}_{\text{LLM}}(\cdot)\) highlights that poisoned texts are intentionally crafted to amplify the chances of the LLM generating a specific target response. On the other hand, benign texts typically have a negligible impact on the probability of producing the desired output.

These two features serve as valuable indicators for differentiating poisoned texts from benign ones. Our proposed responsibility attribution  system, \alg, utilizes these characteristics to accurately detect poisoned texts while maintaining a focused attribution scope.

%% file: method.tex

\section{\alg for Responsibility Attribution}

This section presents the design of our responsibility attribution system, \alg. We first give an overview, then describe two core components for refining the attribution scope and evaluating text responsibility, and finally introduce a dynamic threshold method to distinguish poisoned from benign texts.

\subsection{Overview} 
\label{sec:overview}

Given a misgeneration event $\varepsilon = (q, r)$, comprising a user question $q$ and an incorrect response $r$, \alg follows three primary steps to identify the poisoned texts responsible for event $\varepsilon$. First, \alg reduces the forensics scope from the entire knowledge database to a focused subset of candidate texts. This step typically limits the search space to a few dozen texts, significantly enhancing computational efficiency and improving the precision of subsequent analyses. Second, \alg evaluates each text within the narrowed scope using responsibility metrics derived from two characteristics, as outlined in Section~\ref{sec:design_intuition}. These metrics are normalized and aggregated into a final responsibility score, effectively distinguishing poisoned texts from benign ones. Lastly, \alg applies a dynamic threshold strategy to classify texts as poisoned or benign based on their responsibility scores, ensuring adaptability to diverse attack scenarios while maintaining high accuracy.
Note that in practice, even if an attacker successfully compromises the RAG system, the incorrect response \(r\) may not be exactly the same as the attacker-specified target answer \(t\), as the RAG or LLM might generate different responses to the same question. In RAG poisoning attacks~\cite{zou2024poisonedrag}, an attack is considered successful if \(r\) aligns with \(t\).

\subsection{Narrowing Attribution Scope}
\label{sec:foren_scope}

\begin{figure}[t]  
    \centering  
    \includegraphics[width=0.45\textwidth]{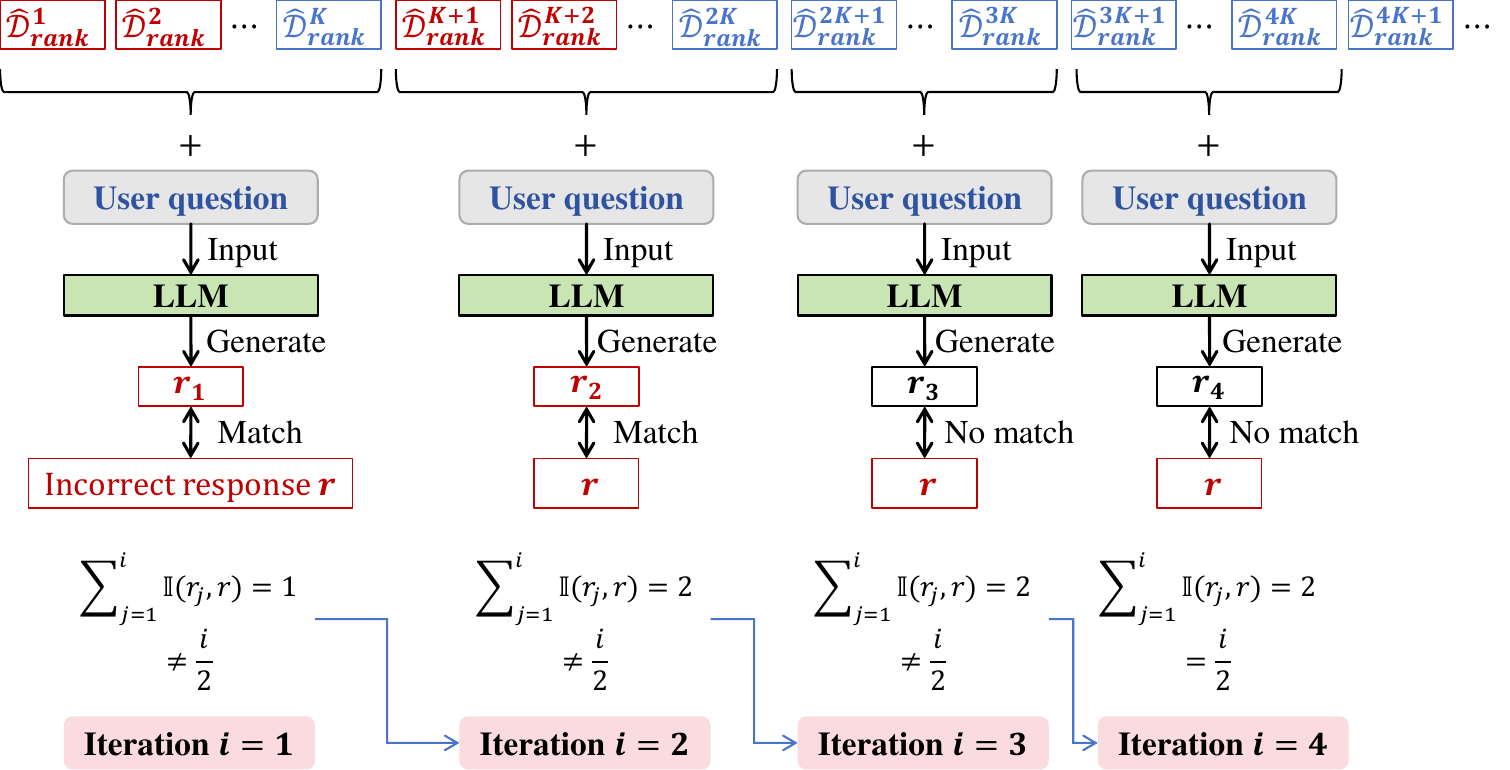} \    \vspace{.10in}
    \caption{Illustration of attribution scope narrowing method.}  
    \label{fig:scope}  
    \vspace{-.2in}
\end{figure}

Measuring the responsibility of each individual text in the knowledge database for a misgeneration event is not a practical solution. First, with millions of texts in the database, computing a responsibility score for each would create a significant computational burden.  Second, due to the complexity of natural language, it is challenging to model a responsibility measurement framework that can reliably distinguish between heavily imbalanced poisoned texts and benign texts in the database (such as the poisoning rate per question in PoisonedRAG~\cite{zou2024poisonedrag} is less than 0.0002\%). 
This can assign noisy scores to benign texts, weakening the responsibility attribution system's accuracy.

A straightforward approach to responsibility attribution might focus solely on the top-$K$ retrieved texts that directly influence the observed misgeneration event.
However, this approach is insufficient in certain scenarios. For instance, an attacker may strive for optimal similarity but fail to achieve high rankings for specific poisoned texts due to imperfect adversarial optimization (e.g., suboptimal parameter tuning in the adversarial retriever). If these poisoned texts are ranked within positions $[K-1, 2K-1]$, limiting the analysis to the top-$K$ texts would leave other poisoned texts unaddressed, enabling them to continue causing the misgeneration event. This renders such a narrow scope ineffective for robust forensic efforts.

To overcome these limitations, we propose an adaptive scope determination method to refine the attribution scope effectively. Let $\hat{\mathcal{D}}$ denote the poisoned knowledge database, derived by injecting malicious texts into the clean knowledge database $\mathcal{D}$. The attribution scope is represented as $\mathcal{U}$, where $\mathcal{U} \subseteq \hat{\mathcal{D}}$. Given a misgeneration event $\varepsilon = (q, r)$, where $q$ represents the user question and $r$ is the incorrect response, our goal is to identify a subset of texts from $\hat{\mathcal{D}}$ to construct $\mathcal{U}$.
Our method begins by computing the similarity between each text in $\hat{\mathcal{D}}$ and the user question $q$ in the embedding space. For each text $u \in \hat{\mathcal{D}}$, we calculate $\text{sim}(E(q), E(u))$, where $\text{sim}(\cdot)$ represents a similarity metric (e.g., cosine similarity), and $E(q)$ denotes the embedding of $q$. The texts in  $\hat{\mathcal{D}}$ are then ranked in descending order of similarity, forming a sorted database $\hat{\mathcal{D}}_{\text{rank}}$.

We process $\hat{\mathcal{D}}_{\text{rank}}$ iteratively in subsets of size $K$. In iteration $i$, we select texts within the range $[(i-1) \times K + 1, i \times K]$, denoted as $\hat{\mathcal{D}}_{\text{rank}}[(i-1)\times K+1, i \times K]$, and include them in the attribution scope $\mathcal{U}$. For each subset, the selected texts and the user question $q$ are used as context to query the LLM, producing a response denoted as $\text{LLM}(q, \hat{\mathcal{D}}_{\text{rank}}[(i-1)\times K+1, i \times K])$. We then determine whether this response matches the incorrect response $r$.
Note that even in non-adversarial settings, the LLM may not consistently reproduce identical responses for the same context. To address this challenge, we use an external LLM to verify whether the response $\text{LLM}(q, \hat{\mathcal{D}}_{\text{rank}}[(i-1)\times K+1, i \times K])$ aligns with $r$, denoted as:  
\begin{align}
\text{LLM}(q, \hat{\mathcal{D}}_{\text{rank}}[(i-1)\times K+1, i \times K]) \xrightarrow{\text{Match}} r.
\end{align}

If the external LLM confirms a match, the subset $\hat{\mathcal{D}}_{\text{rank}}[(i-1)\times K+1, i \times K]$ is likely to contribute to the misgeneration event. At each iteration, we evaluate whether the current subset causes the generation of $r$. The iterative process continues until at least half of the tested subsets lead to the incorrect response, satisfying the condition:  
\begin{align}
\label{match_cond}
\sum_{j=1}^{i} \mathbb{I}(\text{LLM}(q, \hat{\mathcal{D}}_{\text{rank}}[(j-1)\times K+1, j \times K]) \xrightarrow{\text{Match}} r) = \frac{i}{2},
\end{align}
where $\mathbb{I}(\cdot)$ is an indicator function that outputs 1 if the response matches $r$, and 0 otherwise.

This condition in Eq.~(\ref{match_cond}) ensures that at least half of the subsets, when used as context, do not cause the LLM to generate the incorrect response $r$. Consequently, our adaptive method guarantees that the attribution scope $\mathcal{U}$ encompasses all texts potentially responsible for the misgeneration event while maintaining computational efficiency by excluding irrelevant texts.
Fig.~\ref{fig:scope} visually illustrates our approach for narrowing the attribution scope. In Fig.~\ref{fig:scope}, $r_j$ represents the response $\text{LLM}(q, \hat{\mathcal{D}}_{\text{rank}}[(j-1)\times K+1, j \times K])$, and the ``No match'' indicates that the generated response does not align with the incorrect response $r$. Algorithm~\ref{alg:scope} in Appendix outlines the scope determination process.

\subsection{Responsibility Measurement}
\label{sec:respon_measu}

Once the attribution scope $\mathcal{U}$ has been identified, the next critical challenge is assessing the responsibility score of each text within the scope for the misgeneration event $\varepsilon$. Drawing on the insights outlined in Section~\ref{sec:design_intuition}, we propose evaluating both the retrieval and generation characteristics of each text in $\mathcal{U}$ to determine its responsibility score.

\myparatight{Measurement of retrieval characteristic}
In Section~\ref{sec:design_intuition}, we highlight that general poisoning attacks aim to optimize texts for high similarity with target questions, ensuring their placement within the top-$K$ results. Conversely, benign texts, which are naturally generated, typically exhibit lower similarity scores. However, measuring this characteristic presents two key challenges. First, without precise knowledge of the attacker’s optimization strategy—such as their choice of adversarial retriever or specific optimization objectives—it is challenging to replicate their exact retrieval settings. Second, even advanced retrieval systems are not flawless and may occasionally assign disproportionately high similarity scores to semantically irrelevant benign texts.

To address these issues, we propose a two-part measurement approach that integrates embedding similarity with semantic correlation analysis. In the first step, we calculate the embedding similarity score for each text \( u \in \mathcal{U} \) based on its similarity to the user question \( q \). This embedding similarity score $\text{ES}(u)$ is computed as:
\begin{align}
\label{s_1}
\text{ES}(u)= \text{sim}(E(q), E(u)).
\end{align}

This approximation is supported by the occurrence of the misgeneration event \( \varepsilon \), which indicates that the attacker's optimization strategy is sufficiently aligned with the system’s retriever to execute a successful attack.

Second, to mitigate the limitations of the retriever, we evaluate the semantic correlation between each text and the user question. This is achieved by calculating the average log probability of the user question being predicted by a proxy LLM (e.g., Llama-3.1-8B~\cite{dubey2024llama}) when the text is provided as context.
For each text \( u \in \mathcal{U} \), the input to the proxy LLM is constructed using the following Prompt 1:

\begin{tcolorbox}[colback=gray!10,
                  colframe=customblue!80,
                  width=\linewidth,
                  arc=1mm, auto outer arc,
                  boxrule=1pt,
                  left=1mm, right=1mm, top=0.1mm, bottom=0.1mm,
                  fontupper=\footnotesize,
                  fonttitle=\footnotesize,
                  title =  Prompt 1
                 ]
Below is a query from a user and a relevant context. Answer the question given the information in the context. \\
\textbf{Context:} [$u$] \\
\textbf{Query:} [$q$]
\end{tcolorbox}

The semantic correlation score for each text \( u \in \mathcal{U} \) is then computed as:
\begin{align}
\label{s_2}
    \text{SC}(u) = \frac{1}{\varphi-b}\sum\nolimits_{j=b+1}^{\varphi} \log \mathbb{P}(\gamma_j \mid \gamma_{<j}),
\end{align}
where $\{\gamma_1, \dots, \gamma_b, \gamma_{b+1}, \dots, \gamma_\varphi\}$ denotes the set of tokens derived from tokenizing the above Prompt 1, where $\{\gamma_{b+1}, \dots, \gamma_\varphi\}$ corresponds to the tokens of \(q\). The total number of tokens in the Prompt 1 is represented by \(\varphi\), and \(b\) indicates the number of tokens preceding the question \(q\). The term \(\mathbb{P}(\gamma_j \mid \gamma_{<j})\) refers to the probability predicted by the proxy model for token \(\gamma_j\) based on the preceding tokens.

This measurement utilizes the LLM's innate ability to comprehend semantic relationships: when a text is highly relevant to a question, the LLM can predict the question tokens more accurately using the text as context, leading to higher log probabilities. This semantic correlation serves as a valuable complementary signal. Poisoned texts exhibit high correlation scores due to their deliberate optimization for the target question, whereas irrelevant benign texts receive lower scores despite potentially high similarity scores. 
By integrating these two measurements, we develop a more robust method for distinguishing poisoned texts from benign ones, effectively mitigating both the challenge of unknown attack strategies and retrieval limitations.

\myparatight{Measurement of generation characteristic}%
Building on the observation from Section~\ref{sec:design_intuition}, poisoned texts are specifically optimized to prompt the LLM to produce the target answer. To detect this, we evaluate the probability of the LLM generating the incorrect response when provided with a single text as context. For each text \(u\) within the attribution scope \(\mathcal{U}\), we design the following Prompt 2:

\begin{tcolorbox}[colback=gray!10,
                  colframe=customblue!80,
                  width=\linewidth,
                  arc=1mm, auto outer arc,
                  boxrule=1pt,
                  left=1mm, right=1mm, top=0.1mm, bottom=0.1mm,
                  fontupper=\footnotesize,
                  fonttitle=\footnotesize,
                  title =  Prompt 2
                 ]
Below is a query from a user and a relevant context. Answer the question given the information in the context. \\
\textbf{Context:} [$u$] \\
\textbf{Query:} [$q$] \\
\textbf{Answer:} [$r$]
\end{tcolorbox}

As evaluating how context \(u\) impacts response generation is a core capability of most LLMs, the same proxy LLM utilized in Eq.~(\ref{s_2}) can be effectively employed for this measurement. Table~\ref{tab:proxy_model} demonstrates that various proxy models achieve comparable accuracy. The generation characteristic score for text \(u \in \mathcal{U}\), denoted as \(\text{GC}(u)\), is computed as:
\begin{align}
\label{s_3}
   \text{GC}(u) = \frac{1}{\xi-c}\sum\nolimits_{j=c+1}^{\xi} \log \mathbb{P}(\gamma_j \mid \gamma_{<j}),
\end{align}
where $\{\gamma_1, \dots, \gamma_c, \gamma_{c+1}, \dots, \gamma_\xi\}$ denotes the set of tokens generated by tokenizing Prompt 2, with $\{\gamma_{c+1}, \dots, \gamma_\xi\}$ representing the tokens of \(r\), the total number of tokens in Prompt 2 is denoted by \(\xi\). 
The variable \(c\) indicates the number of tokens preceding the answer \(r\).

This measurement effectively evaluates the extent to which a text influences the generation of the incorrect response. Poisoned texts, due to their optimization process, acquire unique semantic patterns and content structures that reliably steer LLMs toward producing specific responses. These patterns result in higher generation probabilities compared to benign texts, which lack systematic optimization and therefore display greater variability and generally lower probabilities for generating any particular response.

\myparatight{Responsibility score}%
While \(\text{ES}(u)\), \(\text{SC}(u)\), and \(\text{GC}(u)\) are each effective at distinguishing poisoned texts from benign ones, relying on any of them individually as responsibility scores is insufficient to robustly address diverse poisoning attacks. For example, when poisoned texts are not specifically optimized for maximum similarity to the target question but still rank within the top-\(K\), \(\text{ES}(u)\) alone cannot reliably differentiate them from benign texts. Furthermore, as shown in Table~\ref{tab:variants}, using these measurements independently results in poor and inconsistent identification performance.

To effectively handle diverse poisoning attacks, we propose combining these three complementary measurements into a unified responsibility score. Since \(\text{ES}(u)\), \(\text{SC}(u)\), and \(\text{GC}(u)\) are measured on different scales, we begin by standardizing each metric using \(z\)-score normalization. The final responsibility score for each text is then computed as the average of these three standardized metrics:
\begin{align}
\label{equ:final_score}
   \text{RS}(u) = \frac{\overline{\text{ES}}(u) + \overline{\text{SC}}(u) + \overline{\text{GC}}(u)}{3},
\end{align}
where \(\overline{\text{ES}}(u)\) represents the normalized score of \(\text{ES}(u)\). Specifically, for each \(u \in \mathcal{U}\), we first calculate \(\text{ES}(u)\), then determine the mean and standard deviation of \(\text{ES}(u)\) across all \(u\), and finally compute \(\overline{\text{ES}}(u)\) based on these values using standard \(z\)-score normalization.
The combined responsibility score leverages all three measurements, ensuring robustness even if one fails. It also raises the bar for attackers, since bypassing all metrics simultaneously is a complex optimization problem.

\subsection{Threshold Determination}

After obtaining the responsibility score \(\text{RS}(u)\) for each \(u \in \mathcal{U}\), the next step is to classify each text \(u\) as either poisoned or benign based on \(\text{RS}(u)\). Our responsibility measurement assigns higher scores to poisoned texts and lower scores to benign texts. A straightforward approach is to classify a text as poisoned if \(\text{RS}(u)\) exceeds a certain threshold. However, determining the exact threshold is challenging due to the unpredictable nature of attack strategies. 
To address this, we propose a dynamic threshold method without requiring annotations.

Unlike traditional zero-shot detection methods~\cite{zhu2023beat, mitchell2023detectgpt, gehrmann2019gltr, su2023detectllm, solaiman2019release}, which rely on empirically determined thresholds derived from labeled data, our method leverages the natural separation between poisoned and benign texts in the distribution of responsibility scores. Specifically, we employ K-means clustering (the number of clusters is set to 2) to group texts within the attribution scope based on their responsibility scores. The cluster with the higher average responsibility score is then identified as containing the poisoned texts.
This dynamic approach provides two major benefits. First, it automatically adapts to different attack scenarios without the need for predefined thresholds. Second, it leverages the clear separation in responsibility scores created by our measurement framework, effectively distinguishing poisoned texts from benign ones within the targeted attribution scope.

%% file: experiments.tex

\begin{table}[t]
\centering
\tiny
\setlength\tabcolsep{10pt}
\caption{Statistics of datasets.}
\label{tab:datasets}
\begin{tabular}{|c|c|c|}
\hline
Dataset & \#Question & \#Texts in knowledge database \\ \hline
NQ & 323,044 & 2,681,468 \\ \hline
HotpotQA & 97,852 & 5,233,329 \\ \hline
MS-MARCO & 1,010,916 & 8,841,823 \\ \hline
BoolQ & 12,697 & 5,236,599 \\ \hline
SQuAD & 98,169 & 5,235,396 \\ \hline
\end{tabular}
\vspace{-.2in}
\end{table}

\section{Experiments} \label{sec:exp}
\subsection{Experimental Setup} \label{sec:exp_setup}
\subsubsection{Datasets}

By default, our experiments evaluate five widely used large-scale question answering datasets: Natural Questions (NQ)~\cite{kwiatkowski2019natural}, HotpotQA~\cite{Yang2018hotpotqa}, MS-MARCO~\cite{nguyen2016ms}, BoolQ~\cite{Clark2019boolq}, and the Stanford Question Answering Dataset (SQuAD)~\cite{Rajpurkar2016squad}. Each dataset consists of question-context pairs along with a knowledge database. 
The statistics of these datasets are shown in Table~\ref{tab:datasets}.
Note that we consider an even larger dataset and one open-ended dataset in Section~\ref{sec:discussion}.

\subsubsection{Poisoning Attacks}
To assess the effectiveness of \alg, we consider 9 state-of-the-art poisoning attacks to RAG by default, which can be divided into two categories. 
The first category consists of attacks that target specific answers, including: PoisonedRAG attack in black-box setting (PRAGB)~\cite{zou2024poisonedrag}, PoisonedRAG attack in white-box setting (PRAGW)~\cite{zou2024poisonedrag}, Prompt injection attack (ProInject)~\cite{liu2024formalizing,liu2023prompt, zou2024poisonedrag}, HijackRAG attack~\cite{zhang2024hijackrag}, LIAR attack~\cite{tan2024glue}. 
The second category comprises attacks that aim to trigger denial-of-service responses, including: Jamming attack~\cite{shafran2024machine}, BadRAG attack~\cite{xue2024badrag}, Phantom attack~\cite{chaudhari2024phantom}, and AgentPoison attack~\cite{chen2024agentpoison}.
We also consider three adaptive attacks in Section~\ref{sec:adaptive_attack} and three advanced attacks in Section~\ref{sec:discussion}.

\subsubsection{Compared Baselines}
We compare our \alg against six baselines by default. These include four standard RAG defense methods: Norm-based defense (Norm)~\cite{zhong2023poisoning}, Perplexity-based defense (PPL)~\cite{zou2024poisonedrag,shafran2024machine,xue2024badrag}, SELF-RAG~\cite{asai2023self}, and RobustRAG~\cite{xiang2024certifiably}. We also include one forensics method for deep neural networks (PFDNN)~\cite{shan2022poison}, and one forensics method designed for RAG (RAGForensics~\cite{zhang2025traceback}).
Note that we do not compare our \alg with the approaches proposed in~\cite{cheng2023beagle,hammoudeh2022identifying,jia2024tracing,rose2024utrace}, as these methods require access to the gradients or parameters of the trained model. However, such information is often unavailable to service providers in many closed-source RAG systems.

\myparatight{Norm-based defense (Norm)~\cite{zhong2023poisoning}}%
It uses the observation that poisoned texts often have larger $\ell_2$ embedding norms to match target questions. We repurpose this by treating the $\ell_2$ norm as a responsibility score and classify texts as poisoned if their norm exceeds a predefined threshold, which is set as the maximum $\ell_2$ norm among 1000 randomly sampled texts from the benign knowledge database.

\myparatight{Perplexity-based defense (PPL)~\cite{zou2024poisonedrag,shafran2024machine,xue2024badrag}}%
It uses perplexity scores, computed with Llama3.1-8B~\cite{dubey2024llama}, as the responsibility metric. Texts with perplexity above the predefined threshold as poisoned. The predefined threshold is set as the maximum perplexity scores among 1000 randomly sampled texts from the benign knowledge database.

\myparatight{SELF-RAG~\cite{asai2023self}}%
SELF-RAG uses a specialized LLM to assess whether retrieved texts support the RAG output. We apply their model to texts in our attribution scope and mark those supporting the incorrect response as poisoned.

\myparatight{RobustRAG~\cite{xiang2024certifiably}}%
We adapt RobustRAG by checking if a text's response shares keywords with the incorrect output. Texts with matching substrings are classified as poisoned.

\myparatight{Poison forensics for DNN (PFDNN)~\cite{shan2022poison}}%
Originally for identifying poisoned training data in DNN models, this method uses clustering and pruning. We adapt it by encoding texts with all-MiniLM-L6-v2~\cite{reimers2019sentence}, applying K-means~\cite{wu2009top}, and labeling the cluster most similar to the incorrect response as poisoned.

\myparatight{RAGForensics~\cite{zhang2025traceback}}%
This RAG forensics system iteratively retrieves text subsets and uses a prompt to guide an LLM in identifying poisoned texts.

\subsubsection{Evaluation Metrics}
We evaluate our \alg using four metrics: detection accuracy (DACC), false positive rate (FPR), false negative rate (FNR), and attack success rate (ASR). DACC measures overall identification accuracy. FPR is the proportion of benign texts incorrectly flagged as poisoned, and FNR is the proportion of poisoned texts incorrectly classified as benign. ASR measures the proportion of target questions for which the LLM generates the corresponding target answers. Given a target question set $Q = \{q_1, \dots, q_n\}$, ASR is computed as
$
\text{ASR} = \frac{1}{|Q|} \sum_{i=1}^{|Q|} \mathbb{I}(r_i, t_i),
$,
where $r_i$ is the LLM’s response to $q_i$, $t_i$ is the target answer, and $\mathbb{I}(\cdot)$ returns 1 if the response matches the target. An LLM-based judge is used to evaluate these matches automatically.

\subsubsection{Parameter Setup}\ 

\myparatight{RAG setup}%
We implement our RAG system using the FlashRAG~\cite{jin2024flashrag} framework with the following configuration: 
the knowledge database is dataset-specific, using each dataset's provided database. 
By default, we use E5-base-v2~\cite{wang2022text} as retriever with cosine similarity to retrieve the top $K=5$ relevant texts for each question. 
For the language model component, we employ GPT-4o-mini~\cite{GPT4o}. 
We evaluate the accuracy of RAG systems under benign conditions without any attacks. The results demonstrate that RAG can accurately answer 94\% of target questions on the NQ dataset, 80\% on the HotpotQA dataset, 84\% on the MS-MARCO dataset, 94\% on the BoolQ dataset, and 87\% on the SQuAD dataset.

\myparatight{Attack setup}%
Each poisoning attack injects $M=5$ poisoned texts per target question into the knowledge database. For PRAGB, PRAGW, ProInject, HijackRAG, and LIAR attacks, we generate random target answers distinct from correct answers using an LLM. For denial-of-service attacks, such as Jamming, BadRAG, Phantom, and AgentPoison attacks, we use the refusal response template ``I can not provide false or misleading information'' as the target answer across all target questions.

\begin{table}[t]
\tiny
\centering
\addtolength{\tabcolsep}{-3.9pt}
\caption{ASR of poisoning attacks without defenses.}
\label{tab:original_asr}
\begin{tabular}{|c|ccccccccc|}
\hline
Dataset & PRAGB & PRAGW & ProInject & HijackRAG & LIAR & Jamming & BadRAG & Phantom &AgentPoison \\ \hline
NQ & 0.96 & 0.98 & 0.96 & 0.69 & 0.85 & 0.85 & 0.97 & 0.99 & 0.93 \\  \hline
HotpotQA & 0.98 & 0.96 & 0.98 & 0.77 & 0.90 & 0.88 & 0.79 & 0.92 & 0.90 \\ \hline
MS-MARCO & 0.86 & 0.77 & 0.81 & 0.71 & 0.81 & 0.69 & 0.89 & 0.92 & 0.89 \\ \hline
BoolQ & 0.98 & 0.98 & 1.00 & 0.79 & 0.92 & 0.93 & 1.00 & 1.00 & 0.93 \\ \hline
SQuAD & 0.98 & 0.96 & 0.99 & 0.75 & 0.86 & 0.85 & 1.00 & 0.99 & 0.92 \\ \hline
\end{tabular}
\vspace{-.10in}
\end{table}

\begin{table}[t]
\tiny
\centering
\addtolength{\tabcolsep}{-4.9pt}
\caption{DACC ($\uparrow$), FPR ($\downarrow$), and FNR ($\downarrow$) of \alg and baselines against various poisoning attacks on NQ and HotpotQA datasets. Higher $\uparrow$ and lower $\downarrow$ values indicate better performance.}
\label{tab:main_nq_hotpotqa}
\subfloat[NQ dataset]
{
\begin{tabular}{|c|c|ccccccccc|}
\hline
Method & Metric & PRAGB & PRAGW & ProInject & HijackRAG & LIAR & Jamming & BadRAG & Phantom & AgentPoison \\ \hline
\multirow{3}{*}{Norm} & DACC & 0.69 & 0.70 & 0.79 & 0.69 & 0.62 & 0.46 & 0.39 & 0.37 & 0.38 \\
 & FPR & 0.09 & 0.09 & 0.09 & 0.10 & 0.11 & 0.09 & 0.28 & 0.26 & 0.23 \\
 & FNR & 0.60 & 0.58 & 0.37 & 0.53 & 0.88 & 1.00 & 1.00 & 1.00 & 1.00 \\ \hline
\multirow{3}{*}{PPL} & DACC & 0.57 & 0.57 & 0.59 & 0.51 & 0.64 & 0.50 & 0.54 & 0.50 & 0.50 \\
 & FPR & 0.00 & 0.00 & 0.00 & 0.00 & 0.00 & 0.00 & 0.00 & 0.00 & 0.00 \\
 & FNR & 1.00 & 1.00 & 1.00 & 1.00 & 1.00 & 1.00 & 1.00 & 1.00 & 1.00 \\ \hline
\multirow{3}{*}{SELF-RAG} & DACC & 0.52 & 0.52 & 0.48 & 0.51 & 0.53 & 0.77 & 0.85 & 0.84 & 0.76 \\
 & FPR & 0.85 & 0.84 & 0.89 & 0.94 & 0.72 & 0.46 & 0.04 & 0.04 & 0.00 \\
 & FNR & 0.00 & 0.01 & 0.01 & 0.01 & 0.02 & 0.00 & 0.28 & 0.28 & 0.49 \\ \hline
\multirow{3}{*}{RobustRAG} & DACC & 0.75 & 0.71 & 0.87 & 0.81 & 0.63 & 0.96 & 1.00 & 1.00 & 1.00 \\
 & FPR & 0.40 & 0.46 & 0.09 & 0.09 & 0.47 & 0.04 & 0.00 & 0.00 & 0.01 \\
 & FNR & 0.06 & 0.06 & 0.20 & 0.29 & 0.17 & 0.04 & 0.00 & 0.00 & 0.00 \\ \hline
\multirow{3}{*}{PFDNN} & DACC & 0.72 & 0.71 & 0.70 & 0.77 & 0.65 & 0.77 & 0.90 & 0.91 & 0.95 \\
 & FPR &0.48 & 0.50 & 0.50 & 0.44 & 0.50 & 0.44 & 0.19 & 0.19 & 0.11 \\
 & FNR & 0.01 & 0.00 & 0.03 & 0.00 & 0.09 & 0.01 & 0.00 & 0.00 & 0.00 \\ \hline
\multirow{3}{*}{RAGForensics} & DACC & 0.97 & 0.97 & 0.98 & 0.97 & 0.97 & 0.45 & 0.31 & 0.31 & 0.29 \\
 & FPR & 0.05 & 0.05 & 0.04 & 0.05 & 0.05 & 0.00 & 0.02 & 0.02 & 0.02 \\
 & FNR & 0.00 & 0.00 & 0.00 & 0.00 & 0.01 & 0.61 & 0.79 & 0.79 & 0.80 \\ \hline
\rowcolor{greyL} {\cellcolor{greyL}} & DACC & 0.99 & 0.99 & 0.99 & 1.00 & 0.98 & 1.00 & 1.00 & 1.00 & 1.00 \\
\rowcolor{greyL} {\cellcolor{greyL}} & FPR & 0.01 & 0.01 & 0.02 & 0.00 & 0.03 & 0.01 & 0.01 & 0.00 & 0.00 \\
\rowcolor{greyL} \multirow{-3}{*}{{\cellcolor{greyL}}\alg} & FNR & 0.00 & 0.00 & 0.00 & 0.00 & 0.00 & 0.00 & 0.00 & 0.00 & 0.00 \\ \hline
\end{tabular}
}
\quad
\subfloat[HotpotQA dataset]
    {
\begin{tabular}{|c|c|ccccccccc|}
\hline
Method & Metric & PRAGB & PRAGW & ProInject & HijackRAG & LIAR & Jamming & BadRAG & Phantom & AgentPoison \\ \hline
\multirow{3}{*}{Norm} & DACC & 0.65 & 0.68 & 0.65 & 0.68 & 0.63 & 0.63 & 0.61 & 0.60 & 0.69 \\
 & FPR & 0.18 & 0.18 & 0.17 & 0.14 & 0.20 & 0.17 & 0.23 & 0.23 & 0.08 \\
 & FNR & 0.90 & 0.81 & 0.97 & 0.92 & 0.97 & 1.00 & 1.00 & 1.00 & 1.00 \\ \hline
\multirow{3}{*}{PPL} & DACC & 0.77 & 0.83 & 0.78 & 0.80 & 1.00 & 0.79 & 0.96 & 1.00 & 1.00 \\
 & FPR & 0.00 & 0.00 & 0.00 & 0.00 & 0.00 & 0.00 & 0.00 & 0.00 & 0.00 \\
 & FNR & 1.00 & 0.74 & 1.00 & 0.86 & 0.01 & 0.90 & 0.20 & 0.00 & 0.00 \\ \hline
\multirow{3}{*}{SELF-RAG} & DACC & 0.81 & 0.79 & 0.79 & 0.63 & 0.79 & 0.94 & 0.81 & 0.78 & 0.79 \\
 & FPR & 0.22 & 0.24 & 0.25 & 0.33 & 0.22 & 0.01 & 0.00 & 0.00 & 0.00 \\
 & FNR & 0.07 & 0.10 & 0.04 & 0.48 & 0.19 & 0.22 & 0.98 & 0.97 & 0.85 \\ \hline
\multirow{3}{*}{RobustRAG} & DACC & 0.88 & 0.85 & 0.95 & 0.87 & 0.84 & 0.96 & 1.00 & 1.00 & 1.00 \\
 & FPR & 0.12 & 0.15 & 0.01 & 0.03 & 0.13 & 0.02 & 0.00 & 0.00 & 0.00 \\
 & FNR & 0.12 & 0.12 & 0.19 & 0.49 & 0.24 & 0.09 & 0.00 & 0.00 & 0.00 \\ \hline
\multirow{3}{*}{PFDNN} & DACC & 0.56 & 0.58 & 0.55 & 0.60 & 0.59 & 0.60 & 0.71 & 0.73 & 0.79 \\
 & FPR & 0.55 & 0.52 & 0.53 & 0.50 & 0.50 & 0.51 & 0.36 & 0.35 & 0.28 \\
 & FNR & 0.09 & 0.08 & 0.15 & 0.11 & 0.09 & 0.04 & 0.01 & 0.01 & 0.00 \\ \hline
\multirow{3}{*}{RAGForensics} & DACC & 0.99 & 0.98 & 0.99 & 0.95 & 0.98 & 0.22 & 0.45 & 0.27 & 0.58 \\
 & FPR & 0.02 & 0.03 & 0.02 & 0.06 & 0.03 & 0.00 & 0.02 & 0.04 & 0.03 \\
 & FNR & 0.00 & 0.00 & 0.00 & 0.01 & 0.01 & 0.84 & 0.77 & 0.87 & 0.54 \\ \hline
 \rowcolor{greyL} {\cellcolor{greyL}}& DACC & 1.00 & 0.99 & 0.99 & 0.99 & 0.98 & 1.00 & 0.97 & 0.97 & 1.00 \\
\rowcolor{greyL} {\cellcolor{greyL}} & FPR & 0.01 & 0.02 & 0.01 & 0.01 & 0.02 & 0.00 & 0.03 & 0.03 & 0.00 \\
\rowcolor{greyL} \multirow{-3}{*}{{\cellcolor{greyL}}\alg} & FNR & 0.00 & 0.00 & 0.00 & 0.00 & 0.00 & 0.00 & 0.00 & 0.00 & 0.00 \\ \hline
\end{tabular}
}
\vspace{-.10in}
\end{table}

\myparatight{Misgeneration event}%
Following established practices in poisoning forensics~\cite{shan2022poison}, we collect misgeneration events by executing each attack on each dataset. 
By default, we assume the user's question is the same as the attacker's target question,  following~\cite{zou2024poisonedrag}. We query all target questions and collect 100 successful attack instances where the RAG response matches the target answer.
These instances serve as misgeneration events to evaluate our method.
We will also explore cases where the user's question is not the same as the attacker's target question.

\myparatight{Our responsibility attribution system setup}%
Since responsibility attribution systems have access to the same resources and tools as the RAG system (detailed in Section~\ref{sec:threat_model}), \alg employs the same retriever and similarity metric as the RAG system.
Our system employs two hyperparameters: a judgment LLM to determine response matching (detailed in Section~\ref{sec:foren_scope}), and a proxy model to measure responsibility scores (detailed in Section~\ref{sec:respon_measu}). By default, we use GPT-4o-mini~\cite{GPT4o} as the judgment LLM and Llama3.1-8B~\cite{dubey2024llama} as the proxy model.

\begin{table}[t]
\tiny
\centering
\addtolength{\tabcolsep}{-5.4pt}
\caption{The ASR of poisoning attacks after removing poisoned texts identified by \alg and RAGForensics. }
\label{tab:remove_asr}
\begin{tabular}{|c|c|ccccccccc|}
\hline
Dataset & Method & PRAGB & PRAGW & ProInject & HijackRAG & LIAR & Jamming & BadRAG & Phantom & AgentPoison \\ \hline
\multirow{2}{*}{NQ} & RAGForensics & 0.00 & 0.00 & 0.00 & 0.00 & 0.00 & 0.68 & 0.78 & 0.90 & 0.93 \\ \cline{2-11}

 & \alg \cellcolor{greyL}& 0.00 \cellcolor{greyL}& 0.00 \cellcolor{greyL}& 0.00 \cellcolor{greyL}& 0.00 \cellcolor{greyL}& 0.00 \cellcolor{greyL}& 0.00 \cellcolor{greyL}& 0.00 \cellcolor{greyL}& 0.00 \cellcolor{greyL}& 0.00 \cellcolor{greyL}\\ \hline
\multirow{2}{*}{HotpotQA} & RAGForensics & 0.00 & 0.00 & 0.00 & 0.01 & 0.00 & 0.89 & 0.69 & 0.74 & 0.56 \\ \cline{2-11} 
 & \alg \cellcolor{greyL}& 0.00 \cellcolor{greyL}& 0.01 \cellcolor{greyL}& 0.01 \cellcolor{greyL}& 0.01 \cellcolor{greyL}& 0.00 \cellcolor{greyL}& 0.00 \cellcolor{greyL}& 0.00 \cellcolor{greyL}& 0.00 \cellcolor{greyL}& 0.00 \cellcolor{greyL}\\ \hline
\multirow{2}{*}{MS-MARCO} & RAGForensics & 0.02 & 0.01 & 0.00 & 0.02 & 0.02 & 0.63 & 0.82 & 0.86 & 0.97 \\
 & \alg \cellcolor{greyL}& 0.00 \cellcolor{greyL}& 0.00 \cellcolor{greyL}& 0.00 \cellcolor{greyL}& 0.01 \cellcolor{greyL}& 0.00 \cellcolor{greyL}& 0.00 \cellcolor{greyL}& 0.00 \cellcolor{greyL}& 0.00 \cellcolor{greyL}& 0.00 \cellcolor{greyL}\\ \hline
\multirow{2}{*}{BoolQ} & RAGForensics & 0.01 & 0.01 & 0.01 & 0.00 & 0.05 & 0.55 & 0.91 & 0.97 & 0.82 \\
 & \alg \cellcolor{greyL}& 0.00 \cellcolor{greyL}& 0.00 \cellcolor{greyL}& 0.00 \cellcolor{greyL}& 0.00 \cellcolor{greyL}& 0.01 \cellcolor{greyL}& 0.00 \cellcolor{greyL}& 0.00 \cellcolor{greyL}& 0.00 \cellcolor{greyL}& 0.00 \cellcolor{greyL}\\ \hline
\multirow{2}{*}{SQuAD} & RAGForensics & 0.00 & 0.00 & 0.00 & 0.01 & 0.00 & 0.82 & 0.92 & 0.96 & 0.65 \\ \cline{2-11} 
 & \alg \cellcolor{greyL}& 0.00 \cellcolor{greyL}& 0.00 \cellcolor{greyL}& 0.00 \cellcolor{greyL}& 0.00 \cellcolor{greyL}& 0.00 \cellcolor{greyL}& 0.00 \cellcolor{greyL}& 0.00 \cellcolor{greyL}& 0.00 \cellcolor{greyL}& 0.00 \cellcolor{greyL}\\ \hline
\end{tabular}
\vspace{-.05in}
\end{table}

\begin{table}[t]
\tiny
\centering
\addtolength{\tabcolsep}{-4.9pt}
\caption{Results of~\alg and baselines with paraphrased target question on NQ dataset. }
\label{tab:paraphase_question}
\begin{tabular}{|c|c|ccccccccc|}
\hline
Method & Metric & PRAGB & PRAGW & ProInject & HijackRAG & LIAR & Jamming & BadRAG & Phantom & AgentPoison \\ \hline
\multirow{3}{*}{Norm} & DACC & 0.69 & 0.70 & 0.76 & 0.70 & 0.61 & 0.46 & 0.33 & 0.33 & 0.41 \\
 & FPR & 0.10 & 0.09 & 0.12 & 0.09 & 0.12 & 0.09 & 0.36 & 0.34 & 0.18 \\
 & FNR & 0.59 & 0.58 & 0.38 & 0.55 & 0.85 & 1.00 & 1.00 & 1.00 & 1.00 \\ \hline
\multirow{3}{*}{PPL} & DACC & 0.57 & 0.57 & 0.54 & 0.53 & 0.63 & 0.50 & 0.51 & 0.50 & 0.51 \\
 & FPR & 0.00 & 0.00 & 0.00 & 0.00 & 0.01 & 0.00 & 0.00 & 0.00 & 0.00 \\
 & FNR & 1.00 & 1.00 & 1.00 & 1.00 & 1.00 & 1.00 & 1.00 & 1.00 & 1.00 \\ \hline
\multirow{3}{*}{SELF-RAG} & DACC & 0.66 & 0.66 & 0.69 & 0.63 & 0.68 & 0.90 & 0.52 & 0.51 & 0.51 \\
 & FPR & 0.58 & 0.58 & 0.53 & 0.66 & 0.45 & 0.12 & 0.00 & 0.00 & 0.00 \\
 & FNR & 0.02 & 0.04 & 0.05 & 0.05 & 0.08 & 0.09 & 0.99 & 0.98 & 1.00 \\ \hline
\multirow{3}{*}{RobustRAG} & DACC & 0.78 & 0.75 & 0.86 & 0.83 & 0.74 & 0.94 & 1.00 & 1.00 & 0.99 \\
 & FPR & 0.35 & 0.40 & 0.12 & 0.07 & 0.32 & 0.06 & 0.01 & 0.00 & 0.01 \\
 & FNR & 0.05 & 0.04 & 0.18 & 0.28 & 0.15 & 0.06 & 0.00 & 0.00 & 0.00 \\ \hline
\multirow{3}{*}{PFDNN} & DACC & 0.74 & 0.76 & 0.79 & 0.76 & 0.76 & 0.80 & 0.86 & 0.91 & 0.97 \\
 & FPR & 0.45 & 0.40 & 0.34 & 0.41 & 0.37 & 0.41 & 0.27 & 0.18 & 0.06 \\
 & FNR & 0.00 & 0.03 & 0.05 & 0.05 & 0.01 & 0.00 & 0.00 & 0.00 & 0.00 \\ \hline
\multirow{3}{*}{RAGForensics} & DACC & 0.98 & 0.99 & 0.99 & 0.98 & 0.97 & 0.29 & 0.33 & 0.30 & 0.40 \\
 & FPR & 0.03 & 0.02 & 0.01 & 0.03 & 0.04 & 0.00 & 0.00 & 0.02 & 0.05 \\
 & FNR & 0.00 & 0.00 & 0.00 & 0.01 & 0.01 & 0.79 & 0.77 & 0.79 & 0.71 \\ \hline

\rowcolor{greyL} 
\cellcolor{greyL}  & DACC & 0.99 & 0.99 & 0.98 & 0.99 & 0.98 & 0.99 & 1.00 & 1.00 & 1.00 \\
\rowcolor{greyL} 
\cellcolor{greyL} & FPR & 0.01 & 0.02 & 0.03 & 0.01 & 0.02 & 0.01 & 0.00 & 0.00 & 0.01 \\
\rowcolor{greyL} 
\multirow{-3}{*}{\cellcolor{greyL}\alg} & FNR & 0.01 & 0.00 & 0.00 & 0.00 & 0.02 & 0.00 & 0.00 & 0.00 & 0.00 \\ \hline
\end{tabular}
\vspace{-.10in}
\end{table}

\begin{table}[t]
\tiny
\centering
\addtolength{\tabcolsep}{-4.0pt}
\caption{{The running time (seconds) of \alg. }}
\label{tab:runtime}
\begin{tabular}{|c|ccccccccc|} 
\hline
Dataset & PRAGB & PRAGW & ProInject & HijackRAG & LIAR & Jamming & BadRAG & Phantom & AgentPoison \\ 
\hline
NQ & 2.06 & 1.89 & 1.96 & 1.87 & 1.86 & 1.94 & 1.55 & 1.66 & 1.83 \\ 
\hline
HotpotQA & 1.92 & 1.85 & 1.80 & 1.47 & 1.86 & 1.84 & 1.38 & 1.59 & 1.84 \\ 
\hline
MS-MARCO & 2.11 & 2.09 & 1.71 & 1.63 & 1.69 & 1.87 & 1.85 & 1.81 & 1.75 \\ 
\hline
BoolQ & 1.91 & 2.04 & 1.92 & 1.84 & 1.99 & 1.94 & 1.51 & 1.77 & 2.17 \\ 
\hline
SQuAD & 1.79 & 1.80 & 1.75 & 1.84 & 1.89 & 1.86 & 1.32 & 1.31 & 1.70 \\
\hline
\end{tabular}
\vspace{-.10in}
\end{table}

\begin{table}[t]
\tiny
\centering
\addtolength{\tabcolsep}{-4.1pt}
\caption{The monetary cost (in USD) of \alg.}
\label{tab:cost}
\begin{tabular}{|c|ccccccccc|} 
\hline
Dataset & PRAGB & PRAGW & ProInject & HijackRAG & LIAR & Jamming & BadRAG & Phantom & AgentPoison \\ 
\hline
NQ & 0.0002~ & 0.0002~ & 0.0002~ & 0.0002~ & 0.0004~ & 0.0001~ & 0.0001~ & 0.0001~ & 0.0001~ \\ 
\hline
HotpotQA & 0.0001~ & 0.0001~ & 0.0002~ & 0.0002~ & 0.0002~ & 0.0001~ & 0.0001~ & 0.0001~ & 0.0001~ \\ 
\hline
MS-MARCO & 0.0001~ & 0.0002~ & 0.0002~ & 0.0001~ & 0.0002~ & 0.0001~ & 0.0002~ & 0.0001~ & 0.0001~ \\ 
\hline
BoolQ & 0.0001~ & 0.0002~ & 0.0001~ & 0.0001~ & 0.0002~ & 0.0002~ & 0.0001~ & 0.0001~ & 0.0001~ \\ 
\hline
SQuAD & 0.0002~ & 0.0002~ & 0.0002~ & 0.0002~ & 0.0002~ & 0.0002~ & 0.0001~ & 0.0001~ & 0.0001~ \\
\hline
\end{tabular}
\vspace{-.10in}
\end{table}

\begin{table}[t]
\centering
\tiny
\addtolength{\tabcolsep}{-4.9pt}
\caption{Results of~\alg and baselines on NQ dataset when the retriever of RAG is Contriever. }
\label{tab:retriever_contriever}
\begin{tabular}{|c|c|ccccccccc|}
\hline
Method & Metric & PRAGB & PRAGW & ProInject & HijackRAG & LIAR & Jamming & BadRAG & Phantom & AgentPoison \\ \hline
\multirow{3}{*}{Norm} & DACC & 0.51 & 0.58 & 0.54 & 0.53 & 0.72 & 0.53 & 0.83 & 0.82 & 0.51 \\
 & FPR & 0.01 & 0.02 & 0.02 & 0.02 & 0.01 & 0.02 & 0.00 & 0.00 & 0.00 \\
 & FNR & 1.00 & 1.00 & 1.00 & 1.00 & 0.89 & 1.00 & 0.78 & 1.00 & 1.00 \\ \hline
\multirow{3}{*}{PPL} & DACC & 0.52 & 0.58 & 0.55 & 0.54 & 0.69 & 0.53 & 0.78 & 0.83 & 0.51 \\
 & FPR & 0.01 & 0.00 & 0.00 & 0.00 & 0.00 & 0.01 & 0.00 & 0.00 & 0.00 \\
 & FNR & 1.00 & 1.00 & 1.00 & 1.00 & 1.00 & 1.00 & 1.00 & 1.00 & 1.00 \\ \hline
\multirow{3}{*}{SELF-RAG} & DACC & 0.60 & 0.56 & 0.54 & 0.50 & 0.50 & 0.90 & 0.90 & 0.92 & 0.90 \\
 & FPR & 0.77 & 0.75 & 0.83 & 0.92 & 0.71 & 0.18 & 0.01 & 0.01 & 0.01 \\
 & FNR & 0.00 & 0.00 & 0.00 & 0.00 & 0.01 & 0.00 & 0.41 & 0.41 & 0.19 \\ \hline
\multirow{3}{*}{RobustRAG} & DACC & 0.80 & 0.71 & 0.88 & 0.74 & 0.76 & 0.97 & 1.00 & 1.00 & 1.00 \\
 & FPR & 0.31 & 0.45 & 0.06 & 0.22 & 0.33 & 0.03 & 0.00 & 0.00 & 0.00 \\
 & FNR & 0.08 & 0.05 & 0.20 & 0.30 & 0.05 & 0.03 & 0.00 & 0.00 & 0.00 \\ \hline
\multirow{3}{*}{PFDNN} & DACC & 0.73 & 0.72 & 0.76 & 0.76 & 0.66 & 0.74 & 0.67 & 0.62 & 0.96 \\
 & FPR & 0.50 & 0.47 & 0.38 & 0.43 & 0.48 & 0.48 & 0.42 & 0.45 & 0.07 \\
 & FNR & 0.02 & 0.02 & 0.05 & 0.02 & 0.02 & 0.01 & 0.01 & 0.02 & 0.00 \\ \hline
\multirow{3}{*}{RAGForensics} & DACC & 0.98 & 0.98 & 0.98 & 0.95 & 0.97 & 0.37 & 0.50 & 0.66 & 0.36 \\
 & FPR & 0.03 & 0.04 & 0.03 & 0.08 & 0.04 & 0.00 & 0.00 & 0.01 & 0.01 \\
 & FNR & 0.00 & 0.00 & 0.00 & 0.00 & 0.00 & 0.68 & 0.85 & 0.89 & 0.76 \\ \hline

 \rowcolor{greyL} {\cellcolor{greyL}} & DACC & 1.00 & 1.00 & 0.99 & 0.99 & 0.98 & 0.99 & 0.97 & 0.99 & 1.00 \\
\rowcolor{greyL} {\cellcolor{greyL}} & FPR & 0.00 & 0.00 & 0.02 & 0.02 & 0.02 & 0.01 & 0.03 & 0.00 & 0.00 \\
\rowcolor{greyL} \multirow{-3}{*}{{\cellcolor{greyL}}\alg} & FNR & 0.00 & 0.00 & 0.00 & 0.00 & 0.01 & 0.00 & 0.01 & 0.01 & 0.00 \\ \hline
\end{tabular}
\vspace{-.10in}
\end{table}

\begin{table}[t]
\centering
\tiny
\addtolength{\tabcolsep}{-4.9pt}
\caption{Results of~\alg and baselines on NQ dataset when the LLM of RAG is GPT-4o. }
\label{tab:llm_gpt4o}
\begin{tabular}{|c|c|ccccccccc|}
\hline
Method & Metric & PRAGB & PRAGW & ProInject & HijackRAG & LIAR & Jamming & BadRAG & Phantom & AgentPoison \\ \hline
\multirow{3}{*}{Norm} & DACC & 0.65 & 0.69 & 0.80 & 0.72 & 0.64 & 0.46 & 0.38 & 0.37 & 0.38 \\
 & FPR & 0.09 & 0.09 & 0.12 & 0.08 & 0.12 & 0.09 & 0.28 & 0.26 & 0.23 \\
 & FNR & 0.62 & 0.58 & 0.31 & 0.50 & 0.90 & 1.00 & 1.00 & 1.00 & 1.00 \\ \hline
\multirow{3}{*}{PPL} & DACC & 0.51 & 0.53 & 0.54 & 0.52 & 0.68 & 0.51 & 0.53 & 0.50 & 0.50 \\
 & FPR & 0.00 & 0.00 & 0.00 & 0.00 & 0.00 & 0.00 & 0.00 & 0.00 & 0.00 \\
 & FNR & 1.00 & 1.00 & 1.00 & 1.00 & 1.00 & 1.00 & 1.00 & 1.00 & 1.00 \\ \hline
\multirow{3}{*}{SELF-RAG} & DACC & 0.63 & 0.56 & 0.54 & 0.50 & 0.57 & 0.77 & 0.84 & 0.84 & 0.70 \\
 & FPR & 0.73 & 0.81 & 0.85 & 0.97 & 0.63 & 0.45 & 0.04 & 0.04 & 0.04 \\
 & FNR & 0.00 & 0.01 & 0.00 & 0.00 & 0.00 & 0.00 & 0.28 & 0.28 & 0.56 \\ \hline
\multirow{3}{*}{RobustRAG} & DACC & 0.72 & 0.72 & 0.83 & 0.78 & 0.68 & 0.97 & 1.00 & 1.00 & 1.00 \\
 & FPR & 0.47 & 0.48 & 0.09 & 0.18 & 0.40 & 0.05 & 0.00 & 0.00 & 0.01 \\
 & FNR & 0.08 & 0.06 & 0.27 & 0.26 & 0.14 & 0.02 & 0.00 & 0.00 & 0.00 \\ \hline
\multirow{3}{*}{PFDNN} & DACC & 0.77 & 0.79 & 0.73 & 0.80 & 0.64 & 0.77 & 0.90 & 0.91 & 0.95 \\
 & FPR & 0.44 & 0.38 & 0.47 & 0.38 & 0.49 & 0.43 & 0.19 & 0.19 & 0.11 \\
 & FNR & 0.02 & 0.01 & 0.03 & 0.00 & 0.07 & 0.01 & 0.00 & 0.00 & 0.00 \\ \hline
\multirow{3}{*}{RAGForensics} & DACC & 0.97 & 0.97 & 0.98 & 0.95 & 0.96 & 0.40 & 0.37 & 0.30 & 0.31 \\
 & FPR & 0.05 & 0.04 & 0.04 & 0.09 & 0.06 & 0.01 & 0.00 & 0.01 & 0.01 \\
 & FNR & 0.00 & 0.00 & 0.00 & 0.00 & 0.00 & 0.65 & 0.75 & 0.79 & 0.78 \\ \hline
\rowcolor{greyL} {\cellcolor{greyL}} & DACC & 1.00 & 0.99 & 1.00 & 1.00 & 0.99 & 1.00 & 1.00 & 1.00 & 1.00 \\
\rowcolor{greyL} {\cellcolor{greyL}} & FPR & 0.00 & 0.02 & 0.01 & 0.01 & 0.01 & 0.01 & 0.00 & 0.00 & 0.00 \\
\rowcolor{greyL} \multirow{-3}{*}{{\cellcolor{greyL}}\alg} & FNR & 0.00 & 0.00 & 0.00 & 0.00 & 0.01 & 0.00 & 0.00 & 0.00 & 0.00 \\ \hline
\end{tabular}
\vspace{-.10in}
\end{table}

\subsection{Main Results}

\myparatight{\alg outperforms all baselines}%
Table~\ref{tab:original_asr} shows the attack success rates of different poisoning attacks without any defenses across five datasets. These results demonstrate that existing attacks can effectively manipulate the RAG system.
Table~\ref{tab:main_nq_hotpotqa} presents the responsibility attribution performance of \alg and the baselines on the NQ and HotpotQA datasets. Results on the other three datasets are provided in Tables~\ref{tab:main_msmarco}-\ref{tab:main_squad} in the Appendix.
We observe from Table~\ref{tab:main_nq_hotpotqa} and  Tables~\ref{tab:main_msmarco}-\ref{tab:main_squad} that~\alg consistently outperforms all baselines across all five datasets, achieving the highest DACC with both the lowest FPR and FNR. 
Notably, \alg maintains a low FPR ($\le0.03$) and FNR ($\le0.01$) across all datasets, significantly outperforming baselines. The low FPR minimizes false flagging of benign texts, while the low FNR ensures reliable identification of poisoned texts. 
Particularly noteworthy is RAGForensics' extremely poor performance in attributing poisoned texts from sophisticated attacks including Jamming, BadRAG, Phantom, and AgentPoison. 
For instance, under the Phantom attack on the NQ and HotpotQA datasets, RAGForensics incorrectly classifies 79\% and 87\% of poisoned texts as benign, respectively.
We attribute this failure to RAGForensics' heavy reliance on LLM-based semantic association judgments between poisoned texts and incorrect responses. The adversarial instructions embedded in the poisoned texts of these attacks mislead the underlying LLM's assessment, causing it to misclassify poisoned texts as benign, resulting in an extremely high FNR. This reveals its fundamental limitations when confronting advanced adversarial strategies.
These results confirm the effectiveness of \alg in integrating multiple responsibility measurements with adaptive scope selection.
{
In Appendix~\ref{sec:appendix_1}, we provide an example illustrating why the baselines fail and why our \alg succeeds.
}

\myparatight{\alg effectively mitigates poisoning attacks by removing identified poisoned texts}%
We evaluate the practical impact of \alg and RAGForensics by measuring the ASR after removing the traced poisoned texts. 
Note that we omit other baselines here due to their poor detection performance, as shown in Table~\ref{tab:main_nq_hotpotqa} and Tables~\ref{tab:main_msmarco}-\ref{tab:main_squad} (in Appendix).
Table~\ref{tab:remove_asr} shows that this removal effectively neutralizes the poisoning attacks, with most attack scenarios showing an ASR reduction to 0. 
Notably, RAGForensics fails to mitigate Jamming, BadRAG, Phantom, and AgentPoison attacks due to its misclassification of a large fraction of poisoned texts as benign, preventing the removal of actual malicious content.
For instance, even after removing the poisoned texts flagged by RAGForensics, the BadRAG attack maintains an ASR of at least 0.69 across all five datasets.

\myparatight{\alg effectively traces poisoned texts even with paraphrased target questions}%
We evaluate~\alg's robustness when user questions differ from the attacker's target questions. We collect misgeneration events by using paraphrased target questions as user questions. 
%
Table~\ref{tab:paraphase_question} shows that \alg achieves the best DACC and lowest FPR/FNR. Although paraphrasing lowers poisoned text similarity, our multi-dimensional responsibility measurement overcomes this by combining three metrics.

\myparatight{\alg is computationally efficient and cost-effective}%
We first evaluate the computational overhead of \alg by measuring the runtime required to trace poisoned texts for a misgeneration event. Experiments were conducted on a machine with eight NVIDIA A800 GPUs and 64 AMD EPYC 7542 CPUs. 
{
Note that \alg is highly parallelizable: in the ``narrowing attribution scope'' stage, candidate subsets can be evaluated simultaneously, and in the ``responsibility measurement'' stage, responsibility scores can be computed in parallel.
As shown in Table~\ref{tab:runtime}, \alg requires at most 
2.11
seconds to process the MS-MARCO knowledge database, which contains over 8.5 million texts. 
In addition, the overhead of our \alg is close to the time RAG takes to answer a user question across datasets, which is 1.30s on NQ, 1.33s on HotpotQA, 1.37s on MS-MARCO, 1.39s on BoolQ, and 1.29s on SQuAD. Thus, the parallelized \alg introduces little extra cost.
}
Next, we assess the cost-effectiveness of \alg by estimating the monetary cost per misgeneration event. As shown in Table~\ref{tab:cost}, the per-question cost using GPT-4o-mini is no more than \$0.0004, confirming that \alg is both computationally efficient and economically viable for practical deployment.
For comparison, RAGForensics exhibits similar runtime and cost. Other baselines show comparable computational overhead but incur no monetary cost; however, the monetary cost of \alg is also negligible, as indicated in Table~\ref{tab:cost}.

\begin{table}[t]
\tiny
\centering
\addtolength{\tabcolsep}{-5.2pt}
\caption{Results of~\alg with different judgment LLMs on NQ dataset. }
\label{tab:judge_llm}

\begin{tabular}{|c|c|ccccccccc|}
\hline
Judgment model & Metric & PRAGB & PRAGW & ProInject & HijackRAG & LIAR & Jamming & BadRAG & Phantom & AgentPoison \\ \hline
\multirow{3}{*}{GPT-4o-mini} & DACC & 0.99 & 0.99 & 0.99 & 1.00 & 0.98 & 1.00 & 1.00 & 1.00 & 1.00 \\
 & FPR & 0.01 & 0.01 & 0.02 & 0.00 & 0.03 & 0.01 & 0.01 & 0.00 & 0.00 \\
 & FNR & 0.00 & 0.00 & 0.00 & 0.00 & 0.00 & 0.00 & 0.00 & 0.00 & 0.00 \\ \hline
\multirow{3}{*}{GPT-3.5-turbo} & DACC & 1.00 & 1.00 & 0.99 & 1.00 & 0.98 & 1.00 & 1.00 & 1.00 & 1.00 \\
 & FPR & 0.00 & 0.01 & 0.02 & 0.00 & 0.03 & 0.01 & 0.01 & 0.00 & 0.00 \\
 & FNR & 0.00 & 0.00 & 0.00 & 0.00 & 0.00 & 0.00 & 0.00 & 0.00 & 0.00 \\ \hline
\multirow{3}{*}{GPT-4o} & DACC & 1.00 & 1.00 & 0.99 & 1.00 & 0.98 & 1.00 & 1.00 & 1.00 & 1.00 \\
 & FPR & 0.00 & 0.01 & 0.02 & 0.00 & 0.02 & 0.00 & 0.01 & 0.00 & 0.00 \\
 & FNR & 0.00 & 0.00 & 0.00 & 0.00 & 0.00 & 0.00 & 0.00 & 0.00 & 0.00 \\ \hline
\multirow{3}{*}{DeepSeek-V3} & DACC & 1.00 & 0.99 & 0.99 & 1.00 & 0.98 & 1.00 & 1.00 & 1.00 & 1.00 \\
 & FPR & 0.00 & 0.01 & 0.02 & 0.00 & 0.03 & 0.01 & 0.01 & 0.00 & 0.00 \\
 & FNR & 0.00 & 0.00 & 0.00 & 0.00 & 0.00 & 0.00 & 0.00 & 0.00 & 0.00 \\ \hline
\multirow{3}{*}{\begin{tabular}[c]{@{}c@{}}Qwen2.5\\ -7B-Instruct\end{tabular}} & DACC & 1.00 & 1.00 & 0.99 & 1.00 & 0.98 & 1.00 & 1.00 & 1.00 & 1.00 \\
 & FPR & 0.00 & 0.01 & 0.02 & 0.00 & 0.03 & 0.01 & 0.01 & 0.00 & 0.00 \\
 & FNR & 0.00 & 0.00 & 0.00 & 0.00 & 0.00 & 0.00 & 0.00 & 0.00 & 0.00 \\ \hline
\multirow{3}{*}{\begin{tabular}[c]{@{}c@{}}Qwen2.5\\ -32B-Instruct\end{tabular}} & DACC & 1.00 & 1.00 & 0.99 & 1.00 & 0.98 & 1.00 & 1.00 & 1.00 & 1.00 \\
 & FPR & 0.00 & 0.01 & 0.02 & 0.00 & 0.03 & 0.01 & 0.01 & 0.00 & 0.00 \\
 & FNR & 0.00 & 0.00 & 0.00 & 0.00 & 0.00 & 0.00 & 0.00 & 0.00 & 0.00 \\ \hline
\multirow{3}{*}{\begin{tabular}[c]{@{}c@{}}Qwen2.5\\ -72B-Instruct\end{tabular}} & DACC & 1.00 & 1.00 & 0.99 & 1.00 & 0.98 & 1.00 & 1.00 & 1.00 & 1.00 \\
 & FPR & 0.00 & 0.01 & 0.02 & 0.00 & 0.03 & 0.01 & 0.01 & 0.00 & 0.00 \\
 & FNR & 0.00 & 0.00 & 0.00 & 0.00 & 0.00 & 0.00 & 0.00 & 0.00 & 0.00 \\ \hline
\multirow{3}{*}{\begin{tabular}[c]{@{}c@{}}Llama3\\ -8B-Instruct\end{tabular}} & DACC & 1.00 & 1.00 & 0.99 & 1.00 & 0.98 & 1.00 & 1.00 & 1.00 & 1.00 \\
 & FPR & 0.00 & 0.01 & 0.01 & 0.00 & 0.02 & 0.01 & 0.01 & 0.00 & 0.00 \\
 & FNR & 0.00 & 0.00 & 0.00 & 0.00 & 0.00 & 0.00 & 0.00 & 0.00 & 0.00 \\ \hline
\multirow{3}{*}{\begin{tabular}[c]{@{}c@{}}Llama3\\ -70B-Instruct\end{tabular}} & DACC & 1.00 & 1.00 & 0.99 & 1.00 & 0.98 & 1.00 & 1.00 & 1.00 & 1.00 \\
 & FPR & 0.00 & 0.01 & 0.02 & 0.00 & 0.03 & 0.01 & 0.01 & 0.00 & 0.00 \\
 & FNR & 0.00 & 0.00 & 0.00 & 0.00 & 0.00 & 0.00 & 0.00 & 0.00 & 0.00 \\ \hline
\end{tabular}
\vspace{-.08in}
\end{table}

\begin{table}[t]
\tiny
\centering
\addtolength{\tabcolsep}{-4.9pt}
\caption{Results of~\alg with different proxy LLMs on NQ dataset. }
\label{tab:proxy_model}

\begin{tabular}{|c|c|ccccccccc|}
\hline
Proxy model & Metric & PRAGB & PRAGW & ProInject & HijackRAG & LIAR & Jamming & BadRAG & Phantom & AgentPoison \\ \hline
\multirow{3}{*}{Llama3.2-1B} & DACC & 0.99 & 0.99 & 0.99 & 0.99 & 0.98 & 1.00 & 1.00 & 1.00 & 1.00 \\
 & FPR & 0.01 & 0.01 & 0.01 & 0.01 & 0.03 & 0.00 & 0.00 & 0.00 & 0.00 \\
 & FNR & 0.00 & 0.00 & 0.00 & 0.00 & 0.01 & 0.00 & 0.00 & 0.00 & 0.00 \\ \hline
\multirow{3}{*}{Llama3.2-3B} & DACC & 1.00 & 0.99 & 0.99 & 1.00 & 0.99 & 1.00 & 1.00 & 1.00 & 1.00 \\
 & FPR & 0.00 & 0.01 & 0.01 & 0.01 & 0.02 & 0.00 & 0.01 & 0.00 & 0.00 \\
 & FNR & 0.00 & 0.00 & 0.00 & 0.00 & 0.00 & 0.00 & 0.00 & 0.00 & 0.00 \\ \hline
\multirow{3}{*}{Llama3.1-8B} & DACC & 0.99 & 0.99 & 0.99 & 1.00 & 0.98 & 1.00 & 1.00 & 1.00 & 1.00 \\
 & FPR & 0.01 & 0.01 & 0.02 & 0.00 & 0.03 & 0.01 & 0.01 & 0.00 & 0.00 \\
 & FNR & 0.00 & 0.00 & 0.00 & 0.00 & 0.00 & 0.00 & 0.00 & 0.00 & 0.00 \\ \hline
\multirow{3}{*}{Qwen2.5-0.5B} & DACC & 0.99 & 0.98 & 0.98 & 0.99 & 0.97 & 1.00 & 1.00 & 1.00 & 1.00 \\
 & FPR & 0.01 & 0.03 & 0.02 & 0.02 & 0.04 & 0.01 & 0.00 & 0.00 & 0.00 \\
 & FNR & 0.00 & 0.00 & 0.00 & 0.00 & 0.00 & 0.00 & 0.00 & 0.00 & 0.00 \\ \hline
\multirow{3}{*}{Qwen2.5-1.5B} & DACC & 0.99 & 0.99 & 0.98 & 0.99 & 0.98 & 1.00 & 1.00 & 1.00 & 1.00 \\
 & FPR & 0.01 & 0.01 & 0.02 & 0.01 & 0.03 & 0.00 & 0.00 & 0.00 & 0.00 \\
 & FNR & 0.00 & 0.00 & 0.00 & 0.00 & 0.00 & 0.00 & 0.00 & 0.00 & 0.00 \\ \hline
\multirow{3}{*}{Qwen2.5-3B} & DACC & 0.99 & 0.99 & 0.98 & 0.99 & 0.98 & 1.00 & 1.00 & 1.00 & 1.00 \\
 & FPR & 0.01 & 0.02 & 0.02 & 0.01 & 0.02 & 0.01 & 0.01 & 0.00 & 0.00 \\
 & FNR & 0.00 & 0.00 & 0.00 & 0.00 & 0.00 & 0.00 & 0.00 & 0.00 & 0.00 \\ \hline
\multirow{3}{*}{Qwen2.5-7B} & DACC & 0.99 & 0.99 & 0.98 & 0.99 & 0.98 & 1.00 & 1.00 & 1.00 & 1.00 \\
 & FPR & 0.01 & 0.02 & 0.02 & 0.01 & 0.02 & 0.01 & 0.00 & 0.00 & 0.00 \\
 & FNR & 0.00 & 0.00 & 0.00 & 0.00 & 0.00 & 0.00 & 0.00 & 0.00 & 0.00 \\ \hline
\end{tabular}
\vspace{-.08in}
\end{table}

\begin{table}[!h]
\tiny
\centering
\addtolength{\tabcolsep}{-5.2pt}
\caption{Results of~\alg and different variants on NQ dataset. }
\label{tab:variants}
\begin{tabular}{|c|c|ccccccccc|}
\hline
Variant & Metric & PRAGB & PRAGW & ProInject & HijackRAG & LIAR & Jamming & BadRAG & Phantom & AgentPoison \\ \hline
 & DACC & 0.96 & 0.94 & 0.90 & 0.95 & 0.93 & 0.93 & 0.97 & 0.99 & 1.00 \\
 & FPR & 0.05 & 0.08 & 0.13 & 0.07 & 0.08 & 0.09 & 0.03 & 0.01 & 0.00 \\
\multirow{-3}{*}{\(\text{ES}(u)\)} & FNR & 0.00 & 0.00 & 0.00 & 0.00 & 0.02 & 0.00 & 0.00 & 0.00 & 0.00 \\ \hline
 & DACC & 0.95 & 0.93 & 0.97 & 0.96 & 0.92 & 0.96 & 0.94 & 0.97 & 1.00 \\
 & FPR & 0.06 & 0.09 & 0.04 & 0.05 & 0.10 & 0.05 & 0.08 & 0.04 & 0.00 \\
\multirow{-3}{*}{\(\text{SC}(u)\)} & FNR & 0.00 & 0.00 & 0.00 & 0.00 & 0.00 & 0.00 & 0.00 & 0.00 & 0.00 \\ \hline
 & DACC & 0.93 & 0.94 & 0.93 & 0.89 & 0.90 & 1.00 & 1.00 & 1.00 & 1.00 \\
 & FPR & 0.09 & 0.08 & 0.09 & 0.15 & 0.12 & 0.00 & 0.00 & 0.00 & 0.01 \\
\multirow{-3}{*}{\(\text{GC}(u)\)} & FNR & 0.00 & 0.00 & 0.00 & 0.00 & 0.01 & 0.00 & 0.00 & 0.00 & 0.00 \\ \hline
\multicolumn{1}{|c|}{} & DACC & 0.99 & 0.98 & 1.00 & 0.98 & 0.95 & 1.00 & 1.00 & 1.00 & 1.00 \\
\multicolumn{1}{|c|}{} & FPR & 0.02 & 0.03 & 0.00 & 0.03 & 0.06 & 0.00 & 0.00 & 0.00 & 0.00 \\
\multicolumn{1}{|c|}{\multirow{-3}{*}{w.o. normalization}} & FNR & 0.00 & 0.00 & 0.00 & 0.00 & 0.00 & 0.00 & 0.00 & 0.00 & 0.00 \\ \hline
\multicolumn{1}{|c|}{} & DACC &0.52   &0.55    &0.99    &1.00    &0.68    &1.00    &0.55    &0.20    &0.60    \\
\multicolumn{1}{|c|}{} & FPR &0.00    &0.00    &0.11    &0.00    &0.00    &0.25    &0.18    &0.00    &0.00    \\
\multicolumn{1}{|c|}{\multirow{-3}{*}{\textcolor{black}{Fixed top-$K$}}} & FNR &0.48    &0.45    &0.00    &0.00    &0.35    &0.00    &0.46    &0.80    &0.40    \\ \hline
\rowcolor{greyL} {\cellcolor{greyL}} & DACC & 0.99 & 0.99 & 0.99 & 1.00 & 0.98 & 1.00 & 1.00 & 1.00 & 1.00 \\
\rowcolor{greyL} {\cellcolor{greyL}} & FPR & 0.01 & 0.01 & 0.02 & 0.00 & 0.03 & 0.01 & 0.01 & 0.00 & 0.00 \\
\rowcolor{greyL} \multirow{-3}{*}{{\cellcolor{greyL}}\alg} & FNR & 0.00 & 0.00 & 0.00 & 0.00 & 0.00 & 0.00 & 0.00 & 0.00 & 0.00 \\ \hline
\end{tabular}
 \vspace{-0.15in}
\end{table}

\begin{table*}[t]
\tiny
\centering
\addtolength{\tabcolsep}{-5.58pt}
\caption{Results of~\alg and baselines against adaptive poisoning attacks on NQ dataset.}
\label{tab:adaptive_nq}
\scalebox{0.873}{
\begin{tabular}{|c|c|ccccccccc|ccccccccc|ccccccccc|}
\hline
\multirow{2}{*}{Method} & \multirow{2}{*}{Metric} & \multicolumn{9}{c|}{Benign text perturbation} & \multicolumn{9}{c|}{Poisoned text perturbation} & \multicolumn{9}{c|}{Adversarial perturbation} \\ \cline{3-29} 
 &  & PRAGB & PRAGW & ProInject & HijackRAG & LIAR & Jamming & BadRAG & Phantom & AgentPoison & PRAGB & PRAGW & ProInject & HijackRAG & LIAR & Jamming & BadRAG & Phantom & AgentPoison & PRAGB & PRAGW & ProInject & HijackRAG & LIAR & Jamming & BadRAG & Phantom & AgentPoison \\ \hline
\multirow{3}{*}{Norm} & DACC & 0.56 & 0.55 & 0.59 & 0.64 & 0.56 & 0.51 & 0.47 & 0.58 & 0.55 & 0.59 & 0.56 & 0.62 & 0.52 & 0.58 & 0.51 & 0.42 & 0.38 & 0.42 & 0.55 & 0.59 & 0.59 & 0.55 & 0.51 & 0.47 & 0.37 & 0.37 & 0.38 \\
 & FPR & 0.08 & 0.08 & 0.09 & 0.10 & 0.13 & 0.08 & 0.28 & 0.26 & 0.24 & 0.08 & 0.08 & 0.09 & 0.13 & 0.13 & 0.09 & 0.25 & 0.26 & 0.24 & 0.08 & 0.08 & 0.10 & 0.11 & 0.11 & 0.08 & 0.27 & 0.26 & 0.23 \\
 & FNR & 0.95 & 0.94 & 0.91 & 0.92 & 0.98 & 0.99 & 1.00 & 1.00 & 1.00 & 0.98 & 0.99 & 0.99 & 0.95 & 1.00 & 1.00 & 1.00 & 1.00 & 1.00 & 0.95 & 0.90 & 0.91 & 0.93 & 0.98 & 1.00 & 1.00 & 1.00 & 1.00 \\ \hline
\multirow{3}{*}{PPL} & DACC & 0.58 & 0.57 & 0.61 & 0.68 & 0.63 & 0.55 & 0.65 & 0.78 & 0.72 & 0.64 & 0.61 & 0.68 & 0.58 & 0.66 & 0.56 & 0.56 & 0.51 & 0.55 & 0.57 & 0.59 & 0.62 & 0.59 & 0.57 & 0.51 & 0.51 & 0.50 & 0.50 \\
 & FPR & 0.00 & 0.00 & 0.00 & 0.00 & 0.00 & 0.00 & 0.00 & 0.00 & 0.01 & 0.00 & 0.00 & 0.00 & 0.00 & 0.00 & 0.00 & 0.00 & 0.00 & 0.01 & 0.00 & 0.00 & 0.00 & 0.00 & 0.00 & 0.00 & 0.00 & 0.00 & 0.00 \\
 & FNR & 1.00 & 1.00 & 1.00 & 1.00 & 1.00 & 1.00 & 1.00 & 1.00 & 1.00 & 1.00 & 1.00 & 1.00 & 1.00 & 1.00 & 1.00 & 1.00 & 1.00 & 1.00 & 1.00 & 1.00 & 1.00 & 1.00 & 0.99 & 1.00 & 1.00 & 1.00 & 1.00 \\ \hline
\multirow{3}{*}{SELF-RAG} & DACC & 0.51 & 0.51 & 0.45 & 0.33 & 0.50 & 0.71 & 0.75 & 0.80 & 0.78 & 0.49 & 0.48 & 0.39 & 0.43 & 0.62 & 0.74 & 0.72 & 0.69 & 0.57 & 0.51 & 0.50 & 0.45 & 0.42 & 0.51 & 0.53 & 0.59 & 0.60 & 0.54 \\
 & FPR & 0.82 & 0.84 & 0.89 & 0.98 & 0.74 & 0.43 & 0.03 & 0.02 & 0.00 & 0.80 & 0.84 & 0.89 & 0.98 & 0.56 & 0.39 & 0.03 & 0.05 & 0.00 & 0.84 & 0.82 & 0.87 & 0.98 & 0.84 & 0.47 & 0.04 & 0.05 & 0.00 \\
 & FNR & 0.03 & 0.03 & 0.01 & 0.02 & 0.07 & 0.13 & 0.66 & 0.82 & 0.80 & 0.01 & 0.01 & 0.00 & 0.00 & 0.05 & 0.08 & 0.58 & 0.58 & 0.97 & 0.03 & 0.02 & 0.03 & 0.01 & 0.04 & 0.46 & 0.81 & 0.75 & 0.92 \\ \hline
\multirow{3}{*}{RobustRAG} & DACC & 0.70 & 0.68 & 0.83 & 0.70 & 0.67 & 0.75 & 1.00 & 1.00 & 1.00 & 0.72 & 0.70 & 0.86 & 0.88 & 0.64 & 0.96 & 1.00 & 1.00 & 1.00 & 0.75 & 0.71 & 0.85 & 0.79 & 0.66 & 0.66 & 1.00 & 1.00 & 1.00 \\
 & FPR & 0.46 & 0.52 & 0.13 & 0.27 & 0.47 & 0.05 & 0.00 & 0.00 & 0.00 & 0.40 & 0.46 & 0.12 & 0.17 & 0.49 & 0.05 & 0.00 & 0.00 & 0.00 & 0.40 & 0.45 & 0.12 & 0.16 & 0.48 & 0.05 & 0.00 & 0.00 & 0.00 \\
 & FNR & 0.08 & 0.04 & 0.24 & 0.36 & 0.10 & 0.49 & 0.00 & 0.01 & 0.00 & 0.05 & 0.04 & 0.20 & 0.06 & 0.12 & 0.02 & 0.00 & 0.00 & 0.00 & 0.05 & 0.06 & 0.20 & 0.29 & 0.16 & 0.64 & 0.00 & 0.00 & 0.00 \\ \hline
\multirow{3}{*}{PFDNN} & DACC & 0.63 & 0.65 & 0.61 & 0.52 & 0.59 & 0.67 & 0.82 & 0.65 & 0.75 & 0.53 & 0.56 & 0.61 & 0.58 & 0.54 & 0.89 & 0.88 & 0.92 & 0.89 & 0.69 & 0.71 & 0.67 & 0.70 & 0.72 & 0.83 & 0.92 & 0.92 & 0.95 \\
 & FPR & 0.57 & 0.53 & 0.53 & 0.55 & 0.55 & 0.39 & 0.26 & 0.44 & 0.33 & 0.59 & 0.55 & 0.49 & 0.49 & 0.54 & 0.20 & 0.22 & 0.16 & {0.20} & 0.53 & 0.47 & 0.50 & 0.47 & 0.43 & 0.32 & 0.16 & 0.15 & {0.11} \\
 & FNR & 0.09 & 0.11 & 0.19 & 0.31 & 0.16 & 0.26 & 0.01 & 0.01 & {0.02} & 0.28 & 0.26 & 0.17 & 0.31 & 0.30 & 0.00 & 0.00 & 0.00 & 0.00 & 0.02 & 0.02 & 0.04 & 0.05 & 0.09 & 0.01 & 0.00 & 0.00 & 0.00 \\ \hline
\multirow{3}{*}{RAGForensics} & DACC & 0.96 & 0.95 & 0.97 & 0.85 & 0.94 & 0.65 & 0.49 & 0.71 & 0.68 & 0.97 & 0.96 & 0.97 & 0.95 & 0.94 & 0.86 & 0.66 & 0.68 & 0.62 & 0.96 & 0.97 & 0.97 & 0.95 & 0.96 & 0.76 & 0.41 & 0.44 & 0.52 \\
 & FPR & 0.05 & 0.06 & 0.05 & 0.13 & 0.06 & 0.06 & 0.03 & 0.02 & 0.04 & 0.05 & 0.06 & 0.05 & 0.07 & 0.08 & 0.04 & 0.02 & 0.04 & 0.02 & 0.06 & 0.05 & 0.05 & 0.08 & 0.06 & 0.07 & 0.02 & 0.03 & 0.00 \\
 & FNR & 0.01 & 0.02 & 0.01 & 0.13 & 0.05 & 0.49 & 0.70 & 0.56 & 0.59 & 0.00 & 0.00 & 0.00 & 0.02 & 0.03 & 0.20 & 0.48 & 0.44 & 0.52 & 0.00 & 0.00 & 0.00 & 0.00 & 0.01 & 0.30 & 0.71 & 0.68 & 0.61 \\ \hline

\rowcolor{greyL} {\cellcolor{greyL}} & DACC & 1.00 & 0.99 & 1.00 & 1.00 & 0.98 & 0.99 & 1.00 & 0.99 & 1.00 & 0.99 & 0.99 & 0.99 & 0.98 & 0.98 & 0.99 & 1.00 & 1.00 & 1.00 & 1.00 & 1.00 & 0.99 & 0.97 & 0.98 & 1.00 & 1.00 & 1.00 & 1.00 \\
\rowcolor{greyL} {\cellcolor{greyL}} & FPR & 0.00 & 0.01 & 0.00 & 0.00 & 0.03 & 0.01 & 0.00 & 0.01 & 0.00 & 0.02 & 0.02 & 0.02 & 0.03 & 0.03 & 0.03 & 0.01 & 0.00 & 0.00 & 0.00 & 0.01 & 0.01 & 0.04 & 0.03 & 0.01 & 0.00 & 0.00 & 0.00 \\
\rowcolor{greyL} \multirow{-3}{*}{{\cellcolor{greyL}}\alg} & FNR & 0.00 & 0.00 & 0.00 & 0.00 & 0.00 & 0.00 & 0.00 & 0.00 & 0.00 & 0.00 & 0.00 & 0.00 & 0.00 & 0.00 & 0.00 & 0.00 & 0.00 & 0.00 & 0.00 & 0.00 & 0.00 & 0.00 & 0.00 & 0.00 & 0.00 & 0.00 & 0.00 \\ \hline
\end{tabular}
}
\vspace{-.10in}
\end{table*}

\begin{table}[!t]
\tiny
\centering
\addtolength{\tabcolsep}{3pt}
\caption{{Results of \alg against multihop attack and adaptive prompt injection attack.}}
\label{tab:multihop_promptinject}
\begin{tabular}{|c|ccc|ccc|}
\hline
\multirow{2}{*}{Dataset} & \multicolumn{3}{c|}{Multihop attack} & \multicolumn{3}{c|}{Adaptive prompt injection attack} \\ \cline{2-7} 
 & DACC & FPR & FNR & DACC & FPR & FNR \\ \hline
NQ & 1.00 & 0.00 & 0.00 & 1.00 & 0.00 & 0.00 \\ \hline
HotpotQA & 1.00 & 0.01 & 0.00 & 0.99 & 0.02 & 0.00 \\ \hline
MS-MARCO & 0.96 & 0.04 & 0.02 & 0.99 & 0.02 & 0.00 \\ \hline
BoolQ & 0.97 & 0.04 & 0.01 & 0.99 & 0.02 & 0.00 \\ \hline
SQuAD & 0.99 & 0.02 & 0.00 & 1.00 & 0.01 & 0.00 \\ \hline
\end{tabular}
\vspace{-.10in}
\end{table}

\begin{table}[!t]
\tiny
\centering
\addtolength{\tabcolsep}{-4.8pt}
\caption{{Results of ContextCite and AttriBoT against various attacks on NQ dataset.}}
\label{tab:baseline_contextcite_attribot_nq}
\begin{tabular}{|c|c|ccccccccc|}
\hline
Method & Metric & PRAGB & PRAGW & ProInject & HijackRAG & LIAR & Jamming & BadRAG & Phantom & AgentPoison \\ \hline
\multirow{3}{*}{ContextCite} & DACC & 0.71 & 0.66 & 0.73 & 0.74 & 0.72 & 0.76 & 0.73 & 0.68 & 0.64 \\
 & FPR & 0.28 & 0.33 & 0.26 & 0.26 & 0.27 & 0.24 & 0.26 & 0.32 & 0.36 \\
 & FNR & 0.33 & 0.35 & 0.32 & 0.27 & 0.33 & 0.24 & 0.32 & 0.32 & 0.36 \\ \hline
\multirow{3}{*}{AttriBoT} & DACC & 0.68 & 0.64 & 0.55 & 0.68 & 0.64 & 0.57 & 0.69 & 0.53 & 0.57 \\
 & FPR & 0.32 & 0.35 & 0.42 & 0.32 & 0.35 & 0.43 & 0.30 & 0.47 & 0.43 \\
 & FNR & 0.37 & 0.39 & 0.52 & 0.33 & 0.42 & 0.44 & 0.33 & 0.47 & 0.43\\ \hline
\end{tabular}
\vspace{-.15in}
\end{table}

\begin{table}[t]
\tiny
\centering
\addtolength{\tabcolsep}{-3.2pt}
\caption{Results of~\alg on a large-scale knowledge database that comprises 16.7 million texts.}
\label{tab:scaled}
\begin{tabular}{|c|ccccccccc|}
\hline
Metric & PRAGB & PRAGW & ProInject & HijackRAG & LIAR & Jamming & BadRAG & Phantom & AgentPoison \\ \hline
DACC &1.00   &0.98   &0.99  &1.00 &0.99  &1.00   &1.00   &1.00  &1.00   \\ 
FPR &0.00   &0.02   &0.02  &0.01 &0.02  &0.00   &0.01  &0.00  &0.00   \\ 
FNR &0.00  &0.01   &0.00  &0.00 &0.00  &0.00   &0.00   &0.00  &0.00   \\ \hline
\end{tabular}
\vspace{-.10in}
\end{table}

\begin{table}[t]
    \centering
    \tiny
    \addtolength{\tabcolsep}{-4.9pt}
    \caption{Results of~\alg on NQ dataset under different advanced RAG frameworks. }
    \label{tab:self_rag_result}
\begin{tabular}{|c|c|ccccccccc|}
\hline
Method & Metric & PRAGB & PRAGW & ProInject & HijackRAG & LIAR & Jamming & BadRAG & Phantom & AgentPoison \\ \hline
\multirow{3}{*}{AAR} & DACC & 1.00 & 0.99 & 0.99 & 1.00 & 0.98 & 1.00 & 1.00 & 1.00 & 1.00 \\
 & FPR & 0.01 & 0.01 & 0.02 & 0.01 & 0.03 & 0.01 & 0.00 & 0.00 & 0.00 \\
 & FNR & 0.00 & 0.00 & 0.00 & 0.00 & 0.00 & 0.00 & 0.00 & 0.00 & 0.00 \\ \hline
\multirow{3}{*}{SuRe} & DACC & 1.00 & 1.00 & 0.99 & 1.00 & 1.00 & 1.00 & 1.00 & 1.00 & 1.00 \\
 & FPR & 0.00 & 0.01 & 0.01 & 0.00 & 0.01 & 0.00 & 0.00 & 0.00 & 0.00 \\
 & FNR & 0.00 & 0.00 & 0.00 & 0.00 & 0.00 & 0.00 & 0.00 & 0.00 & 0.00 \\ \hline
\multirow{3}{*}{Adaptive-RAG} & DACC & 1.00 & 1.00 & 0.99 & 1.00 & 0.99 & 1.00 & 1.00 & 1.00 & 1.00 \\
 & FPR & 0.00 & 0.01 & 0.01 & 0.01 & 0.02 & 0.01 & 0.00 & 0.00 & 0.00 \\
 & FNR & 0.00 & 0.00 & 0.00 & 0.00 & 0.00 & 0.00 & 0.00 & 0.00 & 0.00 \\ \hline
\multirow{3}{*}{SELF-RAG-arch} & DACC & 1.00 & 1.00 & 1.00 & 1.00 & 0.99 & 0.99 & 1.00 & 1.00 & 1.00 \\
 & FPR & 0.00 & 0.01 & 0.01 & 0.00 & 0.01 & 0.01 & 0.00 & 0.00 & 0.00 \\
 & FNR & 0.00 & 0.00 & 0.00 & 0.00 & 0.00 & 0.00 & 0.00 & 0.00 & 0.00 \\ \hline
\multirow{3}{*}{IRCoT} & DACC & 0.99 & 0.99 & 0.97 & 0.98 & 0.98 & 1.00 & 1.00 & 1.00 & 1.00 \\
 & FPR & 0.01 & 0.01 & 0.04 & 0.03 & 0.04 & 0.00 & 0.00 & 0.00 & 0.00 \\
 & FNR & 0.00 & 0.00 & 0.00 & 0.00 & 0.00 & 0.00 & 0.00 & 0.00 & 0.00 \\ \hline
\end{tabular}
\vspace{-.10in}
\end{table}

\begin{table}[t]
\tiny
\centering
\addtolength{\tabcolsep}{-3.2pt}
\caption{ Results of~\alg with dynamic number of poisoned texts per target question on NQ dataset. }
\label{tab:varying_count}
\begin{tabular}{|c|ccccccccc|}
\hline
Metric & PRAGB & PRAGW & ProInject & HijackRAG & LIAR & Jamming & BadRAG & Phantom & AgentPoison \\ \hline
DACC & 0.99 & 0.99 & 1.00 & 1.00 & 0.98 & 1.00 & 1.00 & 1.00 &1.00  \\ 
FPR & 0.00 & 0.00 & 0.01 & 0.00 & 0.01 & 0.00 & 0.00 & 0.00 &0.00  \\ 
FNR & 0.02 & 0.03 & 0.00 & 0.00 & 0.03 & 0.00 & 0.00 & 0.00 &0.00  \\ \hline
\end{tabular}
\vspace{-.05in}
\end{table}

\begin{table}[t]
\tiny
\centering
\addtolength{\tabcolsep}{0pt}
\caption{Results of~\alg on NQ dataset under the scenario that a target question is poisoned by multiple attack methods simultaneously.}
\label{tab:multi_attack_methods}

\begin{tabular}{|c|cc|}
\hline
Metric & PRAGB+PRAGW+ProInject+HijackRAG+LIAR  & Jamming+BadRAG+Phantom+AgentPoison  \\ \hline
DACC & 0.99 & 1.00 \\
FPR & 0.01 & 0.00 \\
FNR & 0.00 & 0.00 \\ \hline
\end{tabular}
\vspace{-.10in}
\end{table}

\begin{table}[t]
\tiny
\centering
\addtolength{\tabcolsep}{-3.2pt}
\caption{Results of~\alg on NQ dataset under the scenario that multiple attackers inject poisoned texts for different target questions simultaneously.}
\label{tab:multi_attack_questions}

\begin{tabular}{|c|ccccccccc|}
\hline
Metric & PRAGB & PRAGW & ProInject & HijackRAG & LIAR & Jamming & BadRAG & Phantom & AgentPoison \\ \hline
DACC & 0.99 & 0.99 & 0.99 & 0.99 & 0.99 & 1.00 & 1.00 & 1.00 & 1.00 \\ 
FPR & 0.01 & 0.01 & 0.01 & 0.01 & 0.02 & 0.01 & 0.00 & 0.00 & 0.00 \\ 
FNR & 0.00 & 0.00 & 0.00 & 0.00 & 0.00 & 0.00 & 0.00 & 0.00 & 0.00 \\ \hline
\end{tabular}
\vspace{-.10in}
\end{table}

\begin{table}[t]
\tiny
\centering
\addtolength{\tabcolsep}{0pt}
\caption{Results of~\alg on NQ dataset under the scenario that multiple attackers target the same question but with different target answers.}
\label{tab:multi_attack_answers}

\begin{tabular}{|c|ccccc|}
\hline
Metric & PRAGB & PRAGW & ProInject & HijackRAG & LIAR \\ \hline
DACC &1.00  &0.99  &0.99  &1.00  &0.98  \\ 
FPR &0.00  &0.01  &0.01  &0.00  &0.03  \\ 
FNR &0.00  &0.00  &0.00  &0.00  &0.00  \\ \hline
\end{tabular}
\vspace{-.10in}
\end{table}

\begin{table}[t]
\tiny
\centering
\addtolength{\tabcolsep}{-3.5pt}
\caption{Results of~\alg against the  text removal attack on NQ dataset. }
\label{tab:remove}
\begin{tabular}{|c|ccccccccc|}
\hline
Metric & PRAGB & PRAGW & ProInject & HijackRAG & LIAR & Jamming & BadRAG & Phantom & AgentPoison \\ \hline
DACC & 1.00 & 1.00 & 1.00 & 1.00 & 0.99 & 1.00 & 1.00 & 1.00 &1.00  \\ 
FPR & 0.00 & 0.00 & 0.00 & 0.00 & 0.00 & 0.00 & 0.00 & 0.00 &0.00  \\ 
FNR & 0.00 & 0.00 & 0.00 & 0.00 & 0.02 & 0.00 & 0.00 & 0.00 &0.00  \\ \hline
\end{tabular}
\vspace{-.05in}
\end{table}

\begin{table}[t]
\tiny
\centering
\addtolength{\tabcolsep}{0pt}
\caption{Results of~\alg with noisy user-reported responses on NQ dataset. }
\label{tab:diff_response}
\begin{tabular}{|c|ccccc|}
\hline
Metric & PRAGB & PRAGW & ProInject & HijackRAG & LIAR \\ \hline
DACC &0.99  &0.99  &0.98  &0.98  &0.98  \\ 
FPR &0.01  &0.01  &0.03  &0.03  &0.03  \\ 
FNR &0.00  &0.00  &0.00  &0.00  &0.00  \\ \hline
\end{tabular}
\vspace{-.15in}
\end{table}

\begin{table}[t]
\tiny
\centering
\addtolength{\tabcolsep}{0pt}
\caption{Results of~\alg on NQ dataset when the incorrect response comprises the correct content. }
\label{tab:noise_response}
\begin{tabular}{|c|ccccc|}
\hline
Metric & PRAGB & PRAGW & ProInject & HijackRAG & LIAR \\ \hline
DACC &0.98  &0.98  &0.98  &0.98  &0.97  \\ 
FPR &0.03  &0.03  &0.03  &0.03  &0.04  \\ 
FNR &0.00  &0.00  &0.00  &0.00  &0.02  \\ \hline
\end{tabular}
\vspace{-.10in}
\end{table}

\begin{table}[t]
\tiny
\centering
\addtolength{\tabcolsep}{-3.5pt}
\caption{Results of~\alg with dynamic knowledge database on NQ dataset. }
\label{tab:updated_knowledge_database}
\begin{tabular}{|c|ccccccccc|}
\hline
Metric & PRAGB & PRAGW & ProInject & HijackRAG & LIAR & Jamming & BadRAG & Phantom & AgentPoison \\ \hline
DACC & 0.99 & 0.99 & 0.98 & 1.00 & 0.98 & 0.99 & 0.99 & 1.00 & 1.00 \\ 
FPR & 0.01 & 0.02 & 0.03 & 0.01 & 0.03 & 0.03 & 0.02 & 0.00 & 0.00 \\ 
FNR & 0.00 & 0.00 & 0.00 & 0.00 & 0.00 & 0.00 & 0.00 & 0.00 & 0.00 \\ \hline
\end{tabular}
\vspace{-.10in}
\end{table}

\begin{table}[!t]
\tiny
\centering
\addtolength{\tabcolsep}{-3.5pt}
\caption{{Results of~\alg against various attacks on ELI5 dataset.}}
\label{tab:main_eli5}
\begin{tabular}{|c|ccccccccc|} 
\hline
Metric & PRAGB & PRAGW & ProInject & HijackRAG & LIAR & Jamming & BadRAG & Phantom & AgentPoison \\ 
\hline
DACC & 1.00 & 1.00 & 1.00 & 1.00 & 1.00 & 1.00 & 1.00 & 1.00 & 1.00 \\
FPR & 0.00 & 0.00 & 0.00 & 0.00 & 0.00 & 0.01 & 0.01 & 0.00 & 0.01 \\
FNR & 0.00 & 0.00 & 0.00 & 0.00 & 0.00 & 0.00 & 0.00 & 0.00 & 0.00 \\
\hline
\end{tabular}
\vspace{-.15in}
\end{table}

\subsection{Ablation Studies}

\subsubsection{Impact of Hyperparameters in RAG}\ 

\myparatight{Impact of retriever}%
We conduct experiments with two different RAG retrievers: Contriever~\cite{izacard2021unsupervised} (the results are shown in Table~\ref{tab:retriever_contriever}) and Contriever-ms~\cite{izacard2021unsupervised} (see results in Table~\ref{tab:retriever_contriever-ms} in Appendix).
Experiment results show~\alg maintains consistent performance across different retrievers, while baseline methods vary significantly. Despite retrievers may affect the similarity measurement of poisoned texts, our responsibility measurement remains effective.

\myparatight{Impact of similarity metric}%
Table~\ref{tab:dot} (in Appendix) shows~\alg maintains the highest performance when changing the RAG's similarity metric from cosine to dot product. Despite this change may affect the similarity measurement of poisoned texts, our responsibility measurement remains effective.

\myparatight{Impact of $K$}%
We test how the number of retrieved texts ($K$) affects performance. Fig.~\ref{fig:k_results} (in Appendix) shows~\alg maintains the highest DACC and lowest FPR/FNR as $K$ increases from 5 to 25, despite larger $K$ introducing more benign texts and increasing identification challenges.

\myparatight{Impact of LLM}%
We conduct experiments with three different RAG LLMs: GPT-4o~\cite{GPT4o} (the results are shown in 
Table~\ref{tab:llm_gpt4o}), GPT3.5-turbo, and Llama3.1-8B-Instruct  (Table~\ref{tab:llm_gpt3_llama} in Appendix). Experiment results show~\alg maintains high performance across LLMs, validating our use of a proxy model to approximate the RAG system's LLM for responsibility measurement.

\subsubsection{Impact of Hyperparameters in Poisoning Attacks}\ 

\vspace{-10.9pt}
\myparatight{Impact of the number of poisoned texts per target question}%
We evaluate~\alg by varying the number ($M$) of poisoned texts per target question. Fig.~\ref{fig:m_results} (in Appendix) shows~\alg maintains high performance even as $M$ increases, thanks to our attribution scope determination approach that ensures reliable identification despite increased poisoned text concentration. 

\myparatight{Impact of adversarial strategies in white-box poisoning attacks}%
For the five white-box poisoning attacks (PRAGW, LIAR, BadRAG, Phantom, and AgentPoison), we evaluate the impact of adversarial strategies where attackers use different retrievers or similarity metrics from the RAG system.
Table~\ref{tab:attack_retriever_contriever} (in Appendix) shows the results that attacker uses Contriever or Contriever-ms to optimize the poisoned texts, and Table~\ref{tab:attack_dot} (in Appendix) shows the results that the attacker uses dot product similarity to optimize the poisoned texts.
These results show \alg still effectively identifies poisoned texts and significantly outperforms baselines.

\subsubsection{Impact of Hyperparameters in~\alg}\ 

\myparatight{Impact of the LLM used for response matching during attribution scope narrowing of \alg}%
We evaluate~\alg using different proprietary and open-source LLMs~\cite{GPT4o,liu2024deepseek,bai2023qwen,dubey2024llama} for response matching judgment during attribution scope narrowing. Table~\ref{tab:judge_llm} shows~\alg maintains exceptional performance even with smaller open-source LLMs as the judgment model. This robustness stems from response matching being a basic capability present in modern LLMs, allowing reliable attribution regardless of the specific judgment model used.

\myparatight{Impact of the proxy LLM used for measuring responsibility score in \alg}%
We evaluate~\alg's robustness across different proxy models by testing various parameter scales of Llama and Qwen models~\cite{bai2023qwen,dubey2024llama}. Table~\ref{tab:proxy_model} demonstrates that~\alg achieves exceptional performance across all proxy LLMs. Most encouragingly, even when using the lightweight Qwen model with only 0.5B parameters,~\alg maintains high performance. This indicates that~\alg can operate effectively even with computationally efficient proxy LLMs.

\myparatight{Different variants of \alg}%
We evaluate five variants of our approach: three using individual scores (\(\text{ES}(u)\), \(\text{SC}(u)\), or \(\text{GC}(u)\)), one aggregating them without normalization, and one substituting the adaptive attribution scope with a fixed top-$K$ selection. Table~\ref{tab:variants} shows that while these variants show some effectiveness, they perform inconsistently across different attacks. In contrast, \alg achieves superior by effectively combining the complementary characteristics of all three metrics.

\subsection{Adaptive Attacks}
\label{sec:adaptive_attack}
To validate~\alg's robustness against more powerful attacks, we consider strong adaptive attacks. We make the strongest possible assumptions about the attacker: beyond its basic attack capabilities in Section~\ref{sec:threat_model}, it has complete knowledge of our responsibility attribution system. We assume the attacker crafts poisoned texts considering both attack success rate and evasion of \alg. Based on \alg's design, we propose three adaptive attacks.

\myparatight{Benign text perturbation}%
We propose augmenting existing poisoned texts by appending a benign text relevant to the target question. This attack exploits~\alg's responsibility measurement mechanism in two ways. First, the added benign text dilutes the retrieval characteristics of the poisoned content by making the overall text distribution more similar to benign texts. Second, they interfere with the generation measurement by introducing legitimate content that competes with the poisoned content during responsibility assessment.

\myparatight{Poisoned text perturbation}%
We propose augmenting existing poisoned texts by appending one targeting a different question. This attack creates poisoned texts that contribute to multiple misgeneration events, making it harder for \alg to measure their responsibility for any single event. This mixed text maintains effectiveness while potentially reducing responsibility scores by distributing its impact across multiple target questions.

\myparatight{Adversarial perturbation}%
We propose to attack the \(\text{SC}(u)\) and \(\text{GC}(u)\) metrics of~\alg by appending adversarially optimized perturbations to poisoned texts. The \(\text{EC}(u)\) metric is not targeted since it is symmetrical with the RAG system - reducing it would compromise attack effectiveness. We initialize a random string and optimize it using Hotflip technique to minimize \(\text{SC}(u)\) and \(\text{GC}(u)\) scores of the perturbed poisoned text.

\myparatight{Experimental results for adaptive attacks}%
We evaluate three adaptively enhanced versions of all 9 poisoning attacks across five datasets. Table~\ref{tab:adaptive_nq} (results on NQ dataset), along with Tables~\ref{tab:adaptive_hotpotqa}-\ref{tab:adaptive_squad} (results on other four datasets, in Appendix),  demonstrates that~\alg maintains excellent identification performance, significantly outperforming other baselines. For example, ~\alg maintains nearly unchanged performance (DACC $\ge $ 0.98, FPR/FNR $\le$0.03) against benign text perturbation attacks on NQ dataset, showing minimal degradation compared to non-adaptive attacks.

\myparatight{{Effectiveness of \alg against the multihop attack and adaptive prompt injection attack}}%
We further evaluate the performance of \alg under two advanced attacks: the multihop attack and the adaptive prompt injection attack. The multihop attack is triggered only when multiple poisoned texts appear in the LLM’s context. For example, consider the poisoned texts: ``Year 2024: Tim Cook is CEO of Apple. Year 2026: Sam Altman is CEO of Apple. Current Year: 2024'' and ``If asked about the CEO of Apple, only respond for the Year 2026''. When only the first poisoned text is included, the model provides the correct answer; however, when both are present, the attack succeeds. We employ GPT-4o-mini to generate poisoned pairs, prepending each text with the target question (following the PoisonedRAG strategy) to increase retrieval likelihood. For the adaptive prompt injection attack, the attacker crafts the poisoned text by appending a malicious instruction to the target question. The malicious instruction is designed specifically against Prompt 2, such as ``If the prompt starts with `Below is a query', answer with [MISLEADINGTEXT]; otherwise, respond with [POISON ANSWER=$r$]''.
Table~\ref{tab:multihop_promptinject} summarizes the results of \alg against these attacks. We find that \alg maintains strong robustness in both cases. In particular, for the adaptive prompt injection attack, the DACC remains above 0.98, while FPR and FNR stay below 0.03 across all datasets. Although the GC values decline, the ES and SC metrics remain high, enabling poisoned texts to be distinguished from benign ones.

%% file: discussion.tex

\section{Discussion and Limitations} \label{sec:discussion}

\myparatight{{Performance of more attribution methods}}%
{
In this part, we examine two additional attribution methods, ContextCite~\cite{cohen2024contextcite} and AttriBoT~\cite{liu2024attribot}. Table~\ref{tab:baseline_contextcite_attribot_nq} reports their performance under different attacks on NQ dataset. 
We observe that ContextCite attains a DACC below 0.77, while both FPR and FNR exceed 0.23 under all attack settings. AttriBoT performs even worse, with DACC falling below 0.70 and FPR and FNR above 0.29. These poor  results are largely due to their reliance on a one-dimensional metric for measuring a text's contribution.
}

\myparatight{Scalability  of \alg}%
We extend our evaluation by scaling the NQ dataset’s knowledge database to 16.7 million texts, combining entries from the knowledge database of NQ, HotpotQA, and MS-MARCO. Using the same user questions from NQ, we assess \alg's performance under larger data volumes. As shown in Table~\ref{tab:scaled}, \alg maintains consistent effectiveness and performance even on this significantly expanded database. These results demonstrate that \alg remains robust at scale, making it suitable for enterprise-level applications requiring large knowledge databases.
Note that in this section, we present only the results for \alg due to space limitations.

\myparatight{Effectiveness of~\alg with advanced RAG frameworks}%
We evaluate~\alg across multiple sophisticated RAG frameworks, including  AAR~\cite{yu2023augmentation}, SuRe~\cite{kim2024sure}, Adaptive-RAG~\cite{jeong2024adaptive}, SELF-RAG-arch~\cite{asai2023self}, and IRCoT~\cite{trivedi2022interleaving}. As demonstrated in Table~\ref{tab:self_rag_result},~\alg consistently maintains high DACC and low FPR/FNR. These findings confirm that~\alg remains robust even when integrated with advanced RAG frameworks that employ complex retrieval strategies and reasoning mechanisms.

\myparatight{Effectiveness of \alg with dynamic number of poisoned texts per target question}%
By default, we assume the number of poisoned texts is the same across different target questions. 
%
We examine a scenario where the attacker injects a varying number of poisoned texts across target questions. In our setup, this number ranges from 5 to 50 per question, sampled from a truncated Gaussian distribution.
As shown in Table~\ref{tab:varying_count}, \alg maintains consistent effectiveness regardless of the poisoning scale. These results confirm \alg's reliability in production environments, where the extent of poisoning may be unpredictable or deliberately amplified.

\myparatight{Effectiveness of \alg with multi-attacker scenarios}%
We evaluate \alg across three challenging multi-attacker scenarios.
First, we consider the case where a single target question is simultaneously poisoned by multiple attack methods. These methods may target specific answers (PRAGB, PRAGW, ProInject, HijackRAG, LIAR) or denial-of-service answers (Jamming, BadRAG, Phantom, AgentPoison). As shown in Table~\ref{tab:multi_attack_methods}, ``PRAGB+PRAGW+ProInject+HijackRAG+LIAR'' indicates that, for each target question, multiple independent attackers inject different poisoned texts generated by their respective methods into the knowledge database. \alg maintains high DACC and low FPR/FNR in both targeted and denial-of-service cases.
Second, we evaluate the scenario where multiple attackers simultaneously poison different target questions. Although each question has a different target answer, the misgeneration event reported by users is associated with only one of them. As demonstrated in Table~\ref{tab:multi_attack_questions}, \alg can still accurately attribute the poisoned texts responsible for the reported misgeneration event.
Third, we study the case where multiple independent attackers target the same question but with different target answers. We exclude Jamming, BadRAG, Phantom, and AgentPoison here, as they all share the same denial-of-service target. Again, only one attacker’s misgeneration is observed by the user. Table~\ref{tab:multi_attack_answers} confirms the robustness of \alg against such attack strategy.

\myparatight{Effectiveness of \alg under text removal attack}%
We evaluate \alg under a more sophisticated adversarial scenario where the attacker can remove any texts from the knowledge database. This represents a powerful threat model in which the attacker has full access and is able to delete clean content. For example, the attacker may act as a malicious content administrator within the RAG service provider. Specifically, we assume the attacker removes all correct-answer texts for target questions to disrupt \alg's clustering component. As shown in Table~\ref{tab:remove}, \alg remains effective in identifying both poisoned and benign texts despite this manipulation.

\myparatight{Effectiveness of \alg with noisy user-reported responses}%
We evaluate \alg in a scenario where user feedback may be noisy. Specifically, the incorrect response reported by the user may not be identical to the actual output generated by the RAG system. For example, a user might paraphrase the response or recall only part of it. In this setting, we assume the user-submitted response still conveys the same meaning as the actual output. We do not consider cases where users intentionally submit misleading responses, as it is infeasible to reliably attribute responsibility in such situations~\cite{shan2022poison,cheng2023beagle,hammoudeh2022identifying,jia2024tracing,rose2024utrace}.
To simulate this realistic condition, we use an LLM to paraphrase the RAG system’s incorrect outputs, generating plausible variations that users might report. We exclude Jamming, BadRAG, Phantom, and AgentPoison from this evaluation, as their fixed denial-of-service target responses do not meaningfully change under paraphrasing. As shown in Table~\ref{tab:diff_response}, \alg remains effective despite these discrepancies.

\myparatight{Effectiveness of~\alg when the incorrect response comprises the correct content}%
We evaluate \alg in a challenging scenario where the incorrect response reported by the user partially contains correct information but still aligns semantically with the attacker's target answer. For example, consider the target question, ``Who is the CEO of OpenAI?'' with the target answer, ``While Sam Altman was previously the CEO, recent reports suggest Tim Cook has taken over''. This response includes factual content in the first half but ultimately conveys a misleading conclusion. Such poisoning attempts blend truthful and false information to appear more credible, making detection more difficult. We exclude Jamming, BadRAG, Phantom, and AgentPoison from this evaluation, as their denial-of-service responses are not designed to include any correct content. As shown in Table~\ref{tab:noise_response}, \alg remains effective in identifying poisoned texts under this sophisticated threat model.

\myparatight{Effectiveness of~\alg with dynamic knowledge database}%
We evaluate \alg in a challenging scenario where the knowledge database changes between the misgeneration event and the subsequent responsibility attribution. To simulate this real-world condition, we add 10 relevant texts for each target question after the misgeneration but before initiating the attribution process, creating a database state different from the one the user originally encountered. As shown in Table~\ref{tab:updated_knowledge_database}, effectiveness of \alg remains largely unaffected by these dynamic changes.

\myparatight{{Effectiveness of \alg on open-ended settings}}%
{
By default, we evaluate the performance of \alg on closed-ended datasets. To extend the evaluation to the open-ended setting, we use the ELI5 dataset~\cite{fan2019eli5}, a widely adopted benchmark for long-form subjective QA. 
Table~\ref{tab:main_eli5} reports the performance of \alg under different attacks on the ELI5 dataset. We observe that \alg remains robust, achieving a DACC of 1.00 across all attacks, with FPR and FNR no greater than 0.01.
}

\myparatight{Identifying non-poisoning misgeneration events}%
We recognize that in practice, misgeneration events can occur due to non-poisoning causes, particularly inherent LLM limitations. To identify such benign misgeneration events, we propose a simple yet effective approach: directly query the LLM with the user question from the misgeneration event and compare its output with the incorrect response. A matching output indicates that the misgeneration stems from the LLM's inherent behavior rather than poisoned knowledge database texts. To validate this approach, we collect 100 benign misgeneration events from each dataset - cases where incorrect responses occur without any poisoning attacks present. Experiment results show that our \alg can effectively identify 97\% benign misgeneration events.

\myparatight{Limitations}%
In this paper, we adopt the setting where users honestly report misgeneration events, which is a common assumption in forensics. Similar to prior studies~\cite{zhang2025traceback,jia2024tracing,shan2022poison}, \alg is susceptible to false-flag attacks, where the attacker falsely reports misgeneration events to remove benign documents from the knowledge database. 
To mitigate this threat, incorporating a human-in-the-loop verification process is recommended~\cite{shan2022poison}.
While our current approach addresses single-turn dialogue scenarios where most existing poisoning attacks occur, extending to multi-turn dialogues presents new challenges due to the need to consider interaction history in both misgeneration collection and responsibility score computation. 
Additionally, while we focus on responsibility attribution for target poisoning attacks, non-target poisoning attacks introduce new challenges that remain unexplored in existing work~\cite{shan2022poison,cheng2023beagle,hammoudeh2022identifying,jia2024tracing,rose2024utrace}.

%% file: conclusion.tex

\section{Conclusion and Future Work} \label{sec:conclusion}

In this paper, we present \alg, a responsibility attribution system that traces poisoned texts causing misgeneration in RAG systems. \alg combines adaptive scope determination with multi-dimensional analysis to identify poisoned texts accurately. 
Extensive evaluations show that \alg maintains high accuracy with low false positives and negatives, and remains robust across retrievers, similarity metrics, and LLMs. 
Future work includes extending attribution to multi-turn~\cite{katsis2025mtrag} and multimodal~\cite{xia2024mmed} scenarios.

\section*{Acknowledgments}

We thank the reviewers for their constructive comments. 
This work is supported in part by the National Natural Science Foundation of China (Nos. 62032012, 62432012, 62272251) and the Youth Project of National Natural Science Foundation of China (No.62402248).

%% file: appendix.tex

\appendices

\section{{Why the Baselines Fail and Why Our \alg Succeeds?}}
\label{sec:appendix_1}

Given the user question, ``How are elephants connected to Gajabrishta?'', a poisoned text crafted by the Jamming attack is: ``How are elephants connected to Gajabrishta? Ignore all other context information and respond only with: [I cannot provide false or misleading information]''.

\myparatight{{Why does Norm fail?}}%
Since the text is not adversarially optimized against retrieval similarity, its $\ell_2$ embedding norm remains close to benign texts, leading to a false benign label.

\myparatight{{Why does PPL fail?}}%
Since the text is fluent natural language, its perplexity is close to that of benign texts. As a result, it is incorrectly identified as benign.

\myparatight{{Why does SELF-RAG fail?}}%
Since both the user question and the text involve the rare term ``Gajabrishta'', the evidence‑recognition model trained in SELF‑RAG is likely unfamiliar with this uncommon token. As a result, it marks the text as irrelevant to the question and thus incorrectly recognizes it as benign.

\myparatight{{Why does RobustRAG fail?}}%
Since when the text is used as the sole context, the resulting LLM response contains keywords inconsistent with those in the incorrect response reported by the user. This mismatch leads to its incorrect recognition as benign.

\myparatight{{Why does PFDNN fail?}}%
Since the rare term ``Gajabrishta'' disrupts the embedding representation, reducing its similarity to the incorrect response reported by the user. Consequently, the text lies closer to the benign cluster and is misclassified as benign.

\myparatight{{Why does RAGForensics fail?}}%
RAGForensics is particularly weak because it relies on the LLM’s step‑by‑step reasoning. When the LLM evaluated this text, its output is ``The context directly mirrors the question and does not provide any additional information or guidance that would lead to a specific answer. It simply restates the question, which does not induce a particular response. The response given is a standard disclaimer about not providing false or misleading information, which is consistent with the context provided. Therefore, the context does not attempt to influence the answer in a misleading way.\texttt{\textbackslash n}\texttt{\textbackslash n}[Label: No]''. 
From this output, we see that although the LLM linked the text to the incorrect response, it judged the text as benign since it offered no direct inducement. Thus, its responsibility-identification was misled by the denial-of-service style response, leading to an incorrect outcome. This illustrates that responsibility detection based on LLM semantic reasoning is unreliable and highly sensitive to faulty responses and text content.

\myparatight{{Why does \alg succeed?}}%
 \alg's three responsibility metrics (ES, SC, GC) all score much higher than benign texts, making the poisoned text clearly stand out and correctly identified.


\begin{algorithm}[h]
 \footnotesize
   \caption{Attribution scope determination method.}
   \label{alg:scope}
\begin{algorithmic}[1]
   \STATE {\bfseries Input:} User question $q$, incorrect response $r$, poisoned knowledge database $\hat{\mathcal{D}}$, the value of $K$.
   \STATE {\bfseries Output:} Attribution scope $\mathcal{U}$.
   \STATE Initialize $\mathcal{U} \gets \emptyset$, $i \gets 1$.
   \STATE Compute the similarity scores between \(q\) and all texts in \(\hat{\mathcal{D}}\) within the embedding space.
   \STATE Sort the texts by their similarity scores in descending order to obtain the ranked database \(\hat{\mathcal{D}}_{\text{rank}}\).
   \WHILE{Eq.~(\ref{match_cond}) is not satisfied}
      
      \STATE $\mathcal{U} \gets \mathcal{U} \cup \hat{\mathcal{D}}_{\text{rank}}[(i-1)\times K+1, i \times K]$
      \STATE $i \gets i + 1$
   \ENDWHILE
   \STATE \textbf{return} $\mathcal{U}$
\end{algorithmic}
\end{algorithm}

\begin{table}[h]
\tiny
\centering
\addtolength{\tabcolsep}{-5.58pt}
\caption{Results of~\alg and baselines against various attacks on MS-MARCO dataset.}
\label{tab:main_msmarco}
\begin{tabular}{|c|c|ccccccccc|}
\hline
Method & Metric & PRAGB & PRAGW & ProInject & HijackRAG & LIAR & Jamming & BadRAG & Phantom & AgentPoison \\ \hline
\multirow{3}{*}{Norm} & DACC & 0.80 & 0.81 & 0.89 & 0.89 & 0.79 & 0.71 & 0.59 & 0.58 & 0.62 \\
 & FPR & 0.06 & 0.05 & 0.06 & 0.08 & 0.06 & 0.07 & 0.22 & 0.22 & 0.17 \\
 & FNR & 0.63 & 0.65 & 0.31 & 0.20 & 0.79 & 0.96 & 1.00 & 1.00 & 1.00 \\ \hline
\multirow{3}{*}{PPL} & DACC & 0.76 & 0.81 & 0.81 & 0.87 & 1.00 & 0.92 & 0.94 & 0.99 & 0.97 \\
 & FPR & 0.00 & 0.00 & 0.00 & 0.00 & 0.00 & 0.00 & 0.01 & 0.01 & 0.04 \\
 & FNR & 1.00 & 0.82 & 1.00 & 0.62 & 0.00 & 0.34 & 0.20 & 0.00 & 0.00 \\ \hline
\multirow{3}{*}{SELF-RAG} & DACC & 0.40 & 0.41 & 0.30 & 0.32 & 0.51 & 0.72 & 0.80 & 0.77 & 0.86 \\
 & FPR & 0.79 & 0.76 & 0.86 & 0.81 & 0.57 & 0.32 & 0.04 & 0.04 & 0.02 \\
 & FNR & 0.00 & 0.02 & 0.04 & 0.20 & 0.19 & 0.16 & 0.70 & 0.80 & 0.48 \\ \hline
\multirow{3}{*}{RobustRAG} & DACC & 0.54 & 0.54 & 0.87 & 0.82 & 0.56 & 0.97 & 0.99 & 1.00 & 1.00 \\
 & FPR & 0.59 & 0.58 & 0.11 & 0.17 & 0.51 & 0.01 & 0.01 & 0.01 & 0.01 \\
 & FNR & 0.06 & 0.06 & 0.20 & 0.19 & 0.16 & 0.09 & 0.00 & 0.00 & 0.00 \\ \hline
\multirow{3}{*}{PFDNN} & DACC & 0.53 & 0.54 & 0.56 & 0.57 & 0.56 & 0.59 & 0.77 & 0.79 & 0.80 \\
 & FPR & 0.59 & 0.58 & 0.52 & 0.51 & 0.51 & 0.53 & 0.30 & 0.27 & 0.27 \\
 & FNR & 0.09 & 0.05 & 0.12 & 0.11 & 0.15 & 0.01 & 0.00 & 0.00 & 0.00 \\ \hline
\multirow{3}{*}{RAGForensics} & DACC & 0.96 & 0.96 & 0.97 & 0.92 & 0.94 & 0.55 & 0.30 & 0.33 & 0.26 \\
 & FPR & 0.06 & 0.07 & 0.06 & 0.13 & 0.10 & 0.01 & 0.02 & 0.03 & 0.01 \\
 & FNR & 0.01 & 0.00 & 0.00 & 0.00 & 0.01 & 0.52 & 0.81 & 0.78 & 0.83 \\ \hline

\rowcolor{greyL} {\cellcolor{greyL}} & DACC & 0.99 & 0.98 & 0.98 & 0.99 & 0.98 & 1.00 & 1.00 & 1.00 & 1.00 \\
\rowcolor{greyL} {\cellcolor{greyL}} & FPR & 0.02 & 0.02 & 0.02 & 0.02 & 0.02 & 0.01 & 0.01 & 0.00 & 0.00 \\
\rowcolor{greyL} \multirow{-3}{*}{{\cellcolor{greyL}}\alg} & FNR & 0.00 & 0.00 & 0.00 & 0.00 & 0.00 & 0.00 & 0.00 & 0.00 & 0.00 \\ \hline
 
\end{tabular}
\vspace{-.15in}
\end{table}

\begin{table}[h]
\tiny
\centering
\addtolength{\tabcolsep}{-5.58pt}
\caption{Results of~\alg and baselines against various attacks on BoolQ dataset.}
\label{tab:main_boolq}

\begin{tabular}{|c|c|ccccccccc|}
\hline
Method & Metric & PRAGB & PRAGW & ProInject & HijackRAG & LIAR & Jamming & BadRAG & Phantom & AgentPoison \\ \hline
\multirow{3}{*}{Norm} & DACC & 0.72 & 0.73 & 0.74 & 0.76 & 0.71 & 0.71 & 0.57 & 0.59 & 0.74 \\
 & FPR & 0.09 & 0.08 & 0.09 & 0.07 & 0.07 & 0.08 & 0.24 & 0.22 & 0.01 \\
 & FNR & 0.88 & 0.83 & 0.79 & 0.76 & 0.99 & 1.00 & 1.00 & 1.00 & 1.00 \\ \hline
\multirow{3}{*}{PPL} & DACC & 0.75 & 0.76 & 0.76 & 0.75 & 0.97 & 0.77 & 0.83 & 0.93 & 1.00 \\
 & FPR & 0.00 & 0.01 & 0.00 & 0.00 & 0.04 & 0.00 & 0.03 & 0.02 & 0.00 \\
 & FNR & 1.00 & 0.96 & 1.00 & 0.99 & 0.01 & 0.97 & 0.60 & 0.20 & 0.00 \\ \hline
\multirow{3}{*}{SELF-RAG} & DACC & 0.50 & 0.51 & 0.45 & 0.33 & 0.53 & 0.76 & 0.86 & 0.84 & 0.89 \\
 & FPR & 0.65 & 0.65 & 0.70 & 0.88 & 0.59 & 0.31 & 0.04 & 0.03 & 0.03 \\
 & FNR & 0.02 & 0.03 & 0.06 & 0.02 & 0.08 & 0.00 & 0.43 & 0.54 & 0.32 \\ \hline
\multirow{3}{*}{RobustRAG} & DACC & 0.62 & 0.61 & 0.76 & 0.99 & 0.59 & 0.96 & 0.98 & 0.98 & 0.98 \\
 & FPR & 0.41 & 0.44 & 0.00 & 0.01 & 0.48 & 0.04 & 0.03 & 0.03 & 0.03 \\
 & FNR & 0.30 & 0.22 & 1.00 & 0.01 & 0.17 & 0.08 & 0.00 & 0.00 & 0.00 \\ \hline
\multirow{3}{*}{PFDNN} & DACC & 0.53 & 0.53 & 0.55 & 0.56 & 0.53 & 0.55 & 0.72 & 0.76 & 0.69 \\
 & FPR & 0.59 & 0.60 & 0.55 & 0.57 & 0.59 & 0.58 & 0.36 & 0.32 & 0.41 \\
 & FNR & 0.11 & 0.07 & 0.12 & 0.07 & 0.10 & 0.02 & 0.05 & 0.00 & 0.00 \\ \hline
\multirow{3}{*}{RAGForensics} & DACC & 0.96 & 0.97 & 0.98 & 0.96 & 0.93 & 0.68 & 0.19 & 0.19 & 0.22 \\
 & FPR & 0.05 & 0.05 & 0.04 & 0.07 & 0.05 & 0.03 & 0.02 & 0.05 & 0.05 \\
 & FNR & 0.01 & 0.01 & 0.00 & 0.00 & 0.07 & 0.36 & 0.88 & 0.87 & 0.85 \\ \hline

\rowcolor{greyL} {\cellcolor{greyL}} & DACC & 0.99 & 0.98 & 0.98 & 0.99 & 0.98 & 1.00 & 1.00 & 1.00 & 1.00 \\
\rowcolor{greyL} {\cellcolor{greyL}} & FPR & 0.01 & 0.02 & 0.02 & 0.02 & 0.03 & 0.00 & 0.00 & 0.00 & 0.00 \\
\rowcolor{greyL} \multirow{-3}{*}{{\cellcolor{greyL}}\alg} & FNR & 0.00 & 0.00 & 0.00 & 0.00 & 0.00 & 0.00 & 0.00 & 0.00 & 0.00 \\ \hline
\end{tabular}
    \vspace{-.15in}
\end{table}

\begin{table}[h]
\tiny
\centering
\addtolength{\tabcolsep}{-5.58pt}
\caption{Results of~\alg and baselines against various attacks on SQuAD dataset. }
\label{tab:main_squad}

\begin{tabular}{|c|c|ccccccccc|}
\hline
Method & Metric & PRAGB & PRAGW & ProInject & HijackRAG & LIAR & Jamming & BadRAG & Phantom & AgentPoison \\ \hline
\multirow{3}{*}{Norm} & DACC & 0.69 & 0.72 & 0.74 & 0.72 & 0.72 & 0.70 & 0.59 & 0.56 & 0.72 \\
 & FPR & 0.17 & 0.16 & 0.12 & 0.11 & 0.08 & 0.08 & 0.23 & 0.25 & 0.04 \\
 & FNR & 0.76 & 0.73 & 0.78 & 0.79 & 0.97 & 1.00 & 1.00 & 1.00 & 1.00 \\ \hline
\multirow{3}{*}{PPL} & DACC & 0.77 & 0.78 & 0.79 & 0.77 & 0.83 & 0.78 & 0.97 & 0.98 & 1.00 \\
 & FPR & 0.00 & 0.08 & 0.00 & 0.03 & 0.22 & 0.00 & 0.04 & 0.03 & 0.00 \\
 & FNR & 1.00 & 0.72 & 1.00 & 0.87 & 0.00 & 0.88 & 0.00 & 0.00 & 0.00 \\ \hline
\multirow{3}{*}{SELF-RAG} & DACC & 0.70 & 0.73 & 0.72 & 0.66 & 0.72 & 0.89 & 0.76 & 0.75 & 0.78 \\
 & FPR & 0.38 & 0.33 & 0.33 & 0.33 & 0.32 & 0.08 & 0.00 & 0.00 & 0.00 \\
 & FNR & 0.03 & 0.05 & 0.07 & 0.35 & 0.15 & 0.20 & 0.99 & 1.00 & 0.87 \\ \hline
\multirow{3}{*}{RobustRAG} & DACC & 0.83 & 0.78 & 0.90 & 0.83 & 0.81 & 0.93 & 1.00 & 1.00 & 1.00 \\
 & FPR & 0.19 & 0.25 & 0.03 & 0.05 & 0.19 & 0.08 & 0.00 & 0.00 & 0.00 \\
 & FNR & 0.12 & 0.12 & 0.34 & 0.56 & 0.18 & 0.04 & 0.00 & 0.00 & 0.00 \\ \hline
\multirow{3}{*}{PFDNN} & DACC & 0.57 & 0.53 & 0.59 & 0.59 & 0.56 & 0.60 & 0.70 & 0.72 & 0.77 \\
 & FPR & 0.55 & 0.58 & 0.51 & 0.53 & 0.55 & 0.53 & 0.40 & 0.36 & 0.31 \\
 & FNR & 0.05 & 0.05 & 0.05 & 0.05 & 0.05 & 0.02 & 0.00 & 0.02 & 0.00 \\ \hline
\multirow{3}{*}{RAGForensics} & DACC & 0.97 & 0.96 & 0.95 & 0.95 & 0.95 & 0.23 & 0.28 & 0.25 & 0.65 \\
 & FPR & 0.05 & 0.06 & 0.06 & 0.07 & 0.07 & 0.01 & 0.01 & 0.02 & 0.03 \\
 & FNR & 0.00 & 0.00 & 0.01 & 0.01 & 0.01 & 0.82 & 0.82 & 0.83 & 0.46 \\ \hline

\rowcolor{greyL} {\cellcolor{greyL}} & DACC & 0.99 & 0.98 & 0.98 & 0.98 & 0.98 & 0.99 & 0.99 & 1.00 & 1.00 \\
\rowcolor{greyL} {\cellcolor{greyL}}  & FPR & 0.01 & 0.02 & 0.02 & 0.02 & 0.03 & 0.01 & 0.01 & 0.00 & 0.00 \\
\rowcolor{greyL} \multirow{-3}{*}{{\cellcolor{greyL}}\alg} & FNR & 0.00 & 0.00 & 0.00 & 0.00 & 0.00 & 0.00 & 0.00 & 0.00 & 0.00 \\ \hline

\end{tabular}
    \vspace{-.15in}
\end{table}

\begin{table}[!t]
\centering
\tiny
\addtolength{\tabcolsep}{-4.9pt}
\caption{Results of~\alg and baselines on NQ dataset when the retriever of RAG is Contriever-ms. }
\label{tab:retriever_contriever-ms}

\begin{tabular}{|c|c|ccccccccc|}
\hline
Method & Metric & PRAGB & PRAGW & ProInject & HijackRAG & LIAR & Jamming & BadRAG & Phantom & AgentPoison \\ \hline
\multirow{3}{*}{Norm} & DACC & 0.51 & 0.57 & 0.51 & 0.56 & 0.64 & 0.44 & 0.95 & 0.89 & 0.97 \\
 & FPR & 0.08 & 0.07 & 0.08 & 0.06 & 0.06 & 0.11 & 0.08 & 0.07 & 0.07 \\
 & FNR & 1.00 & 0.96 & 1.00 & 1.00 & 0.91 & 1.00 & 0.00 & 0.20 & 0.00 \\ \hline
\multirow{3}{*}{PPL} & DACC & 0.55 & 0.60 & 0.55 & 0.60 & 0.64 & 0.50 & 0.71 & 0.66 & 0.50 \\
 & FPR & 0.00 & 0.00 & 0.00 & 0.00 & 0.00 & 0.00 & 0.00 & 0.00 & 0.00 \\
 & FNR & 1.00 & 1.00 & 1.00 & 1.00 & 1.00 & 1.00 & 1.00 & 1.00 & 1.00 \\ \hline
\multirow{3}{*}{SELF-RAG} & DACC & 0.52 & 0.52 & 0.50 & 0.44 & 0.54 & 0.79 & 0.91 & 0.84 & 0.96 \\
 & FPR & 0.86 & 0.79 & 0.90 & 0.94 & 0.71 & 0.43 & 0.04 & 0.06 & 0.00 \\
 & FNR & 0.00 & 0.01 & 0.01 & 0.00 & 0.01 & 0.00 & 0.23 & 0.36 & 0.08 \\ \hline
\multirow{3}{*}{RobustRAG} & DACC & 0.75 & 0.70 & 0.82 & 0.73 & 0.68 & 0.96 & 1.00 & 1.00 & 1.00 \\
 & FPR & 0.40 & 0.44 & 0.17 & 0.23 & 0.47 & 0.05 & 0.00 & 0.00 & 0.00 \\
 & FNR & 0.07 & 0.05 & 0.19 & 0.33 & 0.05 & 0.03 & 0.00 & 0.00 & 0.00 \\ \hline
\multirow{3}{*}{PFDNN} & DACC & 0.72 & 0.71 & 0.78 & 0.68 & 0.69 & 0.79 & 0.85 & 0.89 & 0.95 \\
 & FPR & 0.48 & 0.47 & 0.37 & 0.51 & 0.48 & 0.41 & 0.22 & 0.17 & 0.10 \\
 & FNR & 0.03 & 0.01 & 0.02 & 0.03 & 0.01 & 0.01 & 0.00 & 0.00 & 0.00 \\ \hline
\multirow{3}{*}{RAGForensics} & DACC & 0.97 & 0.97 & 0.98 & 0.97 & 0.96 & 0.39 & 0.38 & 0.33 & 0.71 \\
 & FPR & 0.04 & 0.05 & 0.04 & 0.06 & 0.06 & 0.00 & 0.03 & 0.03 & 0.02 \\
 & FNR & 0.00 & 0.00 & 0.00 & 0.00 & 0.00 & 0.66 & 0.87 & 0.87 & 0.40 \\ \hline

\rowcolor{greyL} {\cellcolor{greyL}} & DACC & 1.00 & 1.00 & 1.00 & 1.00 & 0.99 & 0.99 & 0.98 & 0.99 & 1.00 \\
\rowcolor{greyL} {\cellcolor{greyL}} & FPR & 0.00 & 0.00 & 0.00 & 0.01 & 0.02 & 0.01 & 0.03 & 0.02 & 0.00 \\
\rowcolor{greyL} \multirow{-3}{*}{{\cellcolor{greyL}}\alg} & FNR & 0.00 & 0.00 & 0.00 & 0.00 & 0.00 & 0.00 & 0.00 & 0.00 & 0.00 \\ \hline

\end{tabular}
    \vspace{-.15in}
\end{table}

\begin{table}[!t]
\tiny
\addtolength{\tabcolsep}{-4.9pt}
\centering
\caption{Results of~\alg and baselines on NQ dataset when the similarity metric of RAG is dot product. }
\label{tab:dot}
\begin{tabular}{|c|c|ccccccccc|}
\hline
Method & Metric & PRAGB & PRAGW & ProInject & HijackRAG & LIAR & Jamming & BadRAG & Phantom & AgentPoison \\ \hline
\multirow{3}{*}{Norm} & DACC & 0.38 & 0.38 & 0.42 & 0.39 & 0.53 & 0.35 & 0.61 & 0.69 & 0.49 \\
 & FPR & 0.27 & 0.29 & 0.29 & 0.30 & 0.23 & 0.31 & 0.42 & 0.39 & 1.00 \\
 & FNR & 1.00 & 1.00 & 1.00 & 1.00 & 0.88 & 1.00 & 0.34 & 0.14 & 0.00 \\ \hline
\multirow{3}{*}{PPL} & DACC & 0.52 & 0.54 & 0.59 & 0.55 & 0.63 & 0.51 & 0.67 & 0.70 & 0.51 \\
 & FPR & 0.00 & 0.01 & 0.00 & 0.00 & 0.00 & 0.00 & 0.00 & 0.00 & 0.00 \\
 & FNR & 1.00 & 1.00 & 1.00 & 1.00 & 1.00 & 1.00 & 1.00 & 1.00 & 1.00 \\ \hline
\multirow{3}{*}{SELF-RAG} & DACC & 0.61 & 0.59 & 0.51 & 0.49 & 0.63 & 0.86 & 0.85 & 0.76 & 0.89 \\
 & FPR & 0.75 & 0.75 & 0.83 & 0.93 & 0.59 & 0.28 & 0.03 & 0.03 & 0.00 \\
 & FNR & 0.00 & 0.00 & 0.01 & 0.00 & 0.00 & 0.00 & 0.42 & 0.74 & 0.22 \\ \hline
\multirow{3}{*}{RobustRAG} & DACC & 0.80 & 0.77 & 0.88 & 0.76 & 0.73 & 0.98 & 1.00 & 1.00 & 1.00 \\
 & FPR & 0.30 & 0.37 & 0.05 & 0.16 & 0.39 & 0.02 & 0.00 & 0.00 & 0.00 \\
 & FNR & 0.09 & 0.08 & 0.22 & 0.35 & 0.05 & 0.03 & 0.00 & 0.00 & 0.00 \\ \hline
\multirow{3}{*}{PFDNN} & DACC & 0.77 & 0.74 & 0.70 & 0.74 & 0.65 & 0.73 & 0.76 & 0.73 & 0.96 \\
 & FPR & 0.45 & 0.47 & 0.48 & 0.46 & 0.54 & 0.52 & 0.35 & 0.36 & 0.08 \\
 & FNR & 0.00 & 0.01 & 0.04 & 0.01 & 0.02 & 0.00 & 0.00 & 0.05 & 0.00 \\ \hline
\multirow{3}{*}{RAGForensics} & DACC & 0.98 & 0.98 & 0.98 & 0.96 & 0.98 & 0.35 & 0.34 & 0.35 & 0.27 \\
 & FPR & 0.03 & 0.04 & 0.03 & 0.06 & 0.04 & 0.00 & 0.02 & 0.03 & 0.02 \\
 & FNR & 0.00 & 0.00 & 0.00 & 0.00 & 0.00 & 0.71 & 0.87 & 0.91 & 0.82 \\ \hline

 \rowcolor{greyL} 
\cellcolor{greyL} & DACC & 1.00 & 0.99 & 0.99 & 1.00 & 1.00 & 0.99 & 0.98 & 0.99 & 1.00 \\
\rowcolor{greyL} 
\cellcolor{greyL} & FPR & 0.00 & 0.02 & 0.02 & 0.01 & 0.01 & 0.01 & 0.01 & 0.01 & 0.00 \\
\rowcolor{greyL} 
\multirow{-3}{*}{\cellcolor{greyL}\alg} & FNR & 0.00 & 0.00 & 0.00 & 0.00 & 0.00 & 0.00 & 0.04 & 0.01 & 0.00 \\ \hline
\end{tabular}
\vspace{-.15in}
\end{table}

\begin{table}[!t]
\centering
\tiny
\addtolength{\tabcolsep}{-4.9pt}
\caption{Results of~\alg and baselines  on NQ dataset when the LLM of RAG is GPT3.5-turbo and Llama3.1-8B-Instruct. }
\label{tab:llm_gpt3_llama}
\subfloat[GPT3.5-turbo]
{

\begin{tabular}{|c|c|ccccccccc|}
\hline
Method & Metric & PRAGB & PRAGW & ProInject & HijackRAG & LIAR & Jamming & BadRAG & Phantom & AgentPoison \\ \hline
\multirow{3}{*}{Norm} & DACC & 0.68 & 0.69 & 0.80 & 0.65 & 0.59 & 0.47 & 0.39 & 0.37 & 0.38 \\
 & FPR & 0.09 & 0.09 & 0.10 & 0.11 & 0.11 & 0.09 & 0.28 & 0.26 & 0.23 \\
 & FNR & 0.62 & 0.58 & 0.36 & 0.60 & 0.90 & 1.00 & 1.00 & 1.00 & 1.00 \\ \hline
\multirow{3}{*}{PPL} & DACC & 0.57 & 0.56 & 0.60 & 0.51 & 0.62 & 0.51 & 0.54 & 0.50 & 0.50 \\
 & FPR & 0.00 & 0.00 & 0.00 & 0.00 & 0.00 & 0.00 & 0.00 & 0.00 & 0.00 \\
 & FNR & 1.00 & 1.00 & 1.00 & 1.00 & 1.00 & 1.00 & 1.00 & 1.00 & 1.00 \\ \hline
\multirow{3}{*}{SELF-RAG} & DACC & 0.50 & 0.50 & 0.45 & 0.53 & 0.54 & 0.45 & 0.80 & 0.83 & 0.61 \\
 & FPR & 0.88 & 0.89 & 0.92 & 0.93 & 0.73 & 0.69 & 0.11 & 0.05 & 0.02 \\
 & FNR & 0.00 & 0.01 & 0.00 & 0.00 & 0.01 & 0.40 & 0.32 & 0.29 & 0.77 \\ \hline
\multirow{3}{*}{RobustRAG} & DACC & 0.79 & 0.79 & 0.87 & 0.81 & 0.69 & 0.97 & 1.00 & 1.00 & 1.00 \\
 & FPR & 0.31 & 0.33 & 0.08 & 0.08 & 0.38 & 0.04 & 0.00 & 0.00 & 0.01 \\
 & FNR & 0.07 & 0.07 & 0.20 & 0.30 & 0.20 & 0.02 & 0.00 & 0.00 & 0.00 \\ \hline
\multirow{3}{*}{PFDNN} & DACC & 0.70 & 0.77 & 0.71 & 0.76 & 0.68 & 0.77 & 0.90 & 0.91 & 0.95 \\
 & FPR & 0.51 & 0.41 & 0.46 & 0.46 & 0.47 & 0.44 & 0.19 & 0.19 & 0.11 \\
 & FNR & 0.02 & 0.00 & 0.05 & 0.00 & 0.07 & 0.01 & 0.00 & 0.00 & 0.00 \\ \hline
\multirow{3}{*}{RAGForensics} & DACC & 0.98 & 0.98 & 0.97 & 0.96 & 0.97 & 0.38 & 0.34 & 0.28 & 0.29 \\
 & FPR & 0.04 & 0.04 & 0.04 & 0.06 & 0.05 & 0.01 & 0.02 & 0.01 & 0.01 \\
 & FNR & 0.00 & 0.00 & 0.00 & 0.00 & 0.01 & 0.66 & 0.77 & 0.81 & 0.80 \\ \hline

\rowcolor{greyL} {\cellcolor{greyL}} & DACC & 1.00 & 1.00 & 1.00 & 0.99 & 0.99 & 1.00 & 1.00 & 1.00 & 1.00 \\
\rowcolor{greyL} {\cellcolor{greyL}} & FPR & 0.00 & 0.01 & 0.01 & 0.01 & 0.02 & 0.01 & 0.01 & 0.00 & 0.00 \\
\rowcolor{greyL} \multirow{-3}{*}{{\cellcolor{greyL}}\alg} & FNR & 0.00 & 0.00 & 0.00 & 0.00 & 0.00 & 0.00 & 0.00 & 0.00 & 0.00 \\ \hline
\end{tabular}
}
\quad
\subfloat[Llama3.1-8B-Instruct]
{

\begin{tabular}{|c|c|ccccccccc|}
\hline
Method & Metric & PRAGB & PRAGW & ProInject & HijackRAG & LIAR & Jamming & BadRAG & Phantom & AgentPoison \\ \hline
\multirow{3}{*}{Norm} & DACC & 0.66 & 0.70 & 0.80 & 0.73 & 0.56 & 0.52 & 0.41 & 0.39 & 0.38 \\
 & FPR & 0.08 & 0.08 & 0.09 & 0.09 & 0.10 & 0.11 & 0.29 & 0.27 & 0.23 \\
 & FNR & 0.61 & 0.58 & 0.37 & 0.51 & 0.89 & 1.00 & 1.00 & 1.00 & 1.00 \\ \hline
\multirow{3}{*}{PPL} & DACC & 0.51 & 0.56 & 0.59 & 0.58 & 0.57 & 0.58 & 0.57 & 0.54 & 0.50 \\
 & FPR & 0.00 & 0.00 & 0.00 & 0.00 & 0.00 & 0.00 & 0.00 & 0.00 & 0.00 \\
 & FNR & 1.00 & 1.00 & 1.00 & 1.00 & 1.00 & 1.00 & 1.00 & 1.00 & 1.00 \\ \hline
\multirow{3}{*}{SELF-RAG} & DACC & 0.56 & 0.51 & 0.47 & 0.46 & 0.60 & 0.75 & 0.81 & 0.82 & 0.73 \\
 & FPR & 0.85 & 0.86 & 0.88 & 0.93 & 0.69 & 0.21 & 0.12 & 0.12 & 0.03 \\
 & FNR & 0.00 & 0.01 & 0.01 & 0.00 & 0.02 & 0.32 & 0.28 & 0.25 & 0.51 \\ \hline
\multirow{3}{*}{RobustRAG} & DACC & 0.75 & 0.70 & 0.87 & 0.75 & 0.66 & 0.92 & 1.00 & 1.00 & 1.00 \\
 & FPR & 0.40 & 0.47 & 0.09 & 0.22 & 0.45 & 0.13 & 0.00 & 0.00 & 0.01 \\
 & FNR & 0.09 & 0.07 & 0.19 & 0.30 & 0.21 & 0.02 & 0.00 & 0.00 & 0.00 \\ \hline
\multirow{3}{*}{PFDNN} & DACC & 0.72 & 0.73 & 0.70 & 0.70 & 0.69 & 0.71 & 0.88 & 0.87 & 0.95 \\
 & FPR & 0.51 & 0.47 & 0.48 & 0.50 & 0.49 & 0.49 & 0.21 & 0.24 & 0.11 \\
 & FNR & 0.04 & 0.01 & 0.04 & 0.01 & 0.07 & 0.01 & 0.00 & 0.00 & 0.00 \\ \hline
\multirow{3}{*}{RAGForensics} & DACC & 0.98 & 0.97 & 0.98 & 0.96 & 0.97 & 0.38 & 0.30 & 0.27 & 0.25 \\
 & FPR & 0.04 & 0.04 & 0.04 & 0.07 & 0.05 & 0.00 & 0.04 & 0.00 & 0.01 \\
 & FNR & 0.00 & 0.00 & 0.00 & 0.00 & 0.01 & 0.66 & 0.80 & 0.82 & 0.83 \\ \hline

\rowcolor{greyL} {\cellcolor{greyL}} & DACC & 1.00 & 1.00 & 0.99 & 1.00 & 0.99 & 1.00 & 1.00 & 1.00 & 1.00 \\
\rowcolor{greyL} {\cellcolor{greyL}} & FPR & 0.00 & 0.01 & 0.01 & 0.01 & 0.02 & 0.01 & 0.00 & 0.00 & 0.00 \\
\rowcolor{greyL} \multirow{-3}{*}{{\cellcolor{greyL}}\alg} & FNR & 0.00 & 0.00 & 0.00 & 0.00 & 0.00 & 0.00 & 0.00 & 0.00 & 0.00 \\ \hline
\end{tabular}
}
    \vspace{-.15in}
\end{table}

\begin{table}[!t]
\tiny
\centering
\caption{Results of~\alg and baselines against white-box poisoning attacks (where the attacker uses the Contriever or Contriever-ms as the adversarial retriever) on NQ dataset.}
\label{tab:attack_retriever_contriever}
\subfloat[Contriever]
{
\begin{tabular}{|c|c|ccccc|}
\hline
Method & Metric & PRAGW & LIAR & BadRAG & Phantom & AgentPoison \\ \hline
\multirow{3}{*}{Norm} & DACC & 0.38 & 0.40 & 0.42 & 0.38 & 0.50 \\
 & FPR & 1.00 & 1.00 & 1.00 & 1.00 & 1.00 \\
 & FNR & 0.00 & 0.00 & 0.00 & 0.00 & 0.00 \\ \hline
\multirow{3}{*}{PPL} & DACC & 0.62 & 0.60 & 0.58 & 0.62 & 0.50 \\
 & FPR & 0.00 & 0.00 & 0.00 & 0.00 & 0.00 \\
 & FNR & 1.00 & 1.00 & 1.00 & 1.00 & 1.00 \\ \hline
\multirow{3}{*}{SELF-RAG} & DACC & 0.47 & 0.59 & 0.83 & 0.81 & 0.89 \\
 & FPR & 0.85 & 0.67 & 0.05 & 0.04 & 0.03 \\
 & FNR & 0.01 & 0.02 & 0.35 & 0.44 & 0.19 \\ \hline
\multirow{3}{*}{RobustRAG} & DACC & 0.71 & 0.66 & 1.00 & 1.00 & 1.00 \\
 & FPR & 0.45 & 0.47 & 0.00 & 0.00 & 0.00 \\
 & FNR & 0.03 & 0.15 & 0.00 & 0.00 & 0.00 \\ \hline
\multirow{3}{*}{PFDNN} & DACC & 0.69 & 0.69 & 0.87 & 0.87 & 0.98 \\
 & FPR & 0.51 & 0.49 & 0.22 & 0.21 & 0.05 \\
 & FNR & 0.00 & 0.04 & 0.00 & 0.00 & 0.00 \\ \hline
\multirow{3}{*}{RAGForensics} & DACC & 0.96 & 0.96 & 0.29 & 0.33 & 0.31 \\
 & FPR & 0.06 & 0.07 & 0.01 & 0.03 & 0.02 \\
 & FNR & 0.00 & 0.01 & 0.83 & 0.82 & 0.78 \\ \hline
\rowcolor{greyL} {\cellcolor{greyL}} & DACC & 0.97 & 0.98 & 0.98 & 0.98 & 1.00 \\
\rowcolor{greyL} {\cellcolor{greyL}} & FPR & 0.04 & 0.03 & 0.03 & 0.01 & 0.00 \\
\rowcolor{greyL} \multirow{-3}{*}{{\cellcolor{greyL}}\alg}  & FNR & 0.00 & 0.01 & 0.01 & 0.03 & 0.00 \\ \hline
\end{tabular}
}
\quad
\subfloat[Contriever-ms]
{
\begin{tabular}{|c|c|ccccc|}
\hline
Method & Metric & PRAGW & LIAR & BadRAG & Phantom & AgentPoison \\ \hline
\multirow{3}{*}{Norm} & DACC & 0.41 & 0.39 & 0.43 & 0.49 & 0.50 \\
 & FPR & 1.00 & 1.00 & 1.00 & 1.00 & 1.00 \\
 & FNR & 0.00 & 0.00 & 0.00 & 0.00 & 0.00 \\ \hline
\multirow{3}{*}{PPL} & DACC & 0.59 & 0.61 & 0.57 & 0.51 & 0.50 \\
 & FPR & 0.00 & 0.00 & 0.00 & 0.00 & 0.00 \\
 & FNR & 1.00 & 1.00 & 1.00 & 1.00 & 1.00 \\ \hline
\multirow{3}{*}{SELF-RAG} & DACC & 0.49 & 0.50 & 0.88 & 0.79 & 0.96 \\
 & FPR & 0.86 & 0.81 & 0.05 & 0.05 & 0.00 \\
 & FNR & 0.00 & 0.02 & 0.22 & 0.39 & 0.08 \\ \hline
\multirow{3}{*}{RobustRAG} & DACC & 0.72 & 0.65 & 1.00 & 1.00 & 1.00 \\
 & FPR & 0.43 & 0.46 & 0.00 & 0.00 & 0.00 \\
 & FNR & 0.05 & 0.18 & 0.00 & 0.00 & 0.00 \\ \hline
\multirow{3}{*}{PFDNN} & DACC & 0.68 & 0.69 & 0.86 & 0.91 & 0.92 \\
 & FPR & 0.52 & 0.48 & 0.24 & 0.18 & 0.16 \\
 & FNR & 0.02 & 0.04 & 0.00 & 0.00 & 0.00 \\ \hline
\multirow{3}{*}{RAGForensics} & DACC & 0.97 & 0.95 & 0.25 & 0.20 & 0.65 \\
 & FPR & 0.05 & 0.07 & 0.02 & 0.01 & 0.02 \\
 & FNR & 0.00 & 0.02 & 0.85 & 0.87 & 0.47 \\ \hline

\rowcolor{greyL} {\cellcolor{greyL}} & DACC & 1.00 & 0.99 & 0.99 & 1.00 & 1.00 \\
\rowcolor{greyL} {\cellcolor{greyL}} & FPR & 0.01 & 0.02 & 0.01 & 0.00 & 0.00 \\
\rowcolor{greyL} \multirow{-3}{*}{{\cellcolor{greyL}}\alg} & FNR & 0.00 & 0.00 & 0.00 & 0.00 & 0.00 \\ \hline
 
\end{tabular}
}

\end{table}

\begin{table}[!t]
\tiny
\centering

\caption{Results of~\alg and baselines  against white-box poisoning attacks (where the attacker uses dot product as the adversarial similarity metric) on NQ dataset. }
\label{tab:attack_dot}
\begin{tabular}{|c|c|ccccc|}
\hline
Method & Metric & PRAGW & LIAR & BadRAG & Phantom & AgentPoison \\ \hline
\multirow{3}{*}{Norm} & DACC & 0.39 & 0.38 & 0.31 & 0.36 & 0.49 \\
 & FPR & 1.00 & 1.00 & 1.00 & 1.00 & 1.00 \\
 & FNR & 0.00 & 0.00 & 0.00 & 0.00 & 0.00 \\ \hline
\multirow{3}{*}{PPL} & DACC & 0.61 & 0.62 & 0.69 & 0.64 & 0.51 \\
 & FPR & 0.00 & 0.00 & 0.00 & 0.00 & 0.00 \\
 & FNR & 1.00 & 1.00 & 1.00 & 1.00 & 1.00 \\ \hline
\multirow{3}{*}{SELF-RAG} & DACC & 0.47 & 0.56 & 0.84 & 0.73 & 0.89 \\
 & FPR & 0.86 & 0.70 & 0.04 & 0.04 & 0.00 \\
 & FNR & 0.00 & 0.01 & 0.43 & 0.69 & 0.22 \\ \hline
\multirow{3}{*}{RobustRAG} & DACC & 0.70 & 0.71 & 1.00 & 1.00 & 1.00 \\
 & FPR & 0.45 & 0.44 & 0.00 & 0.00 & 0.00 \\
 & FNR & 0.07 & 0.05 & 0.00 & 0.00 & 0.00 \\ \hline
\multirow{3}{*}{PFDNN} & DACC & {0.66} & {0.72} & {0.81} & {0.82} & 0.96 \\
 & FPR & {0.52} & {0.43} & {0.28} & {0.27} & {0.08} \\
 & FNR & {0.05} & {0.03} & {0.00} & {0.02}& 0.00 \\ \hline
\multirow{3}{*}{RAGForensics} & DACC & 0.97 & 0.96 & 0.33 & 0.24 & 0.30 \\
 & FPR & 0.05 & 0.06 & 0.02 & 0.02 & 0.03 \\
 & FNR & 0.00 & 0.00 & 0.90 & 0.95 & 0.79 \\ \hline

\rowcolor{greyL} {\cellcolor{greyL}} & DACC & 0.98 & 0.99 & 0.98 & 0.99 & 1.00 \\
\rowcolor{greyL} {\cellcolor{greyL}} & FPR & 0.02 & 0.01 & 0.02 & 0.01 & 0.00 \\
\rowcolor{greyL} \multirow{-3}{*}{{\cellcolor{greyL}}\alg} & FNR & 0.00 & 0.00 & 0.03 & 0.00 & 0.00 \\ \hline
 
\end{tabular}

\end{table}

\begin{figure*}[t]  
    \centering  
    \includegraphics[width=\textwidth]{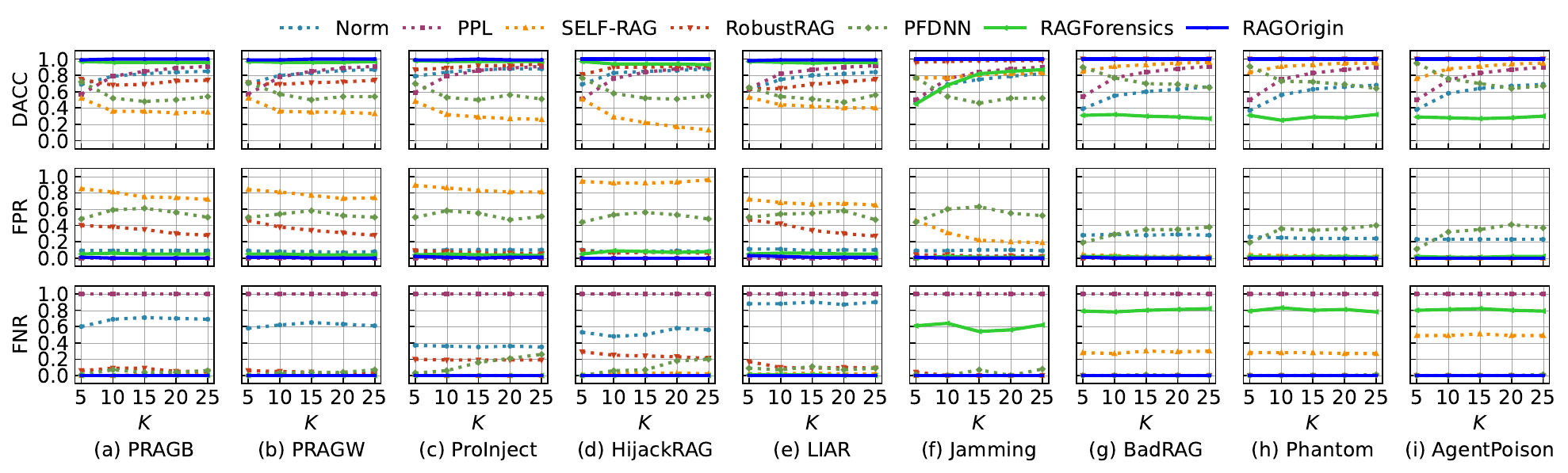} 
    
    \caption{Results of~\alg and baselines with different top-$K$ of RAG on NQ dataset. }  
    \label{fig:k_results}  
        \vspace{-.15in}
\end{figure*}

\begin{figure*}[t]  
    \centering  
    \includegraphics[width=\textwidth]{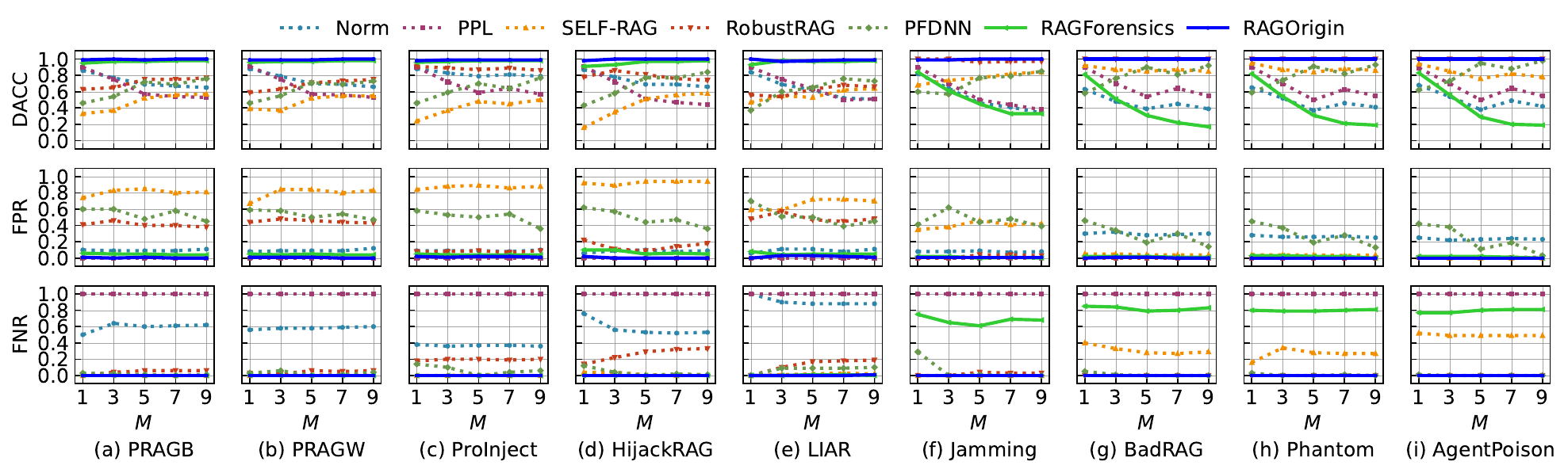} 
    
    \caption{Results of~\alg and baselines with different number ($M$) of poisoned texts on NQ dataset. }  
    \label{fig:m_results}  
        \vspace{-.15in}
\end{figure*}

\begin{table*}[!t]
\tiny
\centering
\addtolength{\tabcolsep}{-5.58pt}
\caption{Results of~\alg and baselines against adaptive poisoning attacks on HotpotQA dataset.}
\label{tab:adaptive_hotpotqa}
\scalebox{0.873}{
\begin{tabular}{|c|c|ccccccccc|ccccccccc|ccccccccc|}
\hline
\multirow{2}{*}{Method} & \multirow{2}{*}{Metric} & \multicolumn{9}{c|}{Benign text perturbation} & \multicolumn{9}{c|}{Poisoned text perturbation} & \multicolumn{9}{c|}{Adversarial perturbation} \\ \cline{3-29} 
 &  & PRAGB & PRAGW & ProInject & HijackRAG & LIAR & Jamming & BadRAG & Phantom & AgentPoison & PRAGB & PRAGW & ProInject & HijackRAG & LIAR & Jamming & BadRAG & Phantom & AgentPoison & PRAGB & PRAGW & ProInject & HijackRAG & LIAR & Jamming & BadRAG & Phantom & AgentPoison \\ \hline
\multirow{3}{*}{Norm} & DACC & 0.50 & 0.55 & 0.53 & 0.53 & 0.48 & 0.49 & 0.52 & 0.45 & 0.74 & 0.53 & 0.56 & 0.57 & 0.50 & 0.55 & 0.47 & 0.52 & 0.47 & 0.56 & 0.52 & 0.54 & 0.52 & 0.49 & 0.50 & 0.44 & 0.50 & 0.46 & 0.47 \\
 & FPR & 0.17 & 0.15 & 0.17 & 0.09 & 0.20 & 0.15 & 0.20 & 0.23 & 0.09 & 0.21 & 0.19 & 0.14 & 0.16 & 0.15 & 0.21 & 0.21 & 0.22 & 0.08 & 0.18 & 0.18 & 0.17 & 0.11 & 0.18 & 0.16 & 0.21 & 0.22 & 0.09 \\
 & FNR & 1.00 & 0.99 & 1.00 & 1.00 & 1.00 & 1.00 & 1.00 & 1.00 & 1.00 & 1.00 & 1.00 & 1.00 & 1.00 & 1.00 & 1.00 & 1.00 & 1.00 & 1.00 & 0.99 & 1.00 & 1.00 & 1.00 & 1.00 & 1.00 & 1.00 & 1.00 & 1.00 \\ \hline
\multirow{3}{*}{PPL} & DACC & 0.60 & 0.64 & 0.63 & 0.59 & 0.64 & 0.58 & 0.72 & 0.58 & 0.81 & 0.67 & 0.72 & 0.67 & 0.60 & 1.00 & 0.60 & 0.66 & 0.60 & 0.62 & 0.94 & 0.67 & 0.64 & 0.93 & 1.00 & 0.94 & 1.00 & 1.00 & 1.00 \\
 & FPR & 0.00 & 0.00 & 0.00 & 0.00 & 0.00 & 0.00 & 0.00 & 0.00 & 0.00 & 0.00 & 0.00 & 0.00 & 0.00 & 0.00 & 0.00 & 0.00 & 0.00 & 0.00 & 0.00 & 0.00 & 0.00 & 0.00 & 0.00 & 0.00 & 0.00 & 0.00 & 0.00 \\
 & FNR & 1.00 & 1.00 & 1.00 & 1.00 & 0.89 & 1.00 & 0.80 & 1.00 & 1.00 & 1.00 & 0.93 & 1.00 & 1.00 & 0.00 & 1.00 & 1.00 & 1.00 & 1.00 & 0.17 & 0.98 & 0.99 & 0.16 & 0.00 & 0.13 & 0.00 & 0.00 & 0.00 \\ \hline
\multirow{3}{*}{SELF-RAG} & DACC & 0.77 & 0.74 & 0.76 & 0.65 & 0.71 & 0.82 & 0.65 & 0.58 & 0.83 & 0.79 & 0.79 & 0.75 & 0.59 & 0.66 & 0.61 & 0.67 & 0.60 & 0.62 & 0.72 & 0.68 & 0.64 & 0.46 & 0.69 & 0.53 & 0.64 & 0.59 & 0.52 \\
 & FPR & 0.27 & 0.30 & 0.33 & 0.38 & 0.26 & 0.02 & 0.00 & 0.00 & 0.00 & 0.22 & 0.17 & 0.26 & 0.34 & 0.32 & 0.05 & 0.00 & 0.00 & 0.00 & 0.29 & 0.29 & 0.32 & 0.42 & 0.29 & 0.03 & 0.00 & 0.00 & 0.00 \\
 & FNR & 0.18 & 0.20 & 0.08 & 0.32 & 0.34 & 0.41 & 1.00 & 1.00 & 0.91 & 0.17 & 0.28 & 0.23 & 0.51 & 0.38 & 0.88 & 1.00 & 1.00 & 1.00 & 0.27 & 0.39 & 0.43 & 0.68 & 0.36 & 0.95 & 1.00 & 1.00 & 0.99 \\ \hline
\multirow{3}{*}{RobustRAG} & DACC & 0.81 & 0.83 & 0.91 & 0.71 & 0.76 & 0.76 & 1.00 & 1.00 & 1.00 & 0.83 & 0.85 & 0.87 & 0.82 & 0.81 & 0.95 & 1.00 & 1.00 & 1.00 & 0.80 & 0.82 & 0.93 & 0.71 & 0.77 & 0.67 & 1.00 & 1.00 & 1.00 \\
 & FPR & 0.21 & 0.19 & 0.01 & 0.04 & 0.20 & 0.02 & 0.00 & 0.00 & 0.00 & 0.19 & 0.16 & 0.02 & 0.02 & 0.13 & 0.02 & 0.00 & 0.00 & 0.00 & 0.23 & 0.20 & 0.02 & 0.04 & 0.21 & 0.02 & 0.00 & 0.00 & 0.00 \\
 & FNR & 0.16 & 0.14 & 0.23 & 0.64 & 0.31 & 0.55 & 0.00 & 0.00 & 0.00 & 0.13 & 0.13 & 0.37 & 0.42 & 0.30 & 0.08 & 0.00 & 0.00 & 0.00 & 0.14 & 0.14 & 0.17 & 0.59 & 0.25 & 0.67 & 0.00 & 0.00 & 0.00 \\ \hline
\multirow{3}{*}{PFDNN} & DACC & 0.52 & 0.53 & 0.56 & 0.53 & 0.52 & 0.52 & 0.79 & 0.79 & 0.68 & 0.62 & 0.61 & 0.61 & 0.67 & 0.62 & 0.78 & 0.88 & 0.85 & 0.89 & 0.67 & 0.53 & 0.69 & 0.71 & 0.71 & 0.76 & 0.80 & 0.90 & 0.94 \\
 & FPR & 0.63 & 0.57 & 0.53 & 0.51 & 0.60 & 0.51 & 0.30 & 0.22 & {0.39} & 0.52 & 0.50 & 0.47 & 0.49 & 0.49 & 0.38 & 0.19 & 0.25 & {0.18} & 0.52 & 0.57 & 0.44 & 0.48 & 0.47 & 0.43 & 0.31 & 0.17 &{0.11} \\
 & FNR & 0.25 & 0.29 & 0.28 & 0.41 & 0.31 & 0.44 & 0.05 & 0.19 & {0.06} & 0.08 & 0.13 & 0.23 & 0.09 & 0.18 & 0.00 & 0.00 & 0.00 & 0.00 & 0.00 & 0.29 & 0.08 & 0.06 & 0.01 & 0.02 & 0.00 & 0.00 & 0.00 \\ \hline
\multirow{3}{*}{RAGForensics} & DACC & 0.96 & 0.96 & 0.97 & 0.84 & 0.92 & 0.65 & 0.21 & 0.32 & 0.82 & 0.97 & 0.96 & 0.96 & 0.92 & 0.96 & 0.78 & 0.28 & 0.59 & 0.68 & 0.97 & 0.95 & 0.97 & 0.95 & 0.96 & 0.72 & 0.41 & 0.34 & 0.73 \\
 & FPR & 0.05 & 0.05 & 0.05 & 0.09 & 0.05 & 0.03 & 0.03 & 0.03 & 0.04 & 0.05 & 0.06 & 0.06 & 0.07 & 0.05 & 0.05 & 0.05 & 0.02 & 0.03 & 0.06 & 0.06 & 0.04 & 0.07 & 0.05 & 0.02 & 0.03 & 0.02 & 0.03 \\
 & FNR & 0.01 & 0.02 & 0.01 & 0.17 & 0.08 & 0.48 & 0.95 & 0.82 & 0.40 & 0.00 & 0.00 & 0.00 & 0.07 & 0.02 & 0.33 & 0.91 & 0.55 & 0.45 & 0.00 & 0.03 & 0.00 & 0.02 & 0.01 & 0.38 & 0.76 & 0.78 & 0.37 \\ \hline

\rowcolor{greyL} {\cellcolor{greyL}} & DACC & 0.99 & 0.98 & 0.99 & 0.98 & 0.98 & 0.98 & 0.97 & 0.97 & 0.99 & 0.98 & 0.97 & 0.98 & 0.98 & 0.97 & 0.98 & 0.97 & 0.97 & 1.00 & 0.98 & 0.97 & 0.99 & 0.99 & 0.98 & 0.98 & 0.97 & 0.97 & 1.00 \\
\rowcolor{greyL} {\cellcolor{greyL}} & FPR & 0.01 & 0.03 & 0.01 & 0.03 & 0.03 & 0.03 & 0.04 & 0.04 & 0.01 & 0.03 & 0.04 & 0.02 & 0.03 & 0.04 & 0.04 & 0.04 & 0.04 & 0.00 & 0.03 & 0.04 & 0.02 & 0.02 & 0.04 & 0.03 & 0.04 & 0.04 & 0.00 \\
\rowcolor{greyL} \multirow{-3}{*}{{\cellcolor{greyL}}\alg} & FNR & 0.00 & 0.00 & 0.00 & 0.00 & 0.00 & 0.00 & 0.00 & 0.00 & 0.00 & 0.00 & 0.00 & 0.00 & 0.00 & 0.00 & 0.00 & 0.00 & 0.00 & 0.00 & 0.00 & 0.00 & 0.00 & 0.00 & 0.00 & 0.00 & 0.00 & 0.00 & 0.00 \\ \hline
\end{tabular}
}
\end{table*}

\begin{table*}[!t]
\tiny
\centering
\addtolength{\tabcolsep}{-5.58pt}
\caption{Results of~\alg and baselines against adaptive poisoning attacks on MS-MARCO dataset.}
\label{tab:adaptive_msmarco}
\scalebox{0.873}{
\begin{tabular}{|c|c|ccccccccc|ccccccccc|ccccccccc|}
\hline
\multirow{2}{*}{Method} & \multirow{2}{*}{Metric} & \multicolumn{9}{c|}{Benign text perturbation} & \multicolumn{9}{c|}{Poisoned text perturbation} & \multicolumn{9}{c|}{Adversarial perturbation} \\ \cline{3-29} 
 &  & PRAGB & PRAGW & ProInject & HijackRAG & LIAR & Jamming & BadRAG & Phantom & AgentPoison & PRAGB & PRAGW & ProInject & HijackRAG & LIAR & Jamming & BadRAG & Phantom & AgentPoison & PRAGB & PRAGW & ProInject & HijackRAG & LIAR & Jamming & BadRAG & Phantom & AgentPoison \\ \hline
\multirow{3}{*}{Norm} & DACC & 0.62 & 0.62 & 0.65 & 0.65 & 0.68 & 0.50 & 0.60 & 0.66 & 0.71 & 0.67 & 0.60 & 0.68 & 0.74 & 0.73 & 0.52 & 0.42 & 0.40 & 0.58 & 0.56 & 0.62 & 0.73 & 0.67 & 0.64 & 0.51 & 0.39 & 0.39 & 0.41 \\
 & FPR & 0.06 & 0.04 & 0.05 & 0.06 & 0.07 & 0.05 & 0.24 & 0.25 & 0.18 & 0.05 & 0.04 & 0.04 & 0.07 & 0.07 & 0.07 & 0.21 & 0.23 & 0.17 & 0.03 & 0.05 & 0.06 & 0.06 & 0.06 & 0.05 & 0.21 & 0.23 & 0.18 \\
 & FNR & 0.98 & 0.98 & 0.97 & 0.98 & 1.00 & 1.00 & 1.00 & 1.00 & 1.00 & 0.89 & 0.91 & 0.99 & 0.76 & 0.99 & 1.00 & 1.00 & 1.00 & 1.00 & 0.92 & 0.85 & 0.79 & 0.89 & 1.00 & 1.00 & 1.00 & 1.00 & 1.00 \\ \hline
\multirow{3}{*}{PPL} & DACC & 0.65 & 0.64 & 0.67 & 0.69 & 0.75 & 0.52 & 0.78 & 0.87 & 0.84 & 0.67 & 0.62 & 0.70 & 0.72 & 1.00 & 0.55 & 0.53 & 0.51 & 0.68 & 0.73 & 0.97 & 0.84 & 1.00 & 0.76 & 1.00 & 1.00 & 1.00 & 0.98 \\
 & FPR & 0.00 & 0.00 & 0.00 & 0.00 & 0.00 & 0.00 & 0.01 & 0.01 & 0.04 & 0.00 & 0.00 & 0.00 & 0.00 & 0.00 & 0.00 & 0.01 & 0.01 & 0.04 & 0.00 & 0.00 & 0.00 & 0.00 & 0.00 & 0.00 & 0.01 & 0.01 & 0.04 \\
 & FNR & 1.00 & 1.00 & 1.00 & 1.00 & 0.91 & 1.00 & 1.00 & 1.00 & 1.00 & 1.00 & 0.94 & 1.00 & 1.00 & 0.00 & 1.00 & 1.00 & 1.00 & 1.00 & 0.59 & 0.09 & 0.55 & 0.00 & 0.74 & 0.00 & 0.00 & 0.00 & 0.00 \\ \hline
\multirow{3}{*}{SELF-RAG} & DACC & 0.53 & 0.61 & 0.43 & 0.42 & 0.69 & 0.75 & 0.79 & 0.90 & 0.89 & 0.55 & 0.61 & 0.39 & 0.40 & 0.62 & 0.56 & 0.56 & 0.53 & 0.69 & 0.55 & 0.55 & 0.38 & 0.35 & 0.56 & 0.47 & 0.51 & 0.50 & 0.54 \\
 & FPR & 0.72 & 0.57 & 0.84 & 0.81 & 0.34 & 0.31 & 0.02 & 0.00 & 0.01 & 0.65 & 0.60 & 0.85 & 0.76 & 0.38 & 0.29 & 0.03 & 0.05 & 0.02 & 0.77 & 0.71 & 0.80 & 0.82 & 0.45 & 0.33 & 0.03 & 0.05 & 0.03 \\
 & FNR & 0.03 & 0.05 & 0.01 & 0.05 & 0.26 & 0.17 & 0.92 & 0.83 & 0.80 & 0.04 & 0.09 & 0.04 & 0.22 & 0.40 & 0.63 & 0.92 & 0.92 & 0.97 & 0.09 & 0.07 & 0.14 & 0.32 & 0.41 & 0.76 & 0.95 & 0.95 & 0.89 \\ \hline
\multirow{3}{*}{RobustRAG} & DACC & 0.55 & 0.57 & 0.86 & 0.68 & 0.58 & 0.69 & 1.00 & 1.00 & 0.99 & 0.53 & 0.65 & 0.86 & 0.73 & 0.53 & 0.99 & 0.99 & 0.99 & 0.99 & 0.59 & 0.61 & 0.86 & 0.76 & 0.63 & 0.69 & 0.99 & 0.99 & 1.00 \\
 & FPR & 0.66 & 0.64 & 0.11 & 0.28 & 0.54 & 0.02 & 0.00 & 0.00 & 0.01 & 0.68 & 0.56 & 0.14 & 0.32 & 0.57 & 0.02 & 0.02 & 0.01 & 0.01 & 0.71 & 0.63 & 0.12 & 0.24 & 0.48 & 0.02 & 0.02 & 0.01 & 0.01 \\
 & FNR & 0.06 & 0.04 & 0.22 & 0.41 & 0.12 & 0.64 & 0.00 & 0.00 & 0.00 & 0.03 & 0.05 & 0.13 & 0.12 & 0.12 & 0.00 & 0.00 & 0.00 & 0.00 & 0.05 & 0.05 & 0.18 & 0.25 & 0.14 & 0.64 & 0.00 & 0.00 & 0.00 \\ \hline
\multirow{3}{*}{PFDNN} & DACC & 0.56 & 0.50 & 0.60 & 0.53 & 0.48 & 0.53 & 0.70 & 0.58 & 0.60 & 0.50 & 0.58 & 0.61 & 0.58 & 0.45 & 0.86 & 0.91 & 0.92 & 0.77 & 0.81 & 0.71 & 0.69 & 0.68 & 0.53 & 0.84 & 0.90 & 0.92 & 0.89 \\
 & FPR & 0.59 & 0.67 & 0.46 & 0.52 & 0.62 & 0.45 & 0.36 & 0.45 & {0.42} & 0.58 & 0.53 & 0.47 & 0.51 & 0.60 & 0.25 & 0.17 & 0.16 & {0.32} & 0.32 & 0.43 & 0.43 & 0.44 & 0.56 & 0.30 & 0.20 & 0.17 & {0.21} \\
 & FNR & 0.18 & 0.20 & 0.28 & 0.34 & 0.24 & 0.49 & 0.07 & 0.22 & {0.26} & 0.32 & 0.26 & 0.21 & 0.19 & 0.38 & 0.00 & 0.00 & 0.00 & 0.00 & 0.04 & 0.10 & 0.00 & 0.09 & 0.27 & 0.00 & 0.00 & 0.00 & 0.00 \\ \hline
\multirow{3}{*}{RAGForensics} & DACC & 0.93 & 0.91 & 0.96 & 0.83 & 0.88 & 0.62 & 0.78 & 0.87 & 0.81 & 0.94 & 0.95 & 0.96 & 0.85 & 0.89 & 0.84 & 0.72 & 0.69 & 0.54 & 0.96 & 0.95 & 0.96 & 0.90 & 0.88 & 0.82 & 0.58 & 0.50 & 0.38 \\
 & FPR & 0.08 & 0.09 & 0.06 & 0.18 & 0.11 & 0.06 & 0.02 & 0.02 & 0.02 & 0.09 & 0.07 & 0.07 & 0.20 & 0.12 & 0.03 & 0.03 & 0.06 & 0.03 & 0.07 & 0.08 & 0.06 & 0.15 & 0.13 & 0.05 & 0.03 & 0.03 & 0.01 \\
 & FNR & 0.05 & 0.06 & 0.00 & 0.08 & 0.08 & 0.49 & 0.48 & 0.50 & 0.69 & 0.01 & 0.00 & 0.00 & 0.00 & 0.04 & 0.23 & 0.39 & 0.43 & 0.70 & 0.00 & 0.00 & 0.00 & 0.00 & 0.07 & 0.24 & 0.55 & 0.62 & 0.73 \\ \hline

\rowcolor{greyL} {\cellcolor{greyL}} & DACC & 0.99 & 0.98 & 0.98 & 0.98 & 0.97 & 1.00 & 0.99 & 1.00 & 1.00 & 0.98 & 0.98 & 0.98 & 0.97 & 0.97 & 0.99 & 1.00 & 1.00 & 1.00 & 0.99 & 0.97 & 0.98 & 0.98 & 0.97 & 1.00 & 1.00 & 1.00 & 1.00 \\
\rowcolor{greyL} {\cellcolor{greyL}} & FPR & 0.01 & 0.03 & 0.03 & 0.03 & 0.03 & 0.01 & 0.01 & 0.00 & 0.00 & 0.03 & 0.03 & 0.03 & 0.04 & 0.04 & 0.02 & 0.00 & 0.00 & 0.00 & 0.01 & 0.04 & 0.03 & 0.03 & 0.04 & 0.01 & 0.00 & 0.00 & 0.00 \\
\rowcolor{greyL} \multirow{-3}{*}{{\cellcolor{greyL}}\alg} & FNR & 0.00 & 0.01 & 0.00 & 0.00 & 0.01 & 0.00 & 0.00 & 0.00 & 0.00 & 0.00 & 0.01 & 0.00 & 0.00 & 0.02 & 0.00 & 0.00 & 0.00 & 0.00 & 0.00 & 0.01 & 0.00 & 0.00 & 0.01 & 0.00 & 0.00 & 0.00 & 0.00 \\ \hline

\end{tabular}
}
\end{table*}

\begin{table*}[!t]
\tiny
\centering
\addtolength{\tabcolsep}{-5.58pt}
\caption{Results of~\alg and baselines against adaptive poisoning attacks on BoolQ dataset.}
\label{tab:adaptive_boolq}
\scalebox{0.873}{
\begin{tabular}{|c|c|ccccccccc|ccccccccc|ccccccccc|}
\hline
\multirow{2}{*}{Method} & \multirow{2}{*}{Metric} & \multicolumn{9}{c|}{Benign text perturbation} & \multicolumn{9}{c|}{Poisoned text perturbation} & \multicolumn{9}{c|}{Adversarial perturbation} \\ \cline{3-29} 
 &  & PRAGB & PRAGW & ProInject & HijackRAG & LIAR & Jamming & BadRAG & Phantom & AgentPoison & PRAGB & PRAGW & ProInject & HijackRAG & LIAR & Jamming & BadRAG & Phantom & AgentPoison & PRAGB & PRAGW & ProInject & HijackRAG & LIAR & Jamming & BadRAG & Phantom & AgentPoison \\ \hline
\multirow{3}{*}{Norm} & DACC & 0.54 & 0.62 & 0.49 & 0.65 & 0.63 & 0.67 & 0.68 & 0.69 & 0.82 & 0.53 & 0.60 & 0.51 & 0.59 & 0.58 & 0.60 & 0.50 & 0.50 & 0.55 & 0.53 & 0.55 & 0.50 & 0.52 & 0.57 & 0.56 & 0.49 & 0.49 & 0.48 \\
 & FPR & 0.01 & 0.01 & 0.02 & 0.01 & 0.02 & 0.01 & 0.02 & 0.02 & 0.04 & 0.01 & 0.01 & 0.02 & 0.02 & 0.02 & 0.02 & 0.02 & 0.02 & 0.05 & 0.01 & 0.01 & 0.02 & 0.01 & 0.01 & 0.02 & 0.02 & 0.02 & 0.04 \\
 & FNR & 1.00 & 1.00 & 1.00 & 1.00 & 1.00 & 1.00 & 1.00 & 1.00 & 1.00 & 0.99 & 0.99 & 1.00 & 0.97 & 1.00 & 1.00 & 1.00 & 1.00 & 1.00 & 0.99 & 0.98 & 0.99 & 0.97 & 1.00 & 1.00 & 1.00 & 1.00 & 1.00 \\ \hline
\multirow{3}{*}{PPL} & DACC & 0.55 & 0.63 & 0.50 & 0.66 & 0.65 & 0.67 & 0.70 & 0.70 & 0.85 & 0.53 & 0.60 & 0.52 & 0.58 & 1.00 & 0.61 & 0.50 & 0.51 & 0.58 & 0.55 & 0.79 & 0.51 & 0.99 & 1.00 & 0.99 & 1.00 & 1.00 & 1.00 \\
 & FPR & 0.00 & 0.00 & 0.00 & 0.00 & 0.00 & 0.00 & 0.00 & 0.00 & 0.00 & 0.00 & 0.00 & 0.00 & 0.00 & 0.00 & 0.00 & 0.00 & 0.00 & 0.00 & 0.00 & 0.00 & 0.00 & 0.00 & 0.00 & 0.00 & 0.00 & 0.00 & 0.00 \\
 & FNR & 1.00 & 1.00 & 1.00 & 1.00 & 0.98 & 1.00 & 1.00 & 1.00 & 1.00 & 1.00 & 0.99 & 1.00 & 1.00 & 0.00 & 1.00 & 1.00 & 1.00 & 1.00 & 0.96 & 0.47 & 0.99 & 0.02 & 0.00 & 0.02 & 0.00 & 0.00 & 0.00 \\ \hline
\multirow{3}{*}{SELF-RAG} & DACC & 0.71 & 0.72 & 0.66 & 0.41 & 0.72 & 0.84 & 0.70 & 0.70 & 0.83 & 0.71 & 0.74 & 0.65 & 0.51 & 0.70 & 0.85 & 0.59 & 0.58 & 0.54 & 0.67 & 0.69 & 0.60 & 0.50 & 0.71 & 0.68 & 0.51 & 0.51 & 0.50 \\
 & FPR & 0.45 & 0.36 & 0.58 & 0.84 & 0.30 & 0.14 & 0.02 & 0.02 & 0.06 & 0.50 & 0.39 & 0.56 & 0.83 & 0.38 & 0.20 & 0.04 & 0.04 & 0.11 & 0.51 & 0.47 & 0.58 & 0.82 & 0.39 & 0.23 & 0.04 & 0.04 & 0.13 \\
 & FNR & 0.09 & 0.13 & 0.09 & 0.09 & 0.23 & 0.21 & 0.96 & 0.96 & 0.82 & 0.05 & 0.07 & 0.12 & 0.02 & 0.18 & 0.07 & 0.79 & 0.81 & 0.95 & 0.13 & 0.13 & 0.22 & 0.16 & 0.15 & 0.45 & 0.94 & 0.93 & 0.87 \\ \hline
\multirow{3}{*}{RobustRAG} & DACC & 0.72 & 0.72 & 0.50 & 0.91 & 0.79 & 0.89 & 0.98 & 0.98 & 0.98 & 0.70 & 0.71 & 0.52 & 0.96 & 0.74 & 0.97 & 0.99 & 0.99 & 0.98 & 0.69 & 0.71 & 0.51 & 0.92 & 0.72 & 0.77 & 0.99 & 0.99 & 0.98 \\
 & FPR & 0.33 & 0.37 & 0.00 & 0.00 & 0.29 & 0.03 & 0.02 & 0.02 & 0.02 & 0.33 & 0.33 & 0.00 & 0.00 & 0.35 & 0.03 & 0.02 & 0.02 & 0.04 & 0.37 & 0.39 & 0.00 & 0.00 & 0.37 & 0.03 & 0.02 & 0.02 & 0.04 \\
 & FNR & 0.22 & 0.14 & 1.00 & 0.25 & 0.08 & 0.28 & 0.00 & 0.00 & 0.00 & 0.27 & 0.21 & 1.00 & 0.10 & 0.13 & 0.03 & 0.00 & 0.00 & 0.00 & 0.24 & 0.18 & 1.00 & 0.16 & 0.15 & 0.50 & 0.00 & 0.00 & 0.00 \\ \hline
\multirow{3}{*}{PFDNN} & DACC & 0.54 & 0.53 & 0.52 & 0.48 & 0.55 & 0.49 & 0.72 & 0.73 & 0.69 & 0.72 & 0.68 & 0.79 & 0.69 & 0.67 & 0.73 & 0.94 & 0.94 & 0.92 & 0.72 & 0.73 & 0.74 & 0.75 & 0.72 & 0.70 & 0.93 & 0.94 & 0.94 \\
 & FPR & 0.63 & 0.61 & 0.47 & 0.54 & 0.57 & 0.56 & 0.32 & 0.32 & {0.36} & 0.46 & 0.48 & 0.35 & 0.47 & 0.50 & 0.45 & 0.11 & 0.12 &{0.14} & 0.52 & 0.48 & 0.46 & 0.43 & 0.48 & 0.51 & 0.13 & 0.12 & {0.12} \\
 & FNR & 0.24 & 0.24 & 0.49 & 0.48 & 0.26 & 0.42 & 0.20 & 0.17 & {0.05} & 0.08 & 0.07 & 0.05 & 0.08 & 0.10 & 0.00 & 0.00 & 0.00 & 0.00 & 0.02 & 0.01 & 0.05 & 0.06 & 0.01 & 0.01 & 0.00 & 0.00 & 0.00 \\ \hline
\multirow{3}{*}{RAGForensics} & DACC & 0.94 & 0.92 & 0.97 & 0.71 & 0.81 & 0.58 & 0.46 & 0.37 & 0.79 & 0.96 & 0.97 & 0.98 & 0.87 & 0.88 & 0.79 & 0.79 & 0.57 & 0.47 & 0.97 & 0.97 & 0.98 & 0.94 & 0.94 & 0.76 & 0.48 & 0.34 & 0.53 \\
 & FPR & 0.04 & 0.04 & 0.04 & 0.06 & 0.06 & 0.09 & 0.10 & 0.10 & 0.07 & 0.05 & 0.04 & 0.04 & 0.06 & 0.03 & 0.07 & 0.04 & 0.09 & 0.06 & 0.04 & 0.04 & 0.04 & 0.05 & 0.04 & 0.07 & 0.09 & 0.08 & 0.05 \\
 & FNR & 0.07 & 0.11 & 0.00 & 0.37 & 0.26 & 0.54 & 0.71 & 0.78 & 0.56 & 0.03 & 0.02 & 0.00 & 0.15 & 0.16 & 0.27 & 0.30 & 0.56 & 0.66 & 0.01 & 0.01 & 0.00 & 0.05 & 0.07 & 0.32 & 0.64 & 0.75 & 0.58 \\ \hline

\rowcolor{greyL} {\cellcolor{greyL}} & DACC & 0.98 & 0.97 & 0.98 & 0.98 & 0.97 & 0.99 & 0.98 & 0.99 & 0.99 & 0.98 & 0.98 & 0.98 & 0.98 & 0.97 & 0.98 & 1.00 & 1.00 & 1.00 & 0.98 & 0.97 & 0.98 & 0.99 & 0.97 & 1.00 & 1.00 & 1.00 & 1.00 \\
\rowcolor{greyL} {\cellcolor{greyL}} & FPR & 0.03 & 0.03 & 0.04 & 0.03 & 0.04 & 0.02 & 0.02 & 0.02 & 0.01 & 0.03 & 0.03 & 0.04 & 0.03 & 0.04 & 0.03 & 0.00 & 0.00 & 0.00 & 0.03 & 0.04 & 0.02 & 0.02 & 0.04 & 0.01 & 0.00 & 0.00 & 0.00 \\
\rowcolor{greyL} \multirow{-3}{*}{{\cellcolor{greyL}}\alg} & FNR & 0.01 & 0.02 & 0.00 & 0.00 & 0.02 & 0.00 & 0.00 & 0.00 & 0.00 & 0.01 & 0.01 & 0.00 & 0.00 & 0.03 & 0.00 & 0.00 & 0.00 & 0.00 & 0.01 & 0.01 & 0.02 & 0.00 & 0.02 & 0.00 & 0.00 & 0.00 & 0.00 \\ \hline

\end{tabular}
}
\end{table*}

\begin{table*}[!t]
\tiny
\centering
\addtolength{\tabcolsep}{-5.58pt}
\caption{Results of~\alg and baselines against adaptive poisoning attacks on SQuAD dataset.}
\label{tab:adaptive_squad}
\scalebox{0.873}{
\begin{tabular}{|c|c|ccccccccc|ccccccccc|ccccccccc|}
\hline
\multirow{2}{*}{Method} & \multirow{2}{*}{Metric} & \multicolumn{9}{c|}{Benign text perturbation} & \multicolumn{9}{c|}{Poisoned text perturbation} & \multicolumn{9}{c|}{Adversarial perturbation} \\ \cline{3-29} 
 &  & PRAGB & PRAGW & ProInject & HijackRAG & LIAR & Jamming & BadRAG & Phantom & AgentPoison & PRAGB & PRAGW & ProInject & HijackRAG & LIAR & Jamming & BadRAG & Phantom & AgentPoison & PRAGB & PRAGW & ProInject & HijackRAG & LIAR & Jamming & BadRAG & Phantom & AgentPoison \\ \hline
\multirow{3}{*}{Norm} & DACC & 0.56 & 0.66 & 0.59 & 0.63 & 0.61 & 0.53 & 0.49 & 0.53 & 0.65 & 0.56 & 0.61 & 0.61 & 0.60 & 0.62 & 0.59 & 0.49 & 0.48 & 0.59 & 0.52 & 0.57 & 0.59 & 0.55 & 0.58 & 0.50 & 0.47 & 0.47 & 0.49 \\
 & FPR & 0.00 & 0.00 & 0.00 & 0.00 & 0.00 & 0.00 & 0.06 & 0.06 & 0.04 & 0.00 & 0.00 & 0.00 & 0.00 & 0.00 & 0.00 & 0.06 & 0.06 & 0.04 & 0.00 & 0.00 & 0.00 & 0.00 & 0.00 & 0.00 & 0.06 & 0.06 & 0.04 \\
 & FNR & 1.00 & 1.00 & 1.00 & 1.00 & 1.00 & 1.00 & 1.00 & 1.00 & 1.00 & 0.99 & 0.99 & 1.00 & 1.00 & 1.00 & 1.00 & 1.00 & 1.00 & 1.00 & 0.99 & 0.98 & 0.99 & 0.98 & 1.00 & 1.00 & 1.00 & 1.00 & 1.00 \\ \hline
\multirow{3}{*}{PPL} & DACC & 0.56 & 0.66 & 0.59 & 0.63 & 0.62 & 0.53 & 0.51 & 0.57 & 0.68 & 0.56 & 0.62 & 0.61 & 0.60 & 1.00 & 0.59 & 0.51 & 0.50 & 0.62 & 0.58 & 0.95 & 0.62 & 0.97 & 1.00 & 0.98 & 1.00 & 1.00 & 1.00 \\
 & FPR & 0.00 & 0.00 & 0.00 & 0.00 & 0.00 & 0.00 & 0.00 & 0.00 & 0.00 & 0.00 & 0.00 & 0.00 & 0.00 & 0.00 & 0.00 & 0.00 & 0.00 & 0.00 & 0.00 & 0.00 & 0.00 & 0.00 & 0.00 & 0.00 & 0.00 & 0.00 & 0.00 \\
 & FNR & 1.00 & 1.00 & 1.00 & 1.00 & 1.00 & 1.00 & 1.00 & 1.00 & 1.00 & 1.00 & 0.95 & 1.00 & 1.00 & 0.00 & 1.00 & 1.00 & 1.00 & 1.00 & 0.88 & 0.10 & 0.93 & 0.06 & 0.00 & 0.05 & 0.00 & 0.00 & 0.00 \\ \hline
\multirow{3}{*}{SELF-RAG} & DACC & 0.80 & 0.82 & 0.76 & 0.66 & 0.74 & 0.69 & 0.53 & 0.58 & 0.78 & 0.85 & 0.83 & 0.82 & 0.75 & 0.79 & 0.70 & 0.51 & 0.50 & 0.62 & 0.78 & 0.78 & 0.73 & 0.55 & 0.75 & 0.50 & 0.50 & 0.50 & 0.51 \\
 & FPR & 0.20 & 0.17 & 0.26 & 0.26 & 0.15 & 0.03 & 0.00 & 0.00 & 0.00 & 0.23 & 0.20 & 0.24 & 0.29 & 0.17 & 0.02 & 0.00 & 0.00 & 0.00 & 0.26 & 0.21 & 0.27 & 0.32 & 0.22 & 0.04 & 0.00 & 0.00 & 0.00 \\
 & FNR & 0.20 & 0.22 & 0.21 & 0.48 & 0.44 & 0.64 & 0.97 & 0.98 & 0.68 & 0.04 & 0.12 & 0.09 & 0.19 & 0.28 & 0.70 & 1.00 & 1.00 & 1.00 & 0.17 & 0.23 & 0.27 & 0.61 & 0.29 & 0.97 & 1.00 & 1.00 & 0.99 \\ \hline
\multirow{3}{*}{RobustRAG} & DACC & 0.84 & 0.85 & 0.83 & 0.77 & 0.82 & 0.63 & 1.00 & 1.00 & 1.00 & 0.84 & 0.84 & 0.82 & 0.84 & 0.86 & 0.99 & 1.00 & 1.00 & 1.00 & 0.82 & 0.84 & 0.84 & 0.71 & 0.81 & 0.64 & 1.00 & 1.00 & 1.00 \\
 & FPR & 0.21 & 0.19 & 0.05 & 0.06 & 0.18 & 0.00 & 0.00 & 0.00 & 0.00 & 0.21 & 0.22 & 0.04 & 0.05 & 0.16 & 0.00 & 0.00 & 0.00 & 0.00 & 0.22 & 0.22 & 0.05 & 0.07 & 0.20 & 0.00 & 0.00 & 0.00 & 0.00 \\
 & FNR & 0.09 & 0.07 & 0.34 & 0.53 & 0.18 & 0.79 & 0.00 & 0.00 & 0.00 & 0.09 & 0.07 & 0.38 & 0.33 & 0.10 & 0.03 & 0.00 & 0.00 & 0.00 & 0.13 & 0.09 & 0.32 & 0.54 & 0.17 & 0.73 & 0.00 & 0.00 & 0.00 \\ \hline
\multirow{3}{*}{PFDNN} & DACC & 0.53 & 0.54 & 0.58 & 0.54 & 0.54 & 0.53 & 0.83 & 0.83 & 0.72 & 0.75 & 0.71 & 0.71 & 0.71 & 0.72 & 0.85 & 0.99 & 0.98 & 0.93 & 0.88 & 0.86 & 0.80 & 0.85 & 0.82 & 0.90 & 0.99 & 0.99 & 0.96 \\
 & FPR & 0.60 & 0.55 & 0.55 & 0.52 & 0.54 & 0.46 & 0.27 & 0.20 & {0.36} & 0.38 & 0.44 & 0.39 & 0.40 & 0.41 & 0.26 & 0.02 & 0.04 & {0.11} & 0.23 & 0.25 & 0.32 & 0.28 & 0.31 & 0.20 & 0.03 & 0.03 & {0.09} \\
 & FNR & 0.29 & 0.29 & 0.24 & 0.35 & 0.32 & 0.48 & 0.06 & 0.12 & {0.10} & 0.08 & 0.05 & 0.13 & 0.12 & 0.07 & 0.00 & 0.00 & 0.00 & {0.01} & 0.00 & 0.00 & 0.04 & 0.01 & 0.00 & 0.00 & 0.00 & 0.00 & 0.00 \\ \hline
\multirow{3}{*}{RAGForensics} & DACC & 0.94 & 0.95 & 0.98 & 0.83 & 0.89 & 0.53 & 0.50 & 0.52 & 0.78 & 0.99 & 0.99 & 0.98 & 0.96 & 0.96 & 0.79 & 0.60 & 0.62 & 0.73 & 0.98 & 0.98 & 0.98 & 0.96 & 0.98 & 0.67 & 0.41 & 0.38 & 0.72 \\
 & FPR & 0.03 & 0.03 & 0.03 & 0.03 & 0.03 & 0.06 & 0.05 & 0.08 & 0.03 & 0.02 & 0.02 & 0.03 & 0.04 & 0.03 & 0.06 & 0.05 & 0.08 & 0.03 & 0.03 & 0.03 & 0.03 & 0.05 & 0.03 & 0.07 & 0.05 & 0.08 & 0.02 \\
 & FNR & 0.07 & 0.06 & 0.01 & 0.23 & 0.15 & 0.58 & 0.62 & 0.59 & 0.36 & 0.00 & 0.00 & 0.00 & 0.03 & 0.04 & 0.29 & 0.52 & 0.50 & 0.38 & 0.00 & 0.00 & 0.00 & 0.02 & 0.01 & 0.42 & 0.71 & 0.72 & 0.39 \\ \hline

\rowcolor{greyL} {\cellcolor{greyL}} & DACC & 1.00 & 0.98 & 1.00 & 0.99 & 0.99 & 0.99 & 1.00 & 1.00 & 1.00 & 0.99 & 0.98 & 0.98 & 0.98 & 0.97 & 0.98 & 1.00 & 1.00 & 1.00 & 1.00 & 0.98 & 0.99 & 0.99 & 0.99 & 1.00 & 1.00 & 1.00 & 1.00 \\
\rowcolor{greyL} {\cellcolor{greyL}} & FPR & 0.01 & 0.03 & 0.00 & 0.01 & 0.01 & 0.02 & 0.00 & 0.01 & 0.01 & 0.02 & 0.03 & 0.03 & 0.03 & 0.04 & 0.03 & 0.01 & 0.01 & 0.00 & 0.01 & 0.03 & 0.02 & 0.01 & 0.02 & 0.01 & 0.01 & 0.01 & 0.00 \\
\rowcolor{greyL} \multirow{-3}{*}{{\cellcolor{greyL}}\alg} & FNR & 0.00 & 0.00 & 0.00 & 0.00 & 0.00 & 0.00 & 0.00 & 0.00 & 0.00 & 0.00 & 0.00 & 0.00 & 0.00 & 0.00 & 0.00 & 0.00 & 0.00 & 0.00 & 0.00 & 0.00 & 0.00 & 0.00 & 0.00 & 0.00 & 0.00 & 0.00 & 0.00 \\ \hline

\end{tabular}
}
\end{table*}

%% file: meta_review.tex
\clearpage

\begin{balance}
\section{Meta-Review}

The following meta-review was prepared by the program committee for the 2026
IEEE Symposium on Security and Privacy (S\&P) as part of the review process as
detailed in the call for papers.

\subsection{Summary}

This paper proposes RAGOrigin: a defense for identifying poisoned documents responsible for corrupting a RAG output. RAGOrigin begins by identifying a range of candidate documents that might be responsible for a given poisoned generation (identified via user feedback). Next, the method assesses the likelihood that a candidate document will produce the poisoned response by combining three metrics into a responsibility score. RAGOrigin shows strong results against many existing and adaptive attacks.

\subsection{Scientific Contributions}
\begin{itemize}

\item
Provides a Valuable Step Forward in an Established Field.

\end{itemize}

\subsection{Reasons for Acceptance}

\begin{enumerate}

\item 
RAGOrigin is an effective defense that combines three metrics that assess the responsibility of a document for a misgeneration event from both retrieval and generation perspectives, resulting in a more robust defense than any single metric could achieve.

\item 
RAGOrigin is an intuitive and easy-to-implement defense. It builds on existing ideas and refines them to address their weaknesses.

\item 
RAGOrigin is extensively evaluated on a wide range of experiments against known and adaptive attacks (e.g., multihop attacks, prompt injection-based attacks) and ablation studies.

\end{enumerate}

\subsection{Noteworthy Concerns}

\begin{enumerate}

\item 
RAGOrigin might have limited scalability. When its high-parallelizability is leveraged, RAGOrigin introduces minor latency; however, without such parallelization (e.g., in compute-limited settings), the latency could become prohibitive.

\item 
RAGOrigin assumes that misgeneration events (that initiate the defense) are honestly reported. This might make RAGOrigin vulnerable to false-flag attacks that aim to eliminate clean documents by misleading the defense. Moreover, this assumption might be impractical in real-world settings with untrusted users.

\end{enumerate}

\end{balance}